\documentclass[fleqn,usenatbib]{mnras}

\usepackage{newtxtext,newtxmath}

\usepackage{amsmath,bm}

\usepackage[T1]{fontenc}

\DeclareRobustCommand{\VAN}[3]{#2}
\let\VANthebibliography\thebibliography
\def\thebibliography{\DeclareRobustCommand{\VAN}[3]{##3}\VANthebibliography}

\usepackage{graphicx}	
\usepackage{amsmath}	

\usepackage[flushleft]{threeparttable}
\usepackage{booktabs,caption}
\usepackage{array}
\usepackage{upgreek}

\newcolumntype{P}[1]{>{\centering\arraybackslash}p{#1}}

\usepackage{svg}

\usepackage{gensymb} 

\usepackage[markup=final]{changes}

\setauthormarkup{\relax}             
\setdeletedmarkup{} 

\definechangesauthor[color=black]{anon}

\usepackage{adjustbox}
\usepackage{booktabs}
\usepackage{siunitx}
\usepackage{tabularx}

\newcommand\HItitle{H{\sc i}}

\newcommand{\orcid}[1]{\href{https://orcid.org/#1}{\includesvg[width=10pt]{orcid}}}

\definecolor{orange}{rgb}{1.0,0.5,0.}

\DeclareMathOperator{\sech}{sech}

\def\fdisc{\ifmmode{\>f_{\rm disc}}\else{$f_{\rm disc}$}\fi}
\def\Rd{\ifmmode{\>R_{\rm d}}\else{$R_{\rm d}$}\fi}
\def\md{\ifmmode{\>m_{\rm d}}\else{$m_{\rm d}$}\fi}

\def\MDM{\ifmmode{\>M_{\textnormal{\sc dm}}}\else{$M_{\textnormal{\sc dm}}$}\fi}

\def\XH{\ifmmode{\>X_{\textnormal{\sc h}}} \else{$X_{\textnormal{\sc h}}$}\fi}
\def\nH{\ifmmode{\>n_{\textnormal{\sc h}}} \else{$n_{\textnormal{\sc h}}$}\fi}

\def\maspyr{\ifmmode{\>\textnormal{mas~yr}^{-1}}\else{mas~yr$^{-1}$}\fi}

\def\mG{\ifmmode{\>\mu\mathrm{G}}\else{$\mu$G}\fi}
\def\erg{\ifmmode{\> {\rm erg}}\else{erg}\fi}
\def\keV{\ifmmode{\> {\rm keV}}\else{keV}\fi}

\def\deg{\ifmmode{\>^{\circ}}\else{$^{\circ}$}\fi}
\def\onedeg{\ifmmode{\>1^{\circ}}\else{$1^{\circ}$}\fi}

\def\xvir{\ifmmode{\>\!x_{vir}}\else{$x_{vir}$}\fi}
\def\Mvir{\ifmmode{\>\!M_{vir} }\else{$M_{vir} $}\fi}
\def\rvir{\ifmmode{\>\!r_{vir}}\else{$r_{vir}$}\fi}
\def\vvir{\ifmmode{\>\!v_{vir}}\else{$v_{vir}$}\fi}
\def\Vvir{\ifmmode{\>\!V_{vir} }\else{$V_{vir} $}\fi}

\def\tratio{\ifmmode{\>\tau}\else{$\tau$}\fi}

\def\rms{\ifmmode{\>r_{\textnormal{\sc ms}}}\else{$r_{\textnormal{\sc ms}}$}\fi}

\def\Mpc{\ifmmode{\>\!{\rm Mpc}} \else{Mpc}\fi}
\def\kpc{\ifmmode{\>\!{\rm kpc}} \else{kpc}\fi}
\def\pc{\ifmmode{\>\!{\rm pc}} \else{pc}\fi}

\def\pkpc{\ifmmode{\>\!{\rm kpc}^{-1}} \else{kpc$^{-1}$}\fi}

\def\Gyr{\ifmmode{\>\!{\rm Gyr}} \else{Gyr}\fi}
\def\Myr{\ifmmode{\>\!{\rm Myr}} \else{Myr}\fi}
\def\yr{\ifmmode{\>\!{\rm yr}} \else{yr}\fi}
\def\pyr{\ifmmode{\>\!{\rm yr}^{-1}}\else{yr $^{-1}$} \fi}
\def\s{\ifmmode{\>\!{\rm s}}\else{s}\fi}
\def\ps{\ifmmode{\>\!{\rm s}^{-1}}\else{s$^{-1}$}\fi}
\def\Hz{\ifmmode{\>\!{\rm Hz}}\else{Hz}\fi}

\def\kms{\ifmmode{\>\!{\rm km\,s}^{-1}}\else{km~s$^{-1}$}\fi}

\def\K{\ifmmode{\>\!{\rm K}}\else{K}\fi}

\def\sr{\ifmmode{\>\!{\rm sr}}\else{sr}\fi}
\def\psr{\ifmmode{\>\!{\rm sr}^{-1}}\else{sr$^{-1}$}\fi}
\def\arcs{\ifmmode{\>\!{\rm arcsec}}\else{arcsec}\fi}
\def\parcs{\ifmmode{\>\!{\rm arcsec}^{-1}}\else{arcsec${-1}$}\fi}
\def\parcss{\ifmmode{\>\!{\rm arcsec}^{-2}}\else{arcsec${-2}$}\fi}

\def\cm{\ifmmode{\>\!{\rm cm}}\else{cm}\fi}
\def\cc{\ifmmode{\>\!{\rm cm}^{3}}\else{cm$^{3}$}\fi}
\def\sqc{\ifmmode{\>\!{\rm cm}^{2}}\else{cm$^{2}$}\fi}
\def\pcc{\ifmmode{\>\!{\rm cm}^{-3}}\else{cm$^{-3}$}\fi}
\def\psc{\ifmmode{\>\!{\rm cm}^{-2}}\else{cm$^{-2}$}\fi}

\def\g{\ifmmode{\>\!{\rm g}}\else{g}\fi}
\def\Msun{\ifmmode{\>\!{\rm M}_{\odot}}\else{M$_{\odot}$}\fi}
\def\hMsun{\ifmmode{\> h^{-1}{\rm M}_{\odot}}\else{$h^{-1}$M$_{\odot}$}\fi}

\def\Zsun{\ifmmode{\>\!{\rm Z}_{\odot}}\else{Z$_{\odot}$}\fi}

\def\Lsun{\ifmmode{\>\!{\rm L}_{\odot}}\else{L$_{\odot}$}\fi}

\def\rayl{\ifmmode{\>\!{\rm R}}\else{R}\fi}
\def\mR{\ifmmode{\>\!{\rm mR}}\else{mR}\fi}

\renewcommand{\ion}[2]{\hbox{#1\,{\sc #2}}}

\def\lya{\ifmmode{\>\!{\rm Ly}\alpha}\else{Ly$\alpha$}\fi}

\def\Ha{\ifmmode{\>\!{\rm H}\alpha}\else{H$\alpha$}\fi}
\def\Hb{\ifmmode{\>\!{\rm H}\beta}\else{H$\beta$}\fi}

\def\HI{\ifmmode{\> \textnormal{\ion{H}{i}}} \else{\ion{H}{i}}\fi}
\def\HII{\ifmmode{\> \textnormal{\ion{H}{ii}}} \else{\ion{H}{ii}}\fi}
\def\CIV{\ifmmode{\> \textnormal{\ion{C}{iv}}} \else{\ion{C}{iv}}\fi}
\def\SiIV{\ifmmode{\> \textnormal{\ion{S}{iv}}} \else{\ion{Si}{iv}}\fi}

\def\NH{\ifmmode{\> {\rm N}_{\rm H}} \else{N$_{\rm H}$}\fi}
\def\Ng{\ifmmode{\> {\rm N}_{\rm gas}} \else{N$_{\rm gas}$}\fi}
\def\NHI{\ifmmode{\> {\rm N}_{\HI}} \else{N$_{\HI}$}\fi}
\def\MHI{\ifmmode{\> {\rm M}_{ \HI}} \else{M$_{\HI}$}\fi}

\def\mua{\ifmmode{\>\mu_{ \textnormal{\Ha}}}\else{$\mu_{ \textnormal{\Ha}}$}\fi}
\def\alphabha{\ifmmode{\>\alpha_{B}^{(\textnormal{\Ha})}}\else{$\alpha_{B}^{(\textnormal{\Ha})}$}\fi}

\newcommand{\nexus}{{\sc Nexus}}
\newcommand{\ramses}{{\sc Ramses}}
\newcommand{\agama}{{\small AGAMA}}
\newcommand{\Rb}{R_{\rm b}}
\newcommand{\gaia}{{\em Gaia}}

\pdfpageattr{/Group << /S /Transparency /I false /K false >>}

\title[Terminal Velocities from N-body/hydro models]{\HItitle\ terminal velocity curves $-$ Lessons learned from N-body/hydrodynamic `surrogate' models of the Milky Way}

\author[]{
Hillary Davis,$^{1}$
\thanks{Corresponding author: Hillary Davis.\newline  Email: hdav7324@uni.sydney.edu.au}
Thor Tepper-Garc\'ia$,^{1}$
Naomi McClure-Griffiths,$^{2}$
Joss Bland-Hawthorn,$^{1}$
\newauthor
and Oscar Agertz$^3$
\\
$^{1}$Sydney Institute for Astronomy, School of Physics, University of Sydney, NSW 2006, Australia\\
$^{2}$Mount Stromlo Observatory, Cotter Road, Weston Creek, ACT 2611\\
$^5$Lund Observatory, Division of Astrophysics, Department of Physics, Lund University, Box 43, SE-22100 Lund, Sweden
}

\date{Accepted XXX. Received YYY; in original form ZZZ}

\pubyear{\the\year{}}

\begin{document}
\label{firstpage}
\pagerange{\pageref{firstpage}--\pageref{lastpage}}
\maketitle

\begin{abstract}
The development of an N-body/hydrodynamic `surrogate' model of the Milky Way (MW) $-$ a model that resembles the MW in several key aspects after many Gyrs of evolution $-$ would be extremely beneficial for Galactic Archaeology. Here we present four new `surrogate' models, all built with the \nexus{} framework. The simulations contain stars, dark matter and gas. Our most sophisticated model allows gas to evolve thermodynamically, and includes star formation, metal production, and stellar feedback. The other three models in this work have an isothermal gas disc. We examine these new simulations in the context of cold gas observations of the Galaxy. Our focus is the so-called `\HI\ terminal velocity curve' $-$ a heliocentric measurement of the maximum $V_{\rm los}$ as a function of Galactic longitude $\ell$, which dates back to the early days of radio astronomy. It is a powerful approach to indirectly estimating the gas dynamics because it does not require knowledge about the distance to individual gas clouds, which is difficult to estimate. A comparison of the terminal velocities and recovered rotation curve values in the simulations against observations suggests that our models are in need of further refinement. The gravitational torques associated with our synthetic bars are too strong, driving excessive streaming motion in the inner gas disc. This causes the simulated terminal velocity curves in the Galactic Quadrant I and IV to deviate substantially from each other, unlike what is seen in observed \HI\ terminal velocities of the MW. We suggest possible ways forward for future models.

\end{abstract}

\begin{keywords}
Galaxy: kinematics and dynamics -- galaxies: bar -- software: simulations -- Galaxy: structure -- hydrodynamics -- ISM: kinematics and dynamics
\end{keywords}



\section{Introduction}\label{sec:intro}

Astronomers have always faced challenges studying the structure of the Milky Way (MW) because of our position embedded in the Galactic disc. When we look along a line-of-sight (LOS), we observe the superposition of different structures, whose individual properties are then difficult to separate \citep{hou2014observed, bland2016galaxy}. Circular rotation curves are one of the best tools we have for probing the Galactic gravitational potential, which in turn determines the nature of stellar orbits and the overarching structure of the Milky Way \citep{mcmillan2016mass, ou2024dark, fich1989rotation, levine2008milky}. Since the first detections of atomic hydrogen (\HI) and carbon monoxide (CO) spectral lines in the years 1951 and 1970, respectively, many authors have used observed gas velocities to derive rotation curves \citep{ewen1951observation, wilson1970carbon, gunn1979global, clemens1985massachusetts, dame1987composite, mcclure2007milky, sofue2009unified, mcclure2023atomic, ou2024dark}. Traditionally, we record observations of gas emission line intensity in $(\ell,V_{\rm los})$ diagrams, where $\ell$ is Galactic longitude and $V_{\rm los}$ is line-of-sight velocity. The terminal velocity $(V_{\rm t})$ curve is given by tracing along the outer envelope of an $(\ell, V_{\rm los})$ distribution, or equivalently by estimating the maximum absolute line-of-sight velocity as a function of longitude \citep{bissantz2003gas}. Unfortunately, there is no means of deprojecting an  $(\ell, V_{\rm los})$ diagram into a unique distribution of gas in $(x,y)$ space \citep{weiner1999properties}.  However, if we assume that gas experiences purely circular motion, then we can directly convert measured $(\ell, V_{\rm t})$ values into a rotation curve $(R, \Theta(R))$. The challenge is that gas does not experience purely circular motion in the Galaxy. For instance, we observe gas with velocities $V_{\rm los} > 0$ when $\ell < 0$, and $V_{\rm los} < 0$ when $\ell > 0$, which is ``forbidden'' by circular motion, i.e. it is inconsistent with gas moving toward and away from us on circular orbits \citep{fux19993d}. \deleted[id = anon]{These forbidden velocities can be explained by roughly elliptic orbits induced in the gas by a barred potential.}
\added[id = anon]{These forbidden velocities can be explained by non-circular orbits of a barred potential, and their magnitude can provide insight into bar strength}. Numerous studies have thus exploited observed \HI\ and CO terminal velocities to constrain the properties of the Galactic bar, such as its `strength,' `length,' and pattern speed\footnote{We use quotes to stress that there is no general agreement on how to define these parameters \citep[e.g.][]{iles2025b}. This is an unresolved issue in all galactic bar studies.} \citep{fux19993d, englmaier1999gas, weiner1999properties, sormani2015gas, li2022gas}.

\subsection{Gas flows and the Galactic bar}

At the moment, we do not know with confidence whether the Milky Way bar is rather long and slow \citep[half-length $\Rb \approx 5$~kpc and pattern speed $\Omega_{\rm p} \approx 35 - 45$~\kms~\pkpc; e.g.][]{wegg2015structure} or short and fast \citep[$\Rb \approx 3.5$~kpc and $\Omega_{\rm p} \approx 50- 60$~\kms~\pkpc;][]{hilmi2020fluctuations,vislosky2024gaia}. \citet{wegg2015structure} used red clump giant (RCG) stars from the United Kingdom Infrared Deep Sky Survey \citep[UKIDSS;][]{2007MNRAS.379.1599L}, the Two Micron All Sky Survey \citep[2MASS;][]{2MASS}, the Vista Variables in the Via Lactea survey \citep[VVV;][]{VVV_ref} and the Galactic Legacy Infrared Mid-Plane Survey Extraordinaire \citep[GLIMPSE;][]{glimpse} to argue that the MW's central box/peanut bulge is the vertical extension of a longer bar with two key components: a thin (scaleheight $z_{\rm b} \approx 180$~\pc) bar with $\Rb = 4.6 \pm 0.3$~kpc, and a superthin ($z_{\rm b} \approx 45$~\pc) component that extends  to $\Rb = 5.0 \pm 0.2$~kpc. However, \citet{hilmi2020fluctuations} explain that a bar can appear to be up to double its actual size when attached to a spiral arm. \citet{vislosky2024gaia} compare simulations against \gaia\ Data Release 3 (DR3) data \citep{vallenari2023gaia} to argue that the MW bar may be as short as $\Rb \approx 3$~kpc with a moderate strength spiral structure in the inner disc, or otherwise, up to $\Rb \approx 5.2$~kpc with weaker spiral arms that are likely in the midst of disconnecting from the bar. Furthermore, \citet{lucey2023dynamically} use orbit integration techniques to suggest that the bar extends to a cylindrical radius $\Rb \approx 3.5$~kpc, with an overdensity of stars out to $\Rb \approx 4.8$~kpc possibly related to an attached spiral arm. \added[id = anon]{Studies of the Hercules stream have also been used in this debate about bar pattern speed. The dynamical origin of Hercules can be explained by: (i) The Outer Lindblad resonance (OLR) of a short, fast bar \citep{dehnen2000effect, antoja2014constraints, fragkoudi2019ridges} (ii) Alternatively, the CR of a longer, slower bar \citep{perez2017revisiting, monari2019signatures, chiba2021tree}. \citet{chiba2021tree} have suggested that Hercules stars have a metallicity gradient that can only be explained by the CR of a slow bar model. Furthermore, numerous other moving groups (eg. Sirius, the Hat and the Horn) have been used to constrain the bar pattern speed \citep{trick2021identifying, trick2022identifying, monari2019signatures}.}

\added [id = anon]{More relevant to this work, there is a long history of comparing hydrodynamical models of gas against observed $(\ell, V_{\rm los})$ distributions of \HI\ and CO in order to study the Galactic bar's properties}. \citet{binney1991understanding} explained the structure of $(\ell, V_{\rm los})$ diagrams in the central longitudes $|\ell| < 10 \degree$ and latitudes $|b| < 2 \degree$ by considering gas flows along closed orbits in a barred potential. \citet{sormani2015gas} use the barred Galactic potential of \citet{binney1991understanding} to perform hydrodynamical simulations of gas flow in the MW, with $\Omega_{\rm p} = 63$~\kms~\pkpc\ and a corotation radius, $R_{\rm CR} = 3.7$~kpc. Similarly, \citet{englmaier1999gas} took a fixed gravitational potential for stars, and used smoothed particle hydrodynamics \citep[SPH;][]{mon92a} to model the gas flow through and around a bar characterised by $\Omega_{\rm p} = 59 \pm 2$~kpc~\pkpc\ and $R_{\rm CR} = 3.5 \pm 0.5$~kpc. 

Notably, the models of \citet{sormani2015gas} and \citet{englmaier1999gas} fail to account for a number of key features: (i) the 3-kpc arm; and (ii) the extent of the forbidden velocities that we observe in \HI. \citet{fux19993d} suggested $\Omega_{\rm p} \approx 50$~\kms~\pkpc\ and $R_{\rm CR} \approx 4 - 4.5$~kpc better describe the bar, by constructing a model using N-body and SPH techniques, and focusing on comparison to high density gas features in the CO $(\ell, V_{\rm los})$ diagram. Meanwhile, \citet{weiner1999properties} simulated gas flow with a Eulerian grid code, and used lower density gas near the extreme velocity contour of \HI\ $(\ell, V_{\rm los})$ diagrams to estimate $\Omega_{\rm p} \approx 41.9$~\kms~\pkpc\ and $R_{\rm CR} = 5$~kpc. They argued that their estimate of the pattern speed is  lower than previous studies likely because they focus on comparison to \HI, which has much larger forbidden velocities than are seen in molecular gas. Additionally, \citet{sormani2015gas3} resolve the inability of the \citet{sormani2015gas} model to account for large forbidden velocities and the 3-kpc arm by varying the pattern speed of the bar, along with the quadrupole length and strength. In doing so, they argue for a longer, slower bar with $\Omega_{\rm p} = 40$~\kms~\pkpc. More recently, \citet{li2022gas} have used the made-to-measure method \citep[M2M;][]{syer1996made, de2007nmagic} to construct a dynamical model of a barred MW galaxy, and also obtained a slower pattern speed value of $\Omega_{\rm p} = 37.5 - 40 $~\kms~\pkpc\ . Clearly, we do not know the exact values for many of the Galactic bar's parameters, we only have a broad range of estimates. Here, we study the gas dynamics in our new `surrogate' models in order to add to the pre-existing literature, and to learn more about the MW bar's properties.

\subsection{The Galactic ISM}

The Milky Way's interstellar medium (ISM) is a multiphase, turbulent medium supported by star formation, but also by magnetic fields and cosmic rays. A `three-phase' model simplifies the structure of the ISM by suggesting that it consists of: a dense cold neutral medium (CNM), a warm medium with each neutral (WNM) and ionized (WIM) components, and a hot ionized medium (HIM) \citep[]{mckee1990three, mckee1977theory, vazquez2012there, mcclure2023atomic}. However, turbulent fluctuations in pressure are one example of a mechanism making the ISM's structure more complex since they cause the characteristic temperature and density of each locally stable phase to vary from one location to another \citep{vazquez2012there, mcclure2023atomic}.

\noindent
\textit{Galactic Ecology}. The cycling of material between star formation and stellar feedback processes drives phase changes and shapes the ISM's structure. This interplay between star formation, turbulence and phase structure is challenging to fully capture in a model. \added[id = anon]{Multiphase ISM models have been adopted in many studies \citep[e.g.][]{holmberg2000local}. However, in the context of constraining bar properties with gas dynamics, much of the historical work \citep{englmaier1999gas, weiner1999properties, sormani2015gas, li2022gas,fux19993d} has assumed an isothermal equation of state $-$ the internal energy of the gas is taken to be constant everywhere in space and for all time; in other words, all the gas in the simulation is set to have a single, constant temperature, which often equals the mean temperature of just one phase of the ISM \citep{kissmann2008local, wadA2001numerical}. In this paper, we improve upon earlier work and compare the effect of star-forming versus isothermal gas in ``surrogate'' MW models, when using these models to constrain the Galactic bar.}  


The structure of the paper is as follows. We begin by discussing in Sec.~\ref{sec:simple_model} a simple schematic model assuming the gas features a perfectly circular, planar geometry in $(x,y)$ configuration space in order to build an intuition as to how the phase space coordinates $(x,y,V_x,V_y)$ of the gas maps onto $(\ell, V_{\rm los})$. In Sec.~\ref{s:sim}, we introduce four new controlled simulations of idealised, isolated MW-like galaxies used in our analysis, all created with the \nexus{} framework \citep{tep24a}. In Sec.~\ref{sec:Results}, we argue why our simulations are a satisfactory representation (i.e. surrogates) of the Galaxy. In turn, we construct $(\ell, V_{\rm los})$ diagrams and terminal velocity curves from our simulations, and compare these against observations. Next, we examine some properties of the bars in our simulations in light of the corresponding gas dynamics, and how these may influence bar evolution. Finally, in Sec.~\ref{sec:future}, we suggest ways forward for future attempts at an improved N-body/hydrodynamical surrogate of the MW, and we summarise our key conclusions in Sec.~\ref{sec:Conclusions}.\\

\section{A Simple Analytic Model}\label{sec:simple_model}

\begin{figure*}
    \centering
    \includegraphics[width = 0.8\textwidth]
    {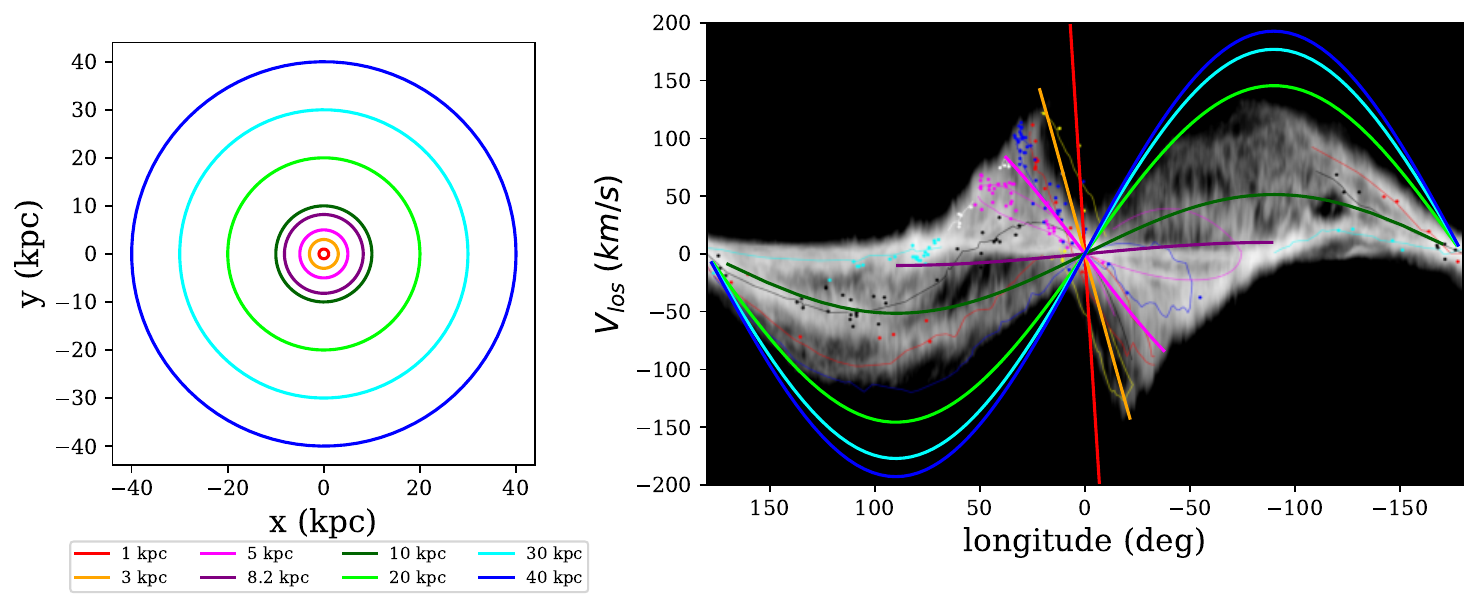}
    \caption{\added[id = anon]{The left panel displays gas moving at $\mathrm{230 \unit{km/s}}$ along circular rings of different radii in $(x,y)$ space}. The sun is located at $R_{\rm 0} = 8.2$~kpc and moves along the solar circle at speed $\Theta_0 = 240$~\kms. The purple ring of gas is positioned at the same radius as the sun. The right panel shows these rings mapped into $(\ell, V_{\rm los})$ space. The rings within the solar circle clearly appear as short diagonal lines confined to a narrow longitude range in the $(\ell, V_{\rm los})$ diagram. Rings outside the solar circle in $(x,y)$ appear as larger sin shaped curves in $(\ell, V_{\rm los})$. \added[id = anon]{The curves in $(\ell, V_{\rm los})$ are overlaid on an observed distribution of \HI\ emission in the Milky Way, adapted from \citet{reid2019trigonometric}. The background greyscale shows the \HI\ brightness temperature, integrated over the central 10 degrees of latitude.}}
   \label{fig:simple_rings}
\end{figure*}

\begin{figure}
    \centering
    \includegraphics[width = 0.5\textwidth]{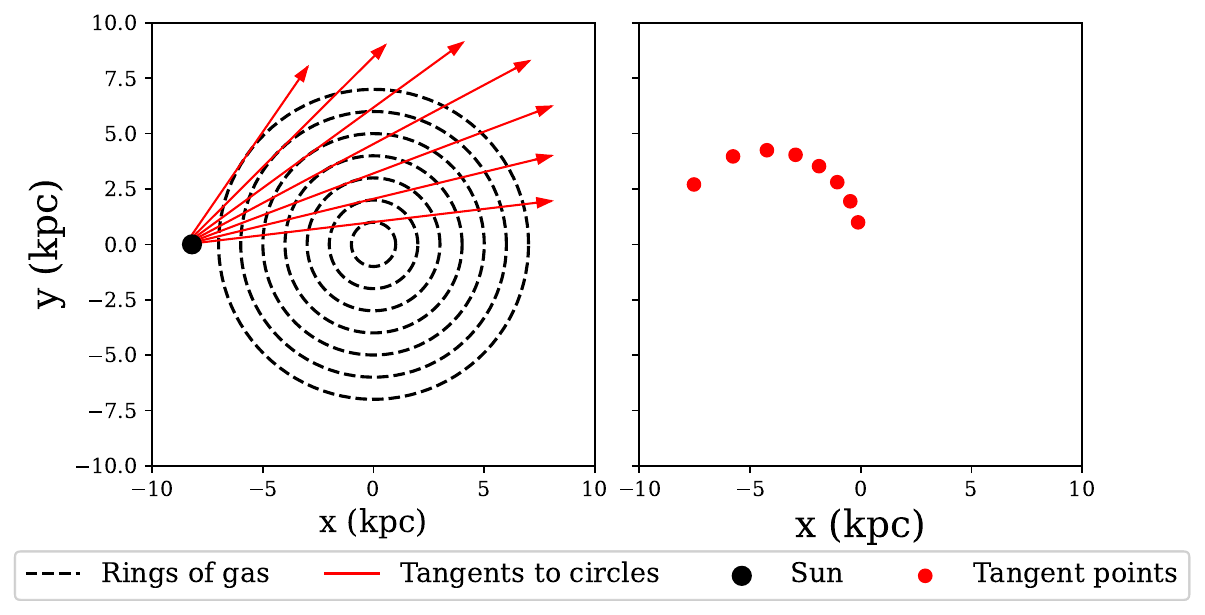}
    \caption{Left: Dashed circles indicate rings of radii $R = 1, 2, 3, 4, 5, 6$ and 7 kpc within the solar circle ($R_{\rm 0} = 8.2$~kpc; the location of the Sun is indicated by the black dot). For gas experiencing purely circular motion the maximum projected (i.e. terminal) velocity occurs at the point where the line-of-sight (red arrows) is a tangent to the relevant circular orbit. Right: The location of the gas with terminal velocities in $(x,y)$ space traces out an arc-like structure (red dots).}
    \label{fig:terminal_points}
\end{figure}

Observations of gas are recorded in $(\ell, V_{\rm los})$ diagrams. There is obviously no unique projection of the $(\ell, V_{\rm los})$ distribution into $(x,y,V_x,V_y)$ phase space. However, we can still use these observations to learn about the MW's structure if we adopt some simplifying assumptions. For instance, if the gas is assumed to move on perfectly circular orbits (i.e. the radial velocity component $V_R \equiv 0$), then the line-of-sight velocity, $V_{\rm los}$, for the gas inside the solar circle ($R < R_{0}$), is given by: 
\begin{equation}\label{eq:vlos}
    V_{los} = \Theta(R)\cdot \frac{R_{0}}{R} \cdot \sin(\ell) \cos(b) - \Theta_{0}\cdot \sin(\ell)\cos(b)
\end{equation}

where $\Theta(R)$ is the circular speed of the gas about its orbit, $\mathrm{\Theta_{0}}$ is the circular speed of the Sun with respect to the Galactic Centre, $R_{\rm 0}$ is the distance of the Sun from the Galactic Centre, $R$ is the radius of gas, $\mathrm{\ell}$ is Galactic longitude and $b$ is Galactic latitude. More specifically, $\Theta(R) = R \omega$, where $\omega$ is the \added[id = anon]{angular velocity of rotation at the given radius \citep{burton1978carbon, clemens1985massachusetts, fich1989rotation}.} 

Let us consider a simple model, shown schematically in Figure ~\ref{fig:simple_rings}. The left panel displays several circular rings of gas in $(x,y)$ space, where each ring has a different radius. We imagine gas to be moving at a constant speed \added[id = anon]{($\Theta(R) = 230$~\kms)} around each ring. In the right panel, we have mapped these rings into $(\ell, V_{\rm los})$ space, and have overlaid the result on the observed distribution of \HI\ gas in the MW \citep{reid2019trigonometric}.

Using this simple model we can readily learn a great deal about the following \citep[cf.][]{englmaier1999gas}
\begin{itemize}
    \item \textit{ Gas within the solar circle $(R < R_{0})$}. This is the case that is most relevant here because much of this paper focuses on comparing simulated terminal velocities against observed data for gas within the solar circle.
    In Fig.~\ref{fig:simple_rings}, we see that gas on rings within the solar circle appear as diagonal lines confined to a short range of longitudes $(- 90 \degree < \ell < 90 \degree)$ in $(\ell, V_{\rm los})$ space. This can be understood by visualising various lines-of-sight at different galactic longitudes. For instance, when $\ell = 0 \degree$, gas on circular orbits will also have $V_{\rm los} = 0$. This is because the line-of-sight vector ($\overrightarrow{los}$) is perpendicular to both the velocity vectors of the gas and the Sun for $\ell = 0 \degree$. From $\ell = 0 \degree$, if we progressively increase $\ell$, $V_{\rm los}$ will increase up until the point where $\overrightarrow{los}$ is tangent to the relevant circular orbit $-$ here a terminal velocity or maximum in the absolute value of the velocity observed along $\overrightarrow{los}$ is reached. After this point, the $V_{\rm los}$ decreases until the gas velocity vector is anti-parallel to $\overrightarrow{los}$, yielding the minimum (most negative) velocity value. Overall, these diagonal lines are confined to a short range of longitudes because the corresponding circular orbits have small radii. We also see that gas on circular orbits within the solar circle are confined to a region where $V_{\rm los} > 0$ when $\ell > 0$, and $V_{\rm los} < 0$ when $\ell < 0$. This is consistent with gas moving toward and away from the Sun on circular orbits. As a result, gas with $V_{\rm los} > 0$ when $\ell < 0$, or $V_{\rm los} < 0$ when $\ell > 0$ is considered ``forbidden'' by circular motion. In this way, we can better understand how forbidden velocities in observed $(\ell, V_{\rm los})$ diagrams may be evidence for \added[id = anon]{non-circular orbits of a barred potential.} \deleted[id = anon]{elliptical orbits, possibly induced by the barred potential of the MW}.  

    \item \textit{Gas on the solar circle $(R = R_{0})$}. \added[id=anon]{In our simple model, we consider a ring of gas at the same radius as the Sun, with the gas travelling at a slightly slower speed than the Sun ($\Theta(R) = 230$~\kms, $\Theta_{\rm 0} = 240$~\kms). This is the purple circular orbit in Fig.~\ref{fig:simple_rings}. This ring maps across to an almost straight line in $(\ell, V_{\rm los})$ space. The line is slightly inclined because the gas is moving slower than the sun. If the gas speed was set equal to the sun speed ($\Theta(R) = \Theta_{\rm 0} = 240$~\kms), this would be a perfectly straight, horizontal line with constant $V_{\rm los} = 0$  because there is zero relative radial velocity between the Sun and the gas on this particular orbit \citep{englmaier1999gas}}.

    
    \item \textit{Gas outside the solar circle $(R > R_{0})$}. In $(\ell, V_{\rm los})$ space, orbits beyond the solar circle appear as sinusoidal curves that extend over a much wider range of longitudes $(-180 \degree \leq \ell \leq 180 \degree)$. \added[id = anon]{At $\ell = \pm 180^\circ$, the line-of-sight vector is perpendicular to the gas' velocity vector, resulting in a line-of-sight velocity of zero ($V_{\rm los} = 0$) at this longitude. In two dimensions, the projection of the gas velocity onto the line-of-sight is given by $\Theta(R)(R_0 / R)\sin \ell$. Consequently, as $\ell$ increases from $90^\circ$ to $180^\circ$, $V_{\rm los}$ decreases and approaches zero.}
    
\end{itemize}

In addition to understanding $(\ell, V_{\rm los})$ distributions, we need to consider terminal velocities ($V_{\rm t}$). A terminal velocity is defined as the maximum in the absolute value of the $V_{\rm los}$ for any given galactic longitude, and they are represented by the outer envelope of the distribution in $(\ell, V_{\rm los})$ space. In this paper, we compare terminal velocity curves extracted from our simulations (henceforth dubbed `simulated' or `synthetic') to {\em observed} \HI\ terminal velocities for the MW. In the case of purely circular motion, the terminal velocity occurs when the line-of-sight is tangent to the relevant circular orbit because $\overrightarrow{los}$ is parallel, or anti-parallel, to the gas velocity vector $\overrightarrow{V}$ at these points.  

This leads us to the question: Where does the gas with these terminal velocity values appear in $(x,y)$ configuration space? Fig.~\ref{fig:terminal_points} provides the answer. If the terminal velocities occur at the tangent points to circular orbits, then the gas giving us these terminal velocities in Quadrant 1 ($0 \degree < \ell < 90 \degree$) will be located along an arc as shown in Fig.~\ref{fig:terminal_points}. However, when gas does not experience purely circular motion, the position of the gas with terminal velocities in $(x,y)$ space will no longer trace out a perfect arc.

Many studies have used observed terminal velocity values when attempting to derive the Galactic circular speed curve \citep{gunn1979global, clemens1985massachusetts, mcclure2007milky, sofue2009unified, mcclure2023atomic, ou2024dark}. These authors simplify the problem, by assuming circular motion and that the terminal velocity occurs at the tangent point, in which case $R = R_{0}\sin(\ell)$. Substituting this into Eq.~\ref{eq:vlos} and rearranging, we arrive at the following expression, which allows us to estimate circular speed values from terminal velocities \citep[e.g.][]{mcclure2016milky}: 

\begin{equation}\label{eq:rot_curve}
    \Theta(R) = |V_{t}| + \Theta_{0}|\sin{\ell}|
\end{equation}
and $R_{\rm t} = R_{\rm 0}\sin(\ell)$.\\

Thus, observed terminal velocity values are useful for probing the Galactic potential. At the same time, this method for deriving the MW's Rotation Curve is limited because it relies on the assumption of gas experiencing purely circular motion, which does not hold true in the MW \citep[e.g.][their appendix]{dri23a}.

\section{N-body/hydrodynamic Simulations}
\label{s:sim}

\subsection{Galaxy model}\label{sec:Galaxy_model}

Our ultimate aim is to understand the flow of gas close the plane of the Galaxy as encoded in the observed $(\ell, V_{\rm los})$ diagram. To this end, we have designed a series of simulations that are as simple as possible while retaining the drivers of gas flows believed to be key, such as a full-scale potential that resembles the MW's both locally (i.e. close to the plane) and globally. Our most sophisticated simulation in addition accounts for the multiphase nature and clumpy structure of the gas, as explained below.

In all our simulations, the MW is approximated by a four-component system: 1) a dark matter (DM) host halo; 2) a central, pre-assembled stellar bulge; 3) a pre-assembled stellar disc; 4) a gas disc; all of these are responsive (`live') and massive (i.e. with self-gravity). Our base galaxy model is defined by the relevant properties of each of these components (mass, structural parameters) as given in Tab.~\ref{t:comp}. This model is our reference model, henceforth referred to as `Model 1'.
\deleted[id = anon]{Initially, the stellar disc's
Toomre Q parameter is $Q(R) \gtrsim 1.3$ at all radii across the disc
}.
The stellar disc's vertical velocity dispersion declines with radius, reaching $\sigma_z \approx 21$~\kms, and a scaleheight $h \approx 0.45$~kpc, both at $R = 8.2$~kpc. The gas disc is initially isothermal with $T = 10^3$~K. We note that this model is similar to the bar unstable,`dry' (i.e. without gas) galaxy model introduced by \citet{tep21v}, but with an additional gaseous disc component. Model 1 is evolved allowing the gas to cool and heat, and to form stars (more details are provided below).

We also consider the following variations of Model 1, which have helped us study the effect of both gas turbulence and temperature (for each the stellar and gaseous discs) on gas flows: 

\begin{enumerate}
    \item Model 2: Is evolved using a strictly isothermal equation of state, i.e. the gas is kept at $T = 10^3$~K at all times, and it is not allowed to form stars.
    \item Model 3: Similar to Model 2, but adopting a hotter gas disc, with a temperature $T = 10^4$~K.
    \item Model 4: Same as Model 3, and additionally a stellar disc with a higher vertical velocity dispersion throughout, reaching $\sigma_z \approx 35$~km/s and a scaleheight $h \approx 0.9$~kpc, both at $R = 8.2$~kpc.
\end{enumerate}

Given the belief that the Galactic bar has a strong effect on the gas dynamics, even
beyond co-rotation \citep{schwarz1981response, combes2004role}, we focus most of our following analysis on Model 1. The other models serve as controls that allow us to isolate the effect that individual variations have on the evolution of the simulations.  

\begin{table*}
\begin{center}
\caption{Relevant parameters of reference model (Model 1). Columns 1 and 2 identify the galactic components and their associated (target) functional forms. The total mass, scale length and cut-off radius are indicated in columns 3, 4, and 5, respectively. Column 6 is the number of collisionless particles used in the simulation (halo, bulge, disc) or to sample the initial gas disc distribution.
}
\label{t:comp}
\begin{tabular}{llccccc}
Component & Profile & Total mass & Radial scalelength & Cut-off radius & Particle count \\
 &  & $M_{\rm tot}$  & $r_s$ & $r_c$ & $N$ \\
 &  & ($10^{10}$~\Msun) & (kpc) & (kpc) & ($10^5$) \\
\hline
DM halo & NFW & 118 & 19 & 250 & 10 \\
Stellar bulge & Hernquist & 1.25 & 0.6 & 2 & 1 \\
Stellar disc & Exp, $\sech^2$ & 4.3 & 2.5 & -- & 10 \\
Gas disc & Exp, $\sech^2$ & 0.46 & 3.5 & -- & 20 \\
\hline
\end{tabular}
\end{center}
\begin{list}{}{}
\item Notes: The NFW and Hernquist functions are defined elsewhere \citep[][respectively]{nav97a,her90a}. \deleted[id = anon]{The stellar disc's radial Toomre parameter is everywhere $Q \gtrsim 1.3$}. The stellar disc's vertical velocity dispersion declines with radius, reaching $\sigma_z \approx 21~\kms$ and a scaleheight $h \approx 0.45$~kpc at $R = 8.2$~kpc. The gas disc is {\em initially} isothermal with $T = 10^3$~K. Neither the stellar disc nor the gas disc need be artificially truncated, given their exponentially declining density profiles. Models 2-4 are variations of this model.
\end{list}
\end{table*}

\subsection{Initial conditions and evolution} \label{s:evol}

Our models are created and evolved with the \nexus{} framework \citep{tep24a}. In brief, the initial conditions (particle positions and velocities) of each of the components making up our synthetic Galaxy  are generated with the self-consistent modelling module (SCM) provided by the Action-based GAlaxy Modelling Architecture software package \citep[\agama; ][]{vas19a}, which have been complemented to include gas in addition to the standard treatment of collisionless components.

The initial conditions are evolved with the adaptive mesh refinement (AMR), N-body/hydrodynamical code \ramses\ \citep{tey02a}, augmented with a proprietary module to account for galaxy formation physics \citep[q.v.][]{age21l}.

All models are evolved in a cubic simulation volume with a length of 600~ckpc per side. The total simulation timespan for each model is about 4 Gyr. At runtime, the AMR grid is maximally refined up to level 13, implying a limiting spatial resolution of \mbox{600~kpc / $2^{13}  \approx 72$ pc}. As a result, the vertical structure of the gas disc is not resolved at $R \lesssim 8.2$~kpc in Model 1 and Model 2, and barely so in Models 3-4. This circumstance is irrelevant in the case of Model 1 (since the gas becomes turbulent as a result of the stellar activity), but it may be somewhat important for Models 2 - 4.

We refer the reader to \citet{tep24a} for more details on \nexus, such as details about the thermodynamic treatment of the gas, or the prescription to include galaxy formation physics.

\section{Results}\label{sec:Results}

Before presenting our main results, we discuss whether our N-body/hydrodynamic models are a reasonable representation of the Milky Way, focusing on a few key aspects that are relevant to the analysis of gas kinematics -- in particular its terminal velocity, such as the mass and structure of each galactic component, and the properties of the emerging central bar.

\begin{table}
    \begin{threeparttable}
        \caption{Comparison of relevant parameters in the MW and Model 1.
        All values in this table are computed at a simulation time of $t = 1.99$~Gyr, and the MW values have been taken from \citet{bland2016galaxy}, unless stated otherwise.} \label{tab:sim_comparison}
            \begin{tabular}{ccccc}
                Component & Observable & Model & MW \\
                \hline
                DM halo \\
                \vspace{0.05cm} & $M_{vir} \; (10^{12} \unit{M_{\odot}})$  & 1.18 & $1.3 \pm 0.3$ \\
                \vspace{0.05cm}  & $R_{vir} \unit{(kpc)}$ & 274 & $282 \pm 30$ \\
                Stellar disc \\
                \vspace{0.05cm} & $M_{tot} (10^{10} \unit{M_{odot}})$ & 4.3 & $3.7 \pm 0.1^{\text{b}}$\\
                \vspace{0.05cm} & $R_{d} \unit{kpc}$ & $2.5$ & $2.6 \pm 0.5$\\
                \vspace{0.05cm} & $h \unit{(kpc)}$ & $0.45^{\text{a}}$ & $0.3 \pm 0.05$ \\
                \vspace{0.05cm} & $\sigma_{R} \unit{(km s^{-1})}$ & $49.29^{\text{a}}$ & $35 \pm 5^{\text{c}}$ \\
                \vspace{0.05cm} & \vspace{0.05cm} & \vspace{0.05cm} & $50 \pm 5$ \\
                \vspace{0.05cm} & $\sigma_{z} \unit{(km s^{-1})}$ & $27.25^{\text{a}}$ & $25 \pm 5^{\text{c}}$ \\
                \vspace{0.05cm} & \vspace{0.05cm} & \vspace{0.05cm} & $50 \pm 5$ \\
                \vspace{0.05cm} & $\Sigma_{*} (M_{\odot}pc^{-2})$ & $60.37$ & $38 \pm 4$\\
                \vspace{0.05cm} & $f_{d}^{\text{d}}$ & $0.47^{\text{e}}$ & $0.53 \pm 0.1$\\
                Bar\\
                \vspace{0.05cm} & Size (kpc) & $2.8 \unit{kpc}^{\text{g}}$ & $-$\\
                \vspace{0.05cm} & $\Omega_{\rm p} \unit{(kms^{-1}kpc^{-1})}$ & \vspace{0.05cm} & \vspace{0.05cm} \\
                \vspace{0.05cm} &  at 2.4 Gyr & $45$ & $-$\\
                \vspace{0.05cm} &  at 4.3 Gyr & $42$ & $-$ \\
                Bar region \\
                $R \lesssim 5 \unit{kpc}$ \\
                \vspace{0.05cm} & Stellar mass $(10^{10} \unit{M_{\odot}})$ & $3.5^{\text{f}}$ & 3.17 \\
                $R \lesssim 2 \unit{kpc}$ \\
                \vspace{0.05cm} & Stellar mass $(10^{10} \unit{M_{\odot}})$ & 2.3 & $-$ \\
                \vspace{0.05cm} & Dynamical mass $(10^{10} \unit{M_{\odot}})$ & $2.7^{\text{f}}$ & $1.85 \pm 0.05$ \\
                \vspace{0.05cm} & Baryon Fraction & $0.85^{\text{f}}$ & $0.83 \pm 0.15$ \\
            \end{tabular}
        \begin{tablenotes}
            \item a) Measured at $R_{0} = 8.2 \unit{kpc}$. b) We have only one stellar disc in our model since we do not need thin-thick disc decomposition to explore the in-plane gas dynamics. The MW value given here includes $6 \pm 3 \times 10^{9} \unit{M_{\odot}}$ from the thick disc \citep{tep21v}. c) Top value is for MW thin disc. Value in row below is for thick disc. d) Measured at $2.2 R_{d}$. e) Measured for $t = 0$~Gyr. f) \citet{portail2017chemodynamical}. g) Computed using the code written by \citet{dehnen2023measuring}. For this snapshot, the \citet{dehnen2023measuring} bar length value is consistent with the \deleted[id = anon]{the} bar semi-major axis length measured by visual inspection from the centre of the simulation to the edge of the cusped $x_{1}$ orbit, as seen in gas $(x,y)$ density projections. 
        \end{tablenotes}
        \end{threeparttable}
\end{table}

\subsection{Quality of models relative to the Milky Way}\label{sec:barred_surrogate}

\begin{table}
    \begin{center}
        \caption{Comparing Models 1 - 4.}
        \begin{tabular}{|P{2 cm}| P{2 cm} |P{2.1 cm}|}
        \hline
        Model Number & When is a Bar present? (Gyr) & Comment \\
        \hline
        \hline
        1 & $ \approx 0.6 - 3.95$ & Bar IS present for most of the sim \\
        \hline 
        2 & $ \approx 0.49 - 1.20$ & Bar gets destroyed quickly \\
        \hline
        3 & $\approx 0.48 - 1.84$ & Bar gets destroyed quickly \\
        \hline 
        4 & no bar & A bar never forms \\
        \hline 
        \end{tabular}\label{tab:bar_comparison}
    \end{center}
\end{table}

\begin{figure*}
    \centering
    \includegraphics[width=0.8\textwidth]{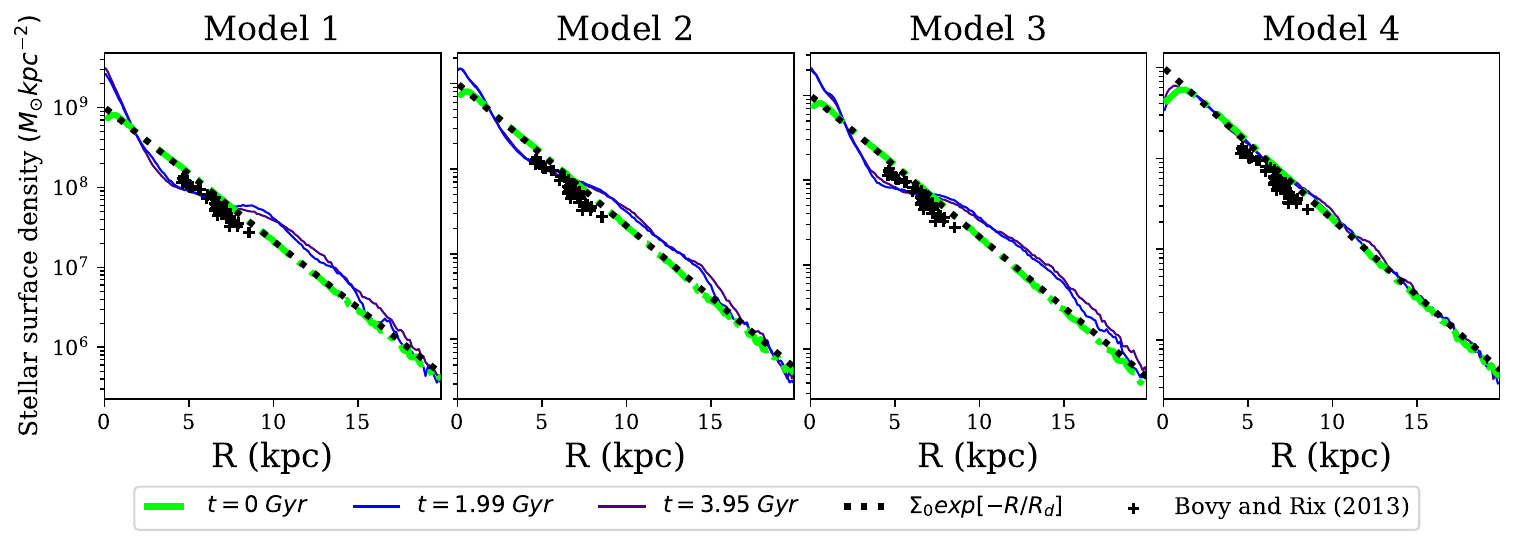}
    \caption{Stellar surface densities for all the simulations evolving over time. We consider three simulation epochs: $t = 0, 2, 4$~Gyr. The black dashed line is identical in all panels, and it represents an exponentially declining surface density profile with $\Sigma_{\rm 0}\exp[-R/R_{d}]$ where $\Sigma_{\rm 0} = 10^{9} \unit{M_{\odot} \; kpc^{-2}}$ and $R_{\rm d} = 2.6 \unit{kpc}$. The black crosses are identical across panels, and they indicate the stellar surface density estimates from \citet{bovy2013direct}, assuming an equal contribution of baryons and dark matter to local acceleration. Clearly, the stellar surface density profile in the simulations is broadly consistent with the observed stellar density profile. }
    \label{fig:surface_density}
\end{figure*}

\subsubsection{Mass and structure of the various galaxy components}

In Tab. \ref{tab:sim_comparison}, we compare some key observed properties of \deleted[id = anon]{the} the MW against equivalent values computed for Model 1. These values are calculated at a simulation time of $t \approx 2$~Gyr of Model 1, at which point a bar has fully formed, and continues to grow between $t \approx 2 - 4$~Gyr in the simulation.\footnote{We can see that a bar is fully formed at this time simply by visual inspection of the $(x,y)$ stellar and gas density distributions. We will address the bar formation process quantitatively in Section \ref{sec:fourier}.} Reassuringly, many of the model's key features are reasonably consistent with the MW's (cf. Tab.~4 in \citealt{tep21v}). The parameter with the largest deviation from the observed MW value is the stellar surface density $\Sigma_\star$, where $\Sigma_{\rm *,model} = 60.37 \unit{M_{\odot} \; pc^{-2}}$ while $\Sigma_{\rm *,MW} = 38 \pm 4 \unit{M_{\odot} \; pc^{-2}}$  \citep[measured at $R_{\odot} = 8.2$~kpc;][]{bovy2013direct}. To investigate this difference further, we consider Fig.~\ref{fig:surface_density} where we plot the stellar surface density profile at three different epochs: $t = 0$~Gyr (green curve), $t \approx 1.99$~Gyr (blue curve), $t \approx 3.95$~Gyr (purple curve). We also plot $\Sigma_{\rm 0}\text{exp}[-R/R_{\rm d}]$, where $\Sigma_{\rm 0} = 10^{9} \unit{M_{\odot} \; kpc^{-2}}$ and $R_{\rm d} = 2.6 \unit{kpc}$, as the black dotted line in Fig.~\ref{fig:surface_density}. \citet{bovy2013direct} have measured the vertical force at heights above the plane $|z| \approx 1$~kpc for $R \approx 4 - 9$~kpc, and used this to estimate the radial surface density near the disc. We include their estimates for stellar surface density using the black crossed markers, assuming an equal contribution of baryons and dark matter to the total local disc potential \citep[cf.][]{bland2016galaxy}; i.e. we scale the \citet{bovy2013direct} measurements by half.

In Fig.~\ref{fig:surface_density}, we can see that the surface density profile at $t = 0$~Gyr agrees well (by design) with the MW's disc. However, as the disc evolves, dynamical (and to some extent, numerical) instabilities kick in, and the disc develops \deleted[id = anon]{substructure} \added[id = anon]{some key structures (eg. spiral arms, and in some models a bar also forms)} \deleted[id = anon]{, with a bar also forming in some of the models}. We see the surface density profile begin to deviate from an exponential profile, and it develops a series of wiggles with systematically lower (higher) densities at radii below (above) $R_0$. This is true for all models. However, overall we find that the surface density profile of the synthetic stellar disc in all models is reasonably consistent with the MW disc's profile at all epochs.

\begin{figure}
    \centering
    \includegraphics[width=0.5\textwidth]
    {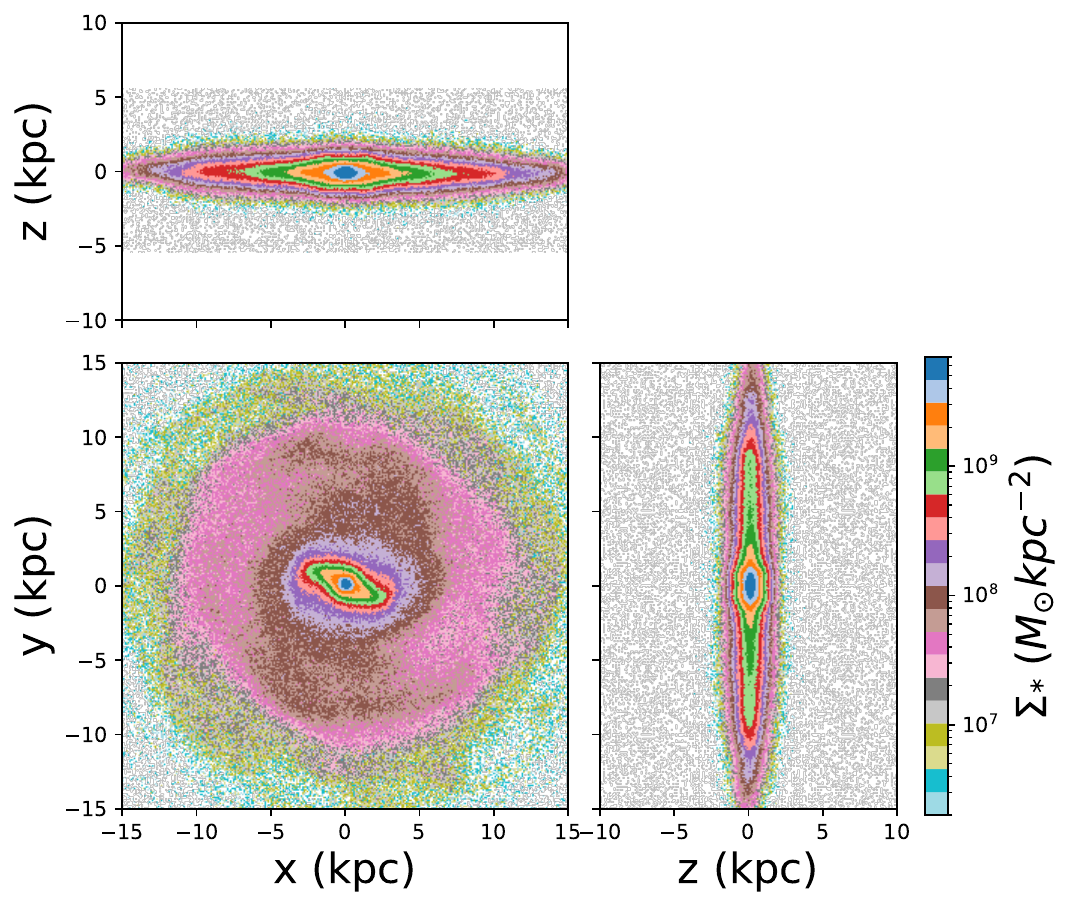}
    \caption{This shows face-on and edge-on stellar density projections of $t = 2.0$~Gyr from Model 1. We can see a distinct bar at the centre of the disc. The bar has already buckled and hence we also observe a boxy-peanut bulge when looking at the edge-on profiles.}
    \label{fig:stellar_density}
\end{figure}

\begin{figure*}
    \centering
    \includegraphics[width=\textwidth]
    {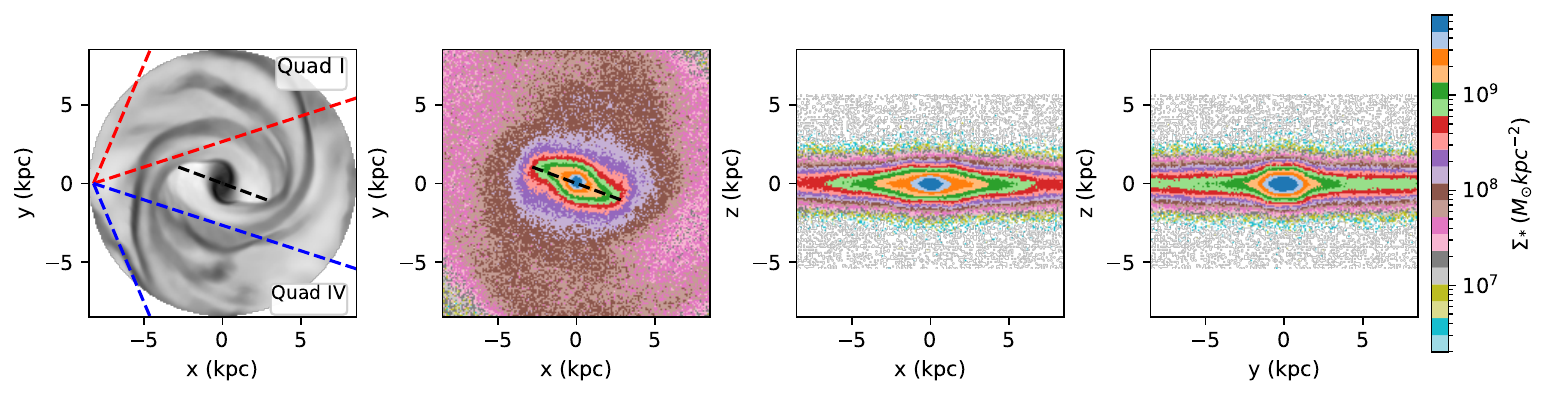}
    \caption{The first panel shows the distribution of gas within the solar circle for $t = 2.0$~Gyr. The stellar bar has been positioned at a $25 \degree$ inclination angle to the line connecting the Sun's position at $(-8.5,0)$~kpc and the galaxy's barycentre (at the origin). The black dotted line indicates the bar's semi-major axis. The dashed red and blue lines mark out the locations of the $18 \degree < \ell < 67 \degree$, and $-67 \degree < \ell < -18 \degree$ regions for Quadrant I and IV, respectively. These are the longitude and radius ranges for which the Quadrant I and IV \HI\ terminal velocities of \citet{mcclure2016milky} have been measured. The second to fourth panels present a zoomed-in view of the stellar density maps from Fig.~\ref{fig:stellar_density}. The bar and its B/P structure are apparent. }
    \label{fig:quadrants}
\end{figure*}

\begin{figure*}
    \centering
    \includegraphics[width=\textwidth]{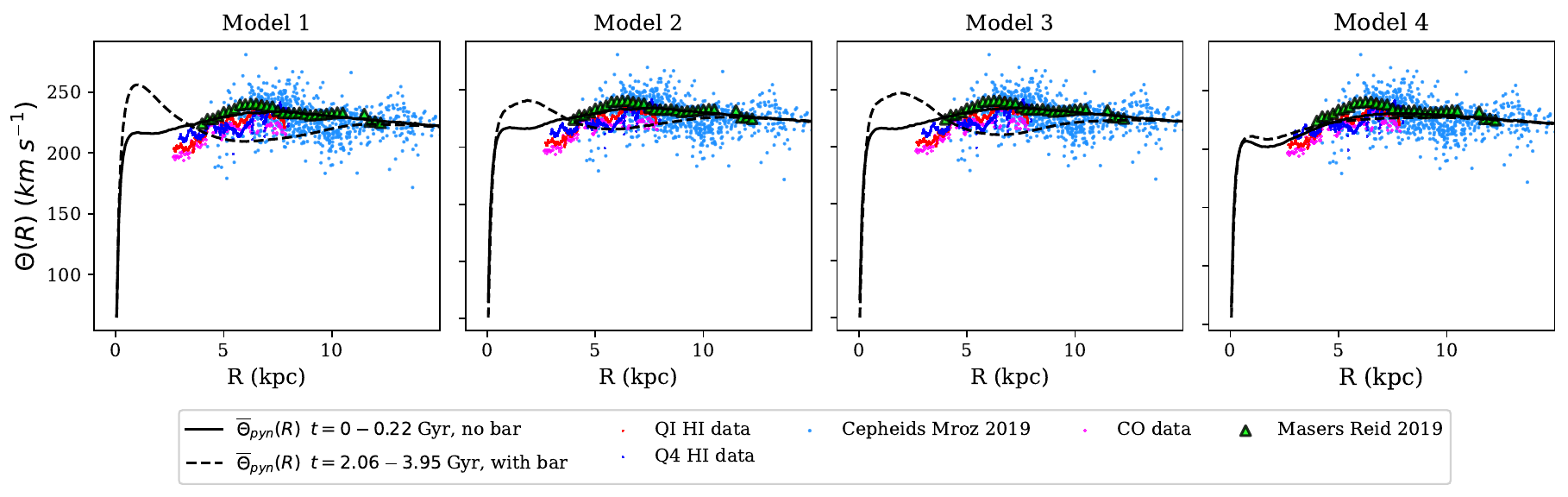}
    \caption{\added[id = anon]{Mean azimuthally-averaged circular rotation speed $\overline{\Theta}_{\rm sim}(R)$ for $t = 0 - 0.22$~Gyr (black solid curve) and $t = 2.06 - 3.95$~Gyr (black dashed curve) in the four simulations}.The colours are various $\Theta (R)$ values from observational data. Models 1 - 3 all form a bar at roughly the same epoch, while Model 4 never forms a bar. We can see that the bar formation process leads to a significant redistribution of mass in the disc such that it is transferred from the outer regions ($R \gtrsim 8$~kpc) to the centre}.
    \label{fig:circ_vel_curves}
\end{figure*}

\subsubsection{The central bar and bulge structure}

Next, we consider the emergent bar. Model 1 forms a central stellar bar (which also leaves a distinct imprint on the gas; see below) early on, and once formed, this remains in place for the duration of the simulation ($\sim 3.35$~Gyr).\footnote{The bar in this model has been discussed in some detail in \citealt[][their sec. 6]{tep24a}.} While Models 2 and 3 both form bars as well, these are relatively short-lived ($\sim 0.7 - 1.4$~Gyr; cf. Tab.~\ref{tab:bar_comparison}). Model 4 does not form a bar over the duration of the simulation. 
The differences in the lifetimes of bars for our models are summarised in Tab.~\ref{tab:bar_comparison}.
These outcomes are intriguing given the apparently small differences between the models (cf. Sec.~\ref{sec:Galaxy_model}), and require an explanation. We will come back to this point in Sec.~\ref{sec:bars}.

In Fig.~\ref{fig:stellar_density}, we present the stellar density distribution of Model 1 at $t \approx 2$~Gyr, with the face-on (bottom-left), and edge-on (top-left, bottom-right) projections shown. We can see a distinct bar at the centre of the disc. The bar has already buckled and hence we also observe a boxy-peanut bulge when looking at the edge-on profiles.

Fig.~\ref{fig:quadrants} includes a zoomed-in view of the panels presented in Fig.~\ref{fig:stellar_density}. The side on stellar-density projections (middle, right) show a boxy/peanut (B/P) structure, which grows as Model 1 evolves.  This feature of Model 1 (and many of the other models) is reassuring because the MW itself has long been known to have a B/P structure. Indeed, the boxy nature of the MW's bulge was first unveiled by COBE satellite data \citep{weiland1994cobe, dwek1995morphology, binney1997photometric},
and later confirmed using star counts from the 2MASS survey \citep{2MASS}. More recently, \citet{wegg2013mapping} used three-dimensional (3D) density maps of red clump giant (RCG) stars from the Vista Variables in VVV survey to show that the MW's central barred region has a distinct peanut shape when viewed side on \citep[cf.][]{bland2016galaxy}.

In the first panel of Fig.~\ref{fig:quadrants}, we present the projected distribution of {\em total} gas within the solar circle at a single snapshot of Model 1. The stellar bar is oriented at a $25 \degree$ angle with respect to the $x = 0$ axis, and its major axis is indicated by the black dotted line. The stellar bar's dynamics leaves a distinct imprint on the gas. We see a dense, nuclear ring at the centre of the gas distribution, surrounded by a low density gas region, which is likely encased by the cusped $x_1$ orbit family. The dashed red and blue lines mark out the locations of the $18 \degree < \ell < 67 \degree$, and $-67 \degree < \ell < -18 \degree$ regions conventionally denoted to be part of `Quadrant I' and `Quadrant IV', respectively. These correspond to the two longitude and radius ranges in which the \HI\ terminal velocities of \citet{mcclure2016milky} have been measured $-$ the latter corresponds the main set of observational data we use in this paper.

\subsubsection{The galactic potential}

In Fig.~\ref{fig:circ_vel_curves}, we compare the circular speed curves ($V_c^2 = R d\Phi/dR$) from the simulations with the observed data, which are effectively a proxy for the galactic potential. The solid black curve displays the circular speed curve averaged over a short time span after the beginning of each simulation ($t = 0 - 0.22$~Gyr), during which time \textit{none of the models have a bar}. The dashed black curve displays the mean circular speed at a later epoch ($t = 2.06 - 3.95$~Gyr), during which time \textit{Model 1 does have a bar, while all the other models do not} (Model 2 and 3 bars have already been destroyed, Model 4 never develops a bar). The coloured points correspond to the observational data: $\Theta(R)$ values derived from Quadrant I and IV neutral atomic hydrogen (\HI) terminal velocities \citep{mcclure2016milky} are shown in red and dark blue respectively; pink dots correspond to carbon-monoxide (CO) data \citep{clemens1985massachusetts, mcclure2016milky}; light blue dots are derived from Cepheids \citep{mroz2019rotation}; and green triangles from Masers \citep{reid2019trigonometric}.

For Models 1-3, the simulated circular velocity curves agree well with the observed data initially. However, as the models evolve, we see the simulated $\Theta(R)$ values begin to deviate more from the observed data: the simulated $\Theta(R)$ values rise to reasonably high values toward the centre and consequentially, the $\Theta(R)$ values also decrease significantly further out in the disc. This change in the circular rotation curve can be explained with respect to bar formation. Once a bar has formed, gravitational torques associated with the bar cause gas to flow inward. This inflowing gas often feeds star formation in the nuclear ring \citep{kormendy2004secular}, particularly relevant for Model 1 \citep[see][their sec.~6]{tep24a}. These processes cause a build up of a central mass concentration. This redistribution of mass into the central regions explains the  changes in the simulated circular rotation curves, where $\Theta(R)$ rises to higher values at small radii $R \lesssim 8$~kpc, and decreases at larger radii. This redistribution of mass in the simulated galaxies (with more mass toward the centre, and mass depleted from further out in the disc) prevails in Model 2 and 3 long after their bars are destroyed, and despite their bars surviving only for a short period of time. The deviation between modelled versus observed circular velocity data that arises as Models 1–3 evolve poses a problem for our models, which we discuss more later in this paper. 

\begin{figure*}
    \centering
    \includegraphics[width=0.8\textwidth]{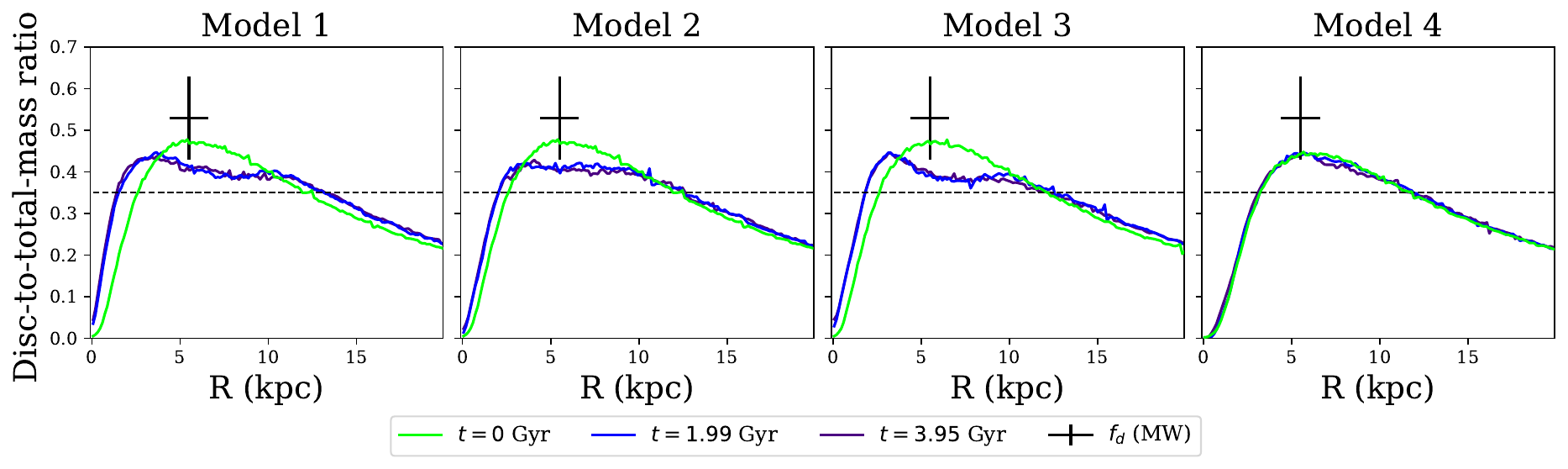}
    \caption{Formally, the disc-to-total mass ratio, $f_{\rm d}$ is defined at a radius of $2.2 R_{d}$. The black cross is the observed MW $f_{\rm d}$ value of $f_{\rm d} = 0.53 \pm 0.1$ measured at $2.2 R_{\rm d} \approx 5.7 \unit{kpc}$, where $R_{\rm d} = 2.6 \pm 0.5 \unit{kpc}$ \citep{bland2016galaxy, tep21v}. For the green, blue and purple curves, we have used Equation \ref{eq:fd} to compute $f_{\rm d} (R)$ at various radii. We directly compare  $f_{\rm d}$  for the simulation against the observations at the $2.2 R_{d}$ radius. The simulated $f_{\rm d}$ values are consistent with this observed data point for $t = 0$~Gyr. However, for Models 1 - 3, $f_{\rm d}$ decreases as the models evolve causing $f_{\rm d}$ to dip slightly below the observed value in these simulations. }
    \label{fig:fd}
\end{figure*}

\begin{figure*}
    \centering
    \includegraphics[width=0.8\textwidth]{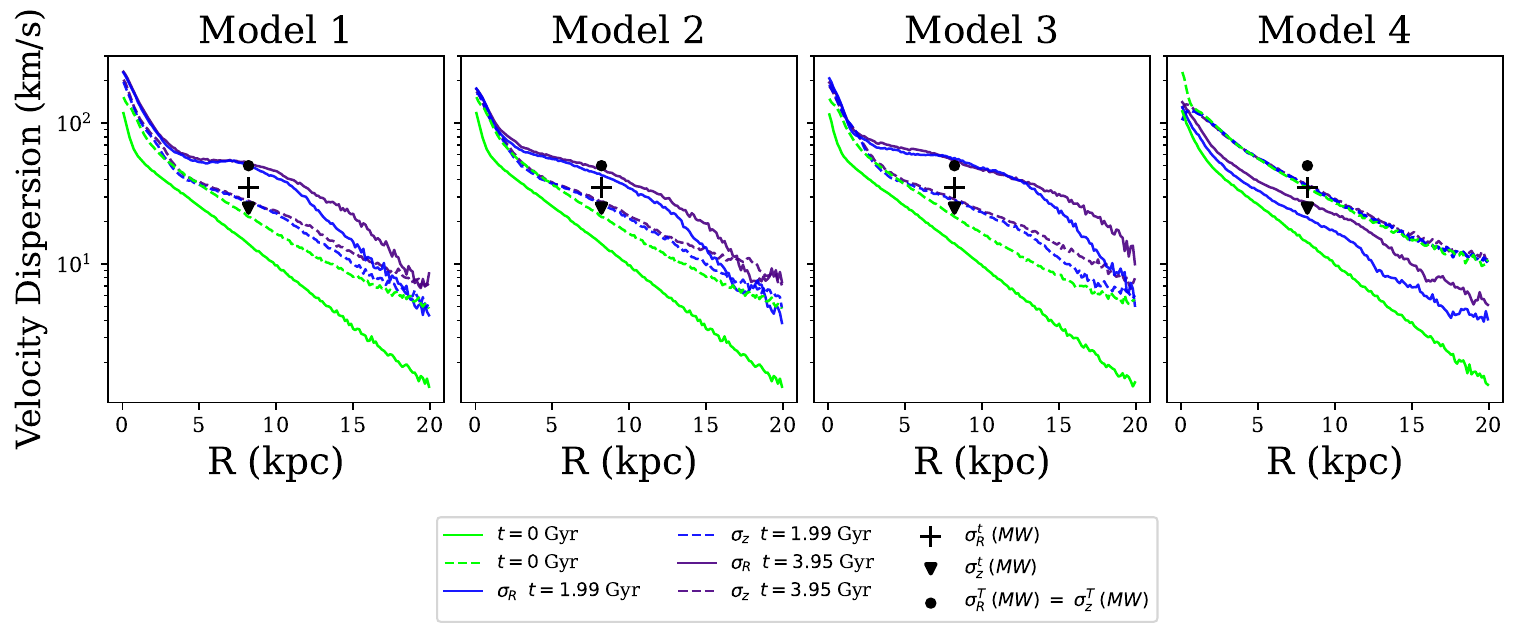}
    \caption{Radial and vertical stellar velocity dispersion for each simulation, evolving over time. The $\sigma_{\rm R}$ values increase significantly as a bar forms and evolves in Models 1-3 as a result of the non-axisymmetric bar transforming ordered rotational kinetic energy into chaotic components of velocity, and hence heating the stellar disc. $\sigma_{\rm R}$ profile changes the least for Model 4 over time since a bar never forms in this model. In all models, the $\sigma_{\rm R}$ values increase first toward the centre of the galaxy, where the bar is, and later increase more at larger radii. \added[id = anon]{$\sigma_{R}^{t}$ and $\sigma_{z}^{t}$ are velocity dispersion values measured for the thin disc of the MW; $\sigma_{R}^{T}$ and $\sigma_{z}^{T}$ are for the thick disc. All MW velocity dispersion values are from \citet{bland2016galaxy}.}}
    \label{fig:vel_dispersions}
\end{figure*}

\subsubsection{Disc-to-total mass ratio}
The mass fraction of disc baryons relative to the total galaxy mass within a radius of $2.2 R_{\rm disc}$, denoted by $f_d$, is defined by \citep{fujii2018dynamics},
\begin{equation} \label{eq:fd}
    f_{\rm d} = \Bigg( \frac{V_{\rm c,disc}(R)}{V_{\rm c,tot}(R)}  \Bigg)^{2}_{R = 2.2R_{\rm disc}} \,
\end{equation}
Here, $V_{\rm c,disc} (R_{\rm s})$ and $V_{\rm c,tot}(R_{\rm s})$ are the circular speeds of the disc and of the entire galaxy.

The $f_d$ parameter has been shown to be of fundamental importance in relation to the stability of discs against bar formation \citep{fujii2018dynamics, bla23a,bland2024turbulent}. Crucially, \citet{fujii2018dynamics} observe an inverse relation between stellar bar formation timescale $\mathrm{\tau_{\rm bar}}$ and $f_{\rm d}$, the so-called `Fujii relation' \citep[cf.][]{bla23a}. The latter suggests that values $f_{\rm d} \lessapprox 0.35$ corresponds to discs which are ``stable'' against bar formation; more precisely, it implies that the disc will not develop a bar within a Hubble time.
The opposite is true for values $f_{\rm d} \gtrapprox 0.35$, with the timescale for bar formation exponentially declining with $f_{\rm d}$ \citep{fujii2018dynamics, bla23a}.

In Fig.~\ref{fig:fd}, we present disc-to-total mass fraction profiles and their corresponding $f_{\rm d}$ values for each of our models. For reference, we include the observed MW $f_{\rm d}$ value of $f_{\rm d} = 0.53 \pm 0.1$ measured at $R \approx 5.7$~kpc, where we have assumed $R_{\rm d} = 2.6 \pm 0.5$~kpc \citep[black cross;][]{bland2016galaxy}.

The threshold $f_{\rm d}$ value for stability is indicated in Fig.~\ref{fig:fd} by the horizontal black dotted line. Initially, the $f_{\rm d}$ values of the synthetic disc in each model is consistent with the MW estimate. However, as the disc evolves (consider e.g. blue and purple curves), its current $f_{\rm d}$ values begin to drop. This is easily understood in terms of the dependence of $f_{\rm d}$ on the circular speed curve (Eq.~\ref{eq:fd}). As we have already seen in Fig.~\ref{fig:circ_vel_curves}, the total circular velocity curve in all models dips too low compared to observations for $R \approx 4.5 - 12$~kpc, which is accompanied by a drop in the corresponding $f_d$. \added[id = anon]{The dip in the rotation curve most likely appears because Model 1 does not have gas accretion $-$ as the bar forms and evolves, gas is funneled toward the galaxy centre, while the gas in the outer disc is depleted and not replenished. Alternatively, a modelled bar that is too strong may independently cause too significant a redistribution of mass, resulting in $V_{c}$ values that are too high at low radii, and $V_{c}$ values that are too low further out in the disc. There may be a combination of effects at play here.}\\

\subsubsection{Velocity Dispersion}
It is well-established that the 3D velocity dispersion of disc stars in the solar neighbourhood increase with stellar age \citep{spitzer1951possible, hanninen2002simulations, sahA2010effect}. Additionally, we know that this observed kinematic ``heating'' of disc stars is anisotropic. Indeed, the radial velocity dispersion of stars tends to increase much more substantially with stellar age compared to the vertical velocity dispersion.
Numerous theories have been proposed in attempt to explain observed heating of disc stars, such as: \added[id = anon]{interactions of disc stars with Giant Molecular Clouds (GMCs) \citep{spitzer1951possible, spitzer1953possible, lacey1984influence}}, spiral arms \citep{barbanis1967orbits, carlberg1985dynamical, fuchs2001density, minchev2006radial}, bars \citep{sahA2010effect, grand2016vertical}, massive dark halo objects (such as massive black holes in the Galactic halo \citep{lacey1985massive}, dark clusters of less massive objects \citep{carr1987dark}), minor mergers \citep{quinn1993heating}, ISM turbulence \citep{van2022giant} or combinations of different effects (eg. GMCs and spirals \citep{jenkins1990spiral}, halo black holes and GMCs \citep{hanninen2002simulations, hanninen2004numerical}). The upshot is that, regardless of the mechanism behind, we believe that any surrogate MW model should, at least broadly, reproduce these observations.

Fig.~\ref{fig:vel_dispersions} presents the radial and vertical velocity dispersions ($\sigma_{\rm R}$ and $\sigma_{\rm z}$) of disc stars in the simulations for $t = 0, 1.99, 3.95$~Gyr. The values for $\sigma_{\rm R}$ and $\sigma_{\rm z}$ values for the MW's thin and thick discs at the solar circle are also presented in Fig.~\ref{fig:vel_dispersions} \citep{bland2016galaxy}.
Comparing the initial dispersion profile with its counterpart at later times, we observe a significant increase in both the in-plane and the vertical directions, with \deleted[id = anon]{the} the in-plane component experiencing more heating, as is qualitatively consistent with observations. Furthermore, heating occurs first close to the centre of the disc, and once the velocity dispersion profile for small $R$ has reached a quasi-stationary state, the heating becomes apparent at larger $R$ as well \citep[cf.][]{khoperskov2003minimum}.

We see that the velocity dispersion values at $R_{\rm 0} = 8.2$~kpc are roughly consistent with their corresponding observed values (black data points). We also see a correlation between the amount of disc heating in the radial direction and the strength of the bar that forms in each simulation. As we will see in Section \ref{sec:fourier}, Model 3 appears to have the strongest bar, even though this bar has a very short lifetime. Model 3 also experiences the largest amount of heating over its evolution compared to the other models, in both the radial direction and in the vertical direction.

Overall, our N-body, hydrodynamic models are a reasonable representation of the Milky Way, although there is of course room for improvement. By design, all models are a good match to observations at early times, relative to the key features discussed above. Crucially, as the simulation evolves, a bar and associated boxy/peanut structure forms, both of which are observed features of the Milky Way. However, the bar in our model causes a significant redistribution of mass in the Galactic disc, with mass building up in the centre of the model while being depleted from the outer disc. Simultaneously, many of the simulation's properties begin to deviate more from the observations. This deviation can be seen in the $\Sigma_{*}$ and $f_{d}$ profiles ($t = 1.99, 3.95$~Gyr in Figures \ref{fig:surface_density}, \ref{fig:fd}), and obviously in the circular speed curve (as seen for $t = 2.06 - 3.95$~Gyr in Figure \ref{fig:circ_vel_curves}). In the future, we want our model to be more consistent with the MW's observed properties after many billions of years of evolution. Obviously, this is challenging to do because the evolution of the model is a non-linear process, making it difficult to predict what will happen to the simulated galaxy's structure after it has been evolved for a long period of time. Nonetheless, as they stand we believe our models are reasonably good to carry on a comparative analysis of the gas kinematics on our simulations and observations.

\begin{figure*}
    \centering
    \includegraphics[width=0.9\textwidth]
    {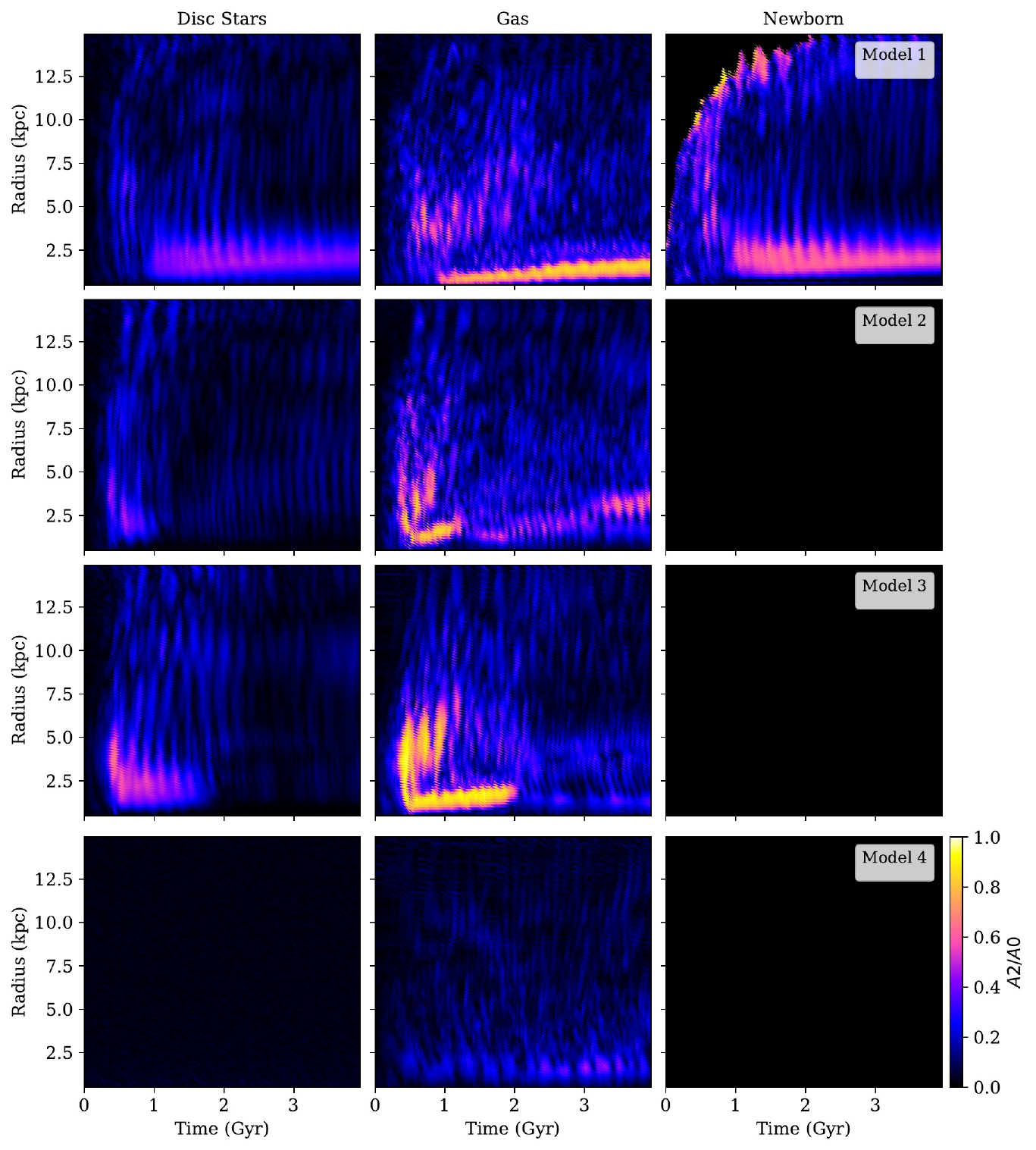}
    \caption{Periodograms of A$_2$ values for all 4 simulations. High A$_2$ values indicate the presence of a strong bisymmetric structure, i.e. spiral arms or a bar; the latter generally in the inner regions.
    }
    \label{fig:A2_fourier}
\end{figure*}

\begin{figure}
    \centering
    \includegraphics[width=0.5\textwidth]{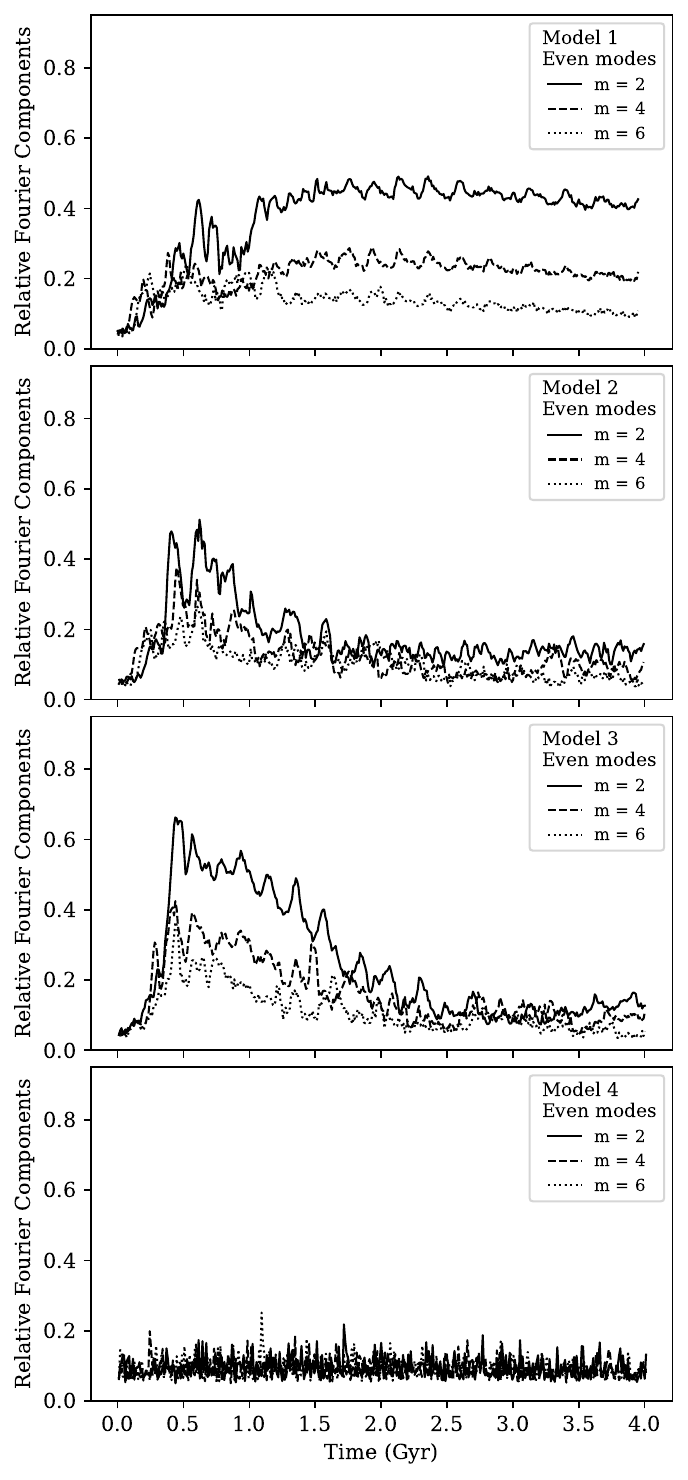}
    \caption{Even fourier modes evolving over time for all four models.}
    \label{fig:fourier_all}
\end{figure}

\begin{table}
    \centering
    \caption{\added[id = anon]{Definitions of key parameters used in this paper.}}
    \label{tab:symbols}
    \begin{tabular}{l>{\raggedright\arraybackslash}p{6.5cm}} 
        \hline
        \textbf{Symbol} & \textbf{Definition} \\
        \hline
        $V_{t}$ & Terminal velocity \\
        $\Theta_{\text{TP}}(R)$ & Rotation curve derived via the tangent-point method \\
        $\Theta_{\text{sim}}(R)$ & Circular velocity curve from \texttt{pynbody}, based on azimuthally averaged ($m = 0$) mass distribution: $\sqrt{R \, \partial \Phi_0 / \partial R}$ \\
        $\delta$ & Relative root-mean-square error between simulated ($\hat{y}_i$) and observed ($y_i$) terminal velocities: $\delta = \sqrt{(1/N) \sum_{i=1}^{N} ((\hat{y}_i - y_i)/y_i)^2}$ \\
        $\alpha$ & Relative root-mean-square error between simulated ($\hat{y}_i$) and observed ($y_i$) circular velocities in the $R_{t} = 2.63 - 7.82$~kpc radius range: $\alpha = \sqrt{(1/N) \sum_{i=1}^{N} ((\hat{y}_i - y_i)/y_i)^2}$\\
        $A_{2}(R)$ & Amplitude of the $m = 2$ Fourier mode \\
        $\tau_{z}$ & Bar strength or vertical component of bar gravitational torque \\
        $a_{\rm bar}$ & Bar semimajor axis length \\
        $R_{\rm CR}$ & Bar corotation radius \\
        $\mathscr{R}$ & Bar rotation rate ($\mathscr{R} = R_{\rm CR} / a_{\rm bar}$) \\
        $\Omega_{\rm p}$ & Bar pattern speed \\
        \hline
    \end{tabular}
\end{table}

\subsection{Properties of the synthetic bar} \label{sec:bars}

\added[id = anon]{One of the main goals of this work is to study properties of the Galactic bar using terminal velocity curves. In Table \ref{tab:symbols}, we define a list of key symbols that are relevant for this analysis.}

\subsubsection{Bar structure}\label{sec:fourier}

The overall structure of the stellar bar is best characterised via Fourier decomposition of its projected mass distribution \citep{kraljic2012two, aguerri2009population, ohta1990surface}. The stellar surface density distribution, $\Sigma(R, \theta)$, in a galaxy snapshot is rewritten as a sum of sinusoidal functions with varying frequencies: 

\begin{equation}
    \Sigma(R, \uptheta) = \frac{a_{0}(R)}{2} + \sum_{m=1}^{\infty}(a_{\rm m}(R)\cos(m\uptheta) + b_{\rm m}(R)\sin(m\uptheta))
\end{equation}

where the Fourier coefficients are given by \citep{aguerri2009population}: 

\begin{equation}
    a_{\rm m}(R) = \frac{1}{\pi}\int_{0}^{2\pi} \Sigma(r, \uptheta)\cos(m\uptheta)d\theta
\end{equation}

\begin{equation}
    b_{\rm m}(R) = \frac{1}{\pi}\int_{0}^{2\pi} \Sigma(r, \uptheta) \sin(m\uptheta) d\uptheta
\end{equation}

Furthermore, the amplitudes of the m-th modes are calculated using \citep{aguerri2009population}: 

\begin{equation}
    \centering 
    \begin{aligned}
        A_{\rm 0}(R) = a_{\rm 0}(R) \; \; \;  \rm m = 0 \\
        A_{\rm i}(R) = \sqrt{a_{\rm m}^{2}(R) + b_{\rm m}^{2}(R)} \; \; \;  \rm m \neq 0
        \end{aligned}
\end{equation}

For galaxies with bars, the even Fourier modes (m = 2, 4, 6 ...) are usually large, and the m = 2 mode, which represents symmetry about two axes, tends to be dominant \citep{ohta1990surface, aguerri2009population}. The A$_2$ profile that we observe for Model 1 of Figures \ref{fig:fourier_all} are typical of a bar that forms and evolves in a MW-like simulation. Initially, we witness A$_2$ values increase quickly over time and with radius. Then, they reach a peak, after which there is a steep decrease in A$_2$. Around the time where A$_2$ peaks, the bar is thought to buckle, which weakens the bar and is thought to cause this rapid decrease in A$_2$ \citep{katariA2022effects}. Finally, there is a period of slow secular evolution, where the A$_2$ values change gradually \citep{lokas2021lopsided, athanassoulA2013bar}. During the secular evolution phase, traditionally a bar will grow in strength and decrease in pattern speed \citep{katariA2022effects}.

In Figure \ref{fig:A2_fourier}, we present periodograms, i.e. individual maps of the A$_2$ values across time and radius for each of our four models \citep[cf.][]{mit23a}. A quick glance at these periodograms provides us with a big picture view of what bars are doing in the models. We focus on the disc stars (column 1 of Fig. \ref{fig:A2_fourier}), but our stellar bars also contain many newborn stars and leave an imprint on the gas, making the other periodograms we have included here (columns 2 and 3 of Fig. \ref{fig:A2_fourier}) valuable.

We use A$_2$ values as a starting point when considering bar strength, although we do not recommend using A$_2$ in isolation (bar torques should be computed, to be discussed in Sec. \ref{sec:torque}). For example, in the periodograms for Model 3, A$_2$ rises to higher values over a larger range of radii than it does for any other simulation, which suggests that Model 3 has the strongest bar. The approximate lifetimes of bars in Table \ref{tab:bar_comparison} can be more easily visualised using the periodograms of Figure \ref{fig:A2_fourier}. We see that the Model 1 bar lives for the longest time, with A$_2$ values in the stellar disc rising to higher values (pink and purple in the periodogram) around $t = 0.6$~Gyr, and then remaining reasonably high at short radii for the rest of the simulation's duration; meanwhile, the Model 2 bar has the shortest lifetime, with A$_2$ rising to high values at small radii for a very short period of time before decreasing back down (blue and black in the periodogram). The periodogram for disc stars in Model 4 is mostly black reflecting the fact that Model 4 forms very little structure over the course of its evolution.

Apparently, the disc in Model 2 does not feature a bar during $t = 2.06 - 3.95$~Gyr (Figure \ref{fig:fourier_all}). While the A$_2$ values for this model are very low in this time period, they are still mildly more dominant than the other Fourier modes. A high $m=2$ mode can also arise from a two-armed spiral structure. Similarly, $m = 1$ is often more dominant for a one-armed spiral, and $m =3$ tends to increase when we have multiple-armed spirals (greater than 2) \citep{yu2020statistical}. The $m = 2$ mode is probably mildly dominant at times $t = 2.06 - 3.95$~Gyr for Model 2 because of the distinct two-armed spiral structure that it has at its centre after the bar has dissolved. \added[id = anon]{These oscillations in A$_2$ have been observed in earlier studies  \citep{lokas2021lopsided, katariA2024importance, anderson2022secular}, and are often said to be a result of spiral arms attaching and detaching from the ends of the bar \citep{debattistA2000constraints, pfenniger2023five}. However, the bar's physical size when measured with visual inspection (see Sec. \ref{sec:constraining_bars}) also fluctuates with time, which may partially contribute to the oscillations in A$_2$.} \deleted[id = anon]{The $A_2$ oscillation in the bar region can also produce errors when using Fourier analysis to estimate the bar length.}



\subsubsection{Bar strength and gravitational torques}\label{sec:torque}

In the context of this work, we argue that the bar torque is the best means of \deleted[id = anon]{quantifying} \added[id = anon]{assessing} bar strength, and is preferred over the use of $m=2$ Fourier components. A bar's torque funnels gas inward, fuels star formation in the centre of the galaxy, influences the angular momentum exchange between the stellar disc and DM halo, and is the primary driver of the streaming and non-circular motions of gas around the bar, with the latter being the main focus of this study. Specifically, we are interested in the vertical component of the torque ${\bm \tau_z}$, which directly influences the angular momentum ${\bm L}_z$ of the gas or stars. A measurement of bar torque directly considers the non-axisymmetric, time-varying gravitational force and the gradient in the gravitational potential \citep{block2001gravitational}. On the other hand, while $m = 2$ Fourier methods can be a good entry point for considering bar strength (as seen in Sec. \ref{sec:fourier}), they are an indirect approach because they are based on surface density contrasts between the bar versus interbar regions \citep{butA2001dust}. 

We compute a rough estimate of bar torque for $t = 3.80$~Gyr in Model 1, which is toward the end of the simulation, once the bar has grown in length and slowed down in pattern speed significantly. We use the following general expression for bar torque: 
\begin{equation}
    {{\bm \tau}_z = R_{\rm bar} \times \bm{F_{\rm bar}} = \frac{d{\bm L}_{z}}{dt} \approx M_{\rm bar}R_{\rm bar}^{2}\Omega_{\rm p}^{2}}
\end{equation}
where $R_{\rm bar}$ is the radius of the bar's semimajor axis, $L_{z}$ is the angular momentum about the disc's axis of rotation, $M_{\rm bar}$ is the estimated stellar mass of the bar. For $t = 3.80$~Gyr in Model 1, $M_{\rm bar} \approx 1\times 10^{10} \unit{M_{\odot}}$, $a_{\rm bar} \approx 3.2 \unit{kpc}$, $r_{\rm bar} = a_{\rm bar}/2 = 1.6 \unit{kpc}$, $\Omega_{\rm p} \approx 42 \unit{km \; s^{-1}\; kpc^{-1}}$. Hence, we estimate
\begin{equation}
{\bm \tau}_z \approx 5 \times 10^{-5}\; \rm kpc^{2}\; M_{\odot}\; yr^{-2}
\end{equation}
for the bar at $t = 3.80$~Gyr.

\begin{figure}
    \centering
    \includegraphics[width=0.5\textwidth]{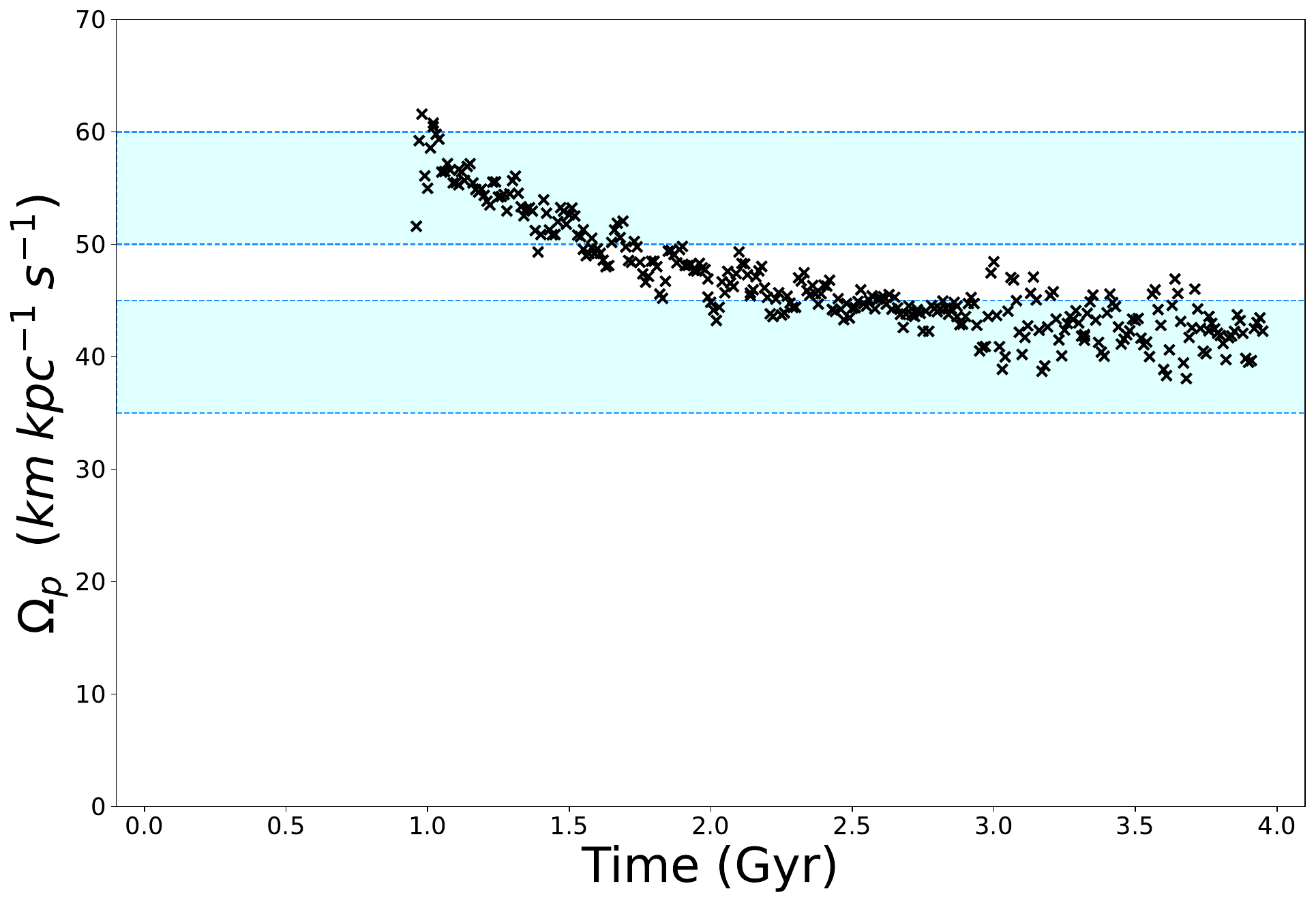}
    \caption{Pattern speed of the Model 1 bar, found using the code from \citet{dehnen2023measuring}. The blue shaded regions represent some estimated pattern speeds for a fast ($\Omega_{\rm p} \approx 35 - 45 \unit{km \; s^{-1} \; kpc^{-1}}$) versus slower bar ($\Omega_{\rm p} \approx 50 - 60 \unit{km \; s^{-1} \; kpc^{-1}}$) in the Milky Way.}
    \label{fig:pattern_speed}
\end{figure}

\begin{figure}
    \centering
    \includegraphics[width=0.5\textwidth]{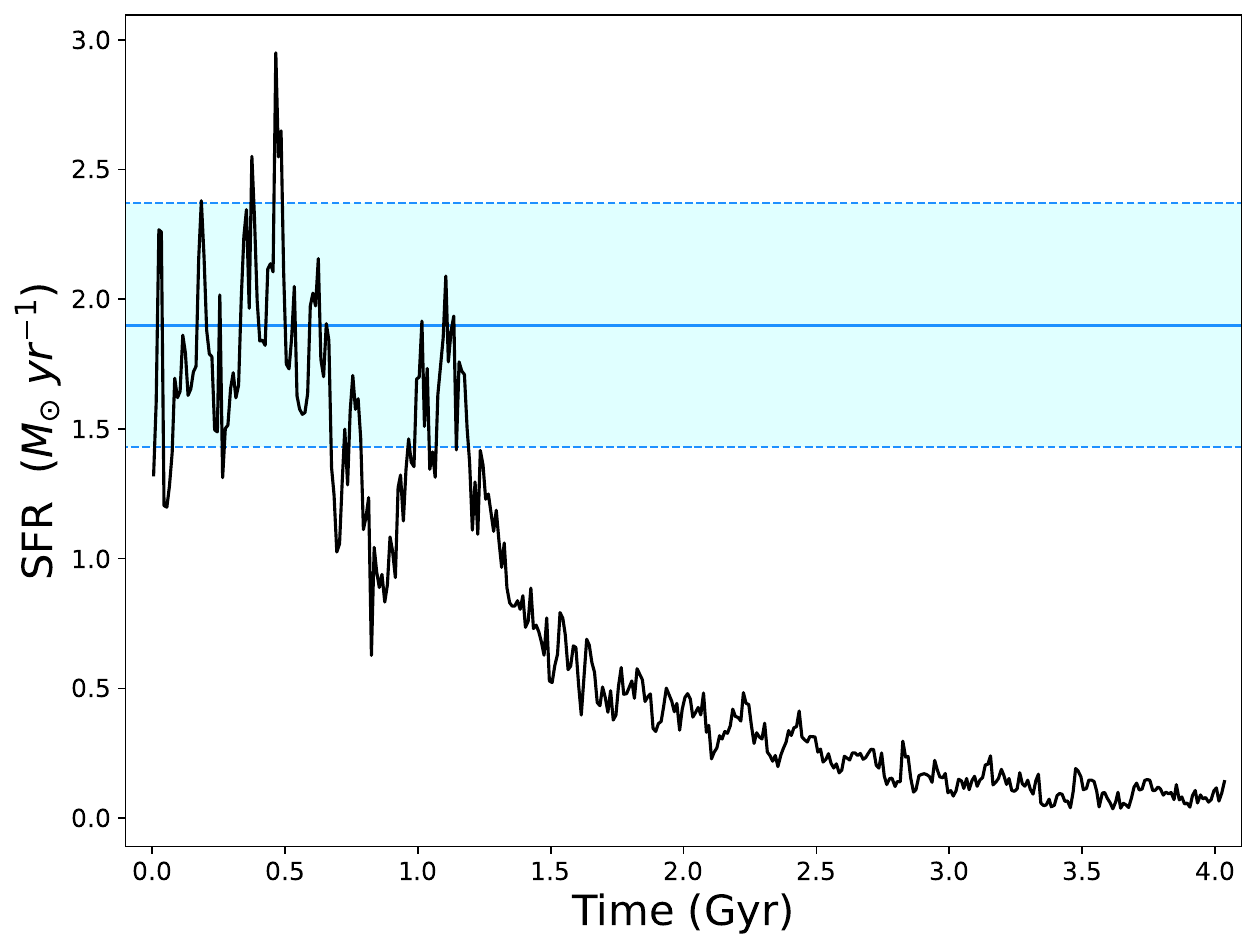}
    \caption{Star formation history for Model 1. Star formation peaks early on and then significantly declines later in the model. Star formation rates (SFR) are quite low for those snapshots where both a bar is present and we also do our terminal velocity analysis ($t = 1.17 - 3.95$~Gyr). The blue line and shaded region represents the estimated SFR for the Milky Way from \citet{chomiuk2011toward}: $\mathrm{1.9 \pm \unit{M_{\odot}yr^{-1}}}$.} 
    \label{fig:SFH}
\end{figure}

\subsubsection{Bar pattern speed}\label{sec:pattern_speed}

The evolution of the bar's pattern speed in Model 1 is shown in Fig.~\ref{fig:pattern_speed}. The pattern speed values were computed using the approach (and corresponding code) developed by \citet{dehnen2023measuring}. We note that it is not always possible to retrieve a meaningful value since there are instances where the code fails to consistently detect a bar. We observe that the bar slows down over time. It reaches a maximum of $61.6 \; \unit{km \; s^{-1} \; kpc^{-1}}$ near the beginning of the selected time period, and slows down to a minimum of $38 \; \unit{km \; s^{-1} \; kpc^{-1}}$ near the end of the simulation.

\subsubsection{The effect of star-forming, multiphase gas on the bar }\label{sec:SFH}

Fig.~\ref{fig:SFH} displays the star formation history of Model 1. Star formation rates (SFR) peak early on, and then decline to low values by the end of the simulation. Ultimately, we want SFR values that are consistent with the Milky Way. Estimates for the MW's SFR are: $\mathrm{1.9 \pm 0.4 \; M_{\odot} \; yr^{-1}}$ \citep{chomiuk2011toward}, $\mathrm{2.0 \pm 0.7 \; M_{\odot}\; yr^{-1}}$ \citep{eliA2022star} and $\mathrm{1.94 \pm 0.47 \; M_{\odot}\; yr^{-1}}$ \citep{tuntipong2024sami}. Our terminal velocity analysis is performed within the $t = 1.17 - 3.95$~Gyr time period because Model 1 has a bar that is fully formed and consistently detectable with \citet{dehnen2023measuring}'s code at these times. From $t = 1.20$~Gyr onwards, Model 1's SFR has dropped to values that are consistently lower than the MW's. Unfortunately, this means that Model 1 does not have a SFR compatible with the MW's for almost all of the time period for which we do the terminal velocity analysis. In the future, we want to add sustained accretion of gas to the model and adjust the SFR so that we have both a bar present and an SFR that agrees with the MW at the same time.


Our simulations strongly indicate that star formation significantly influences bar formation and evolution. Models 1 and 2 differ only in their gas treatment: Model 1 includes multiphase gas with star formation and feedback, while Model 2 uses a purely isothermal gas. In Model 1, the bar persists throughout the simulation; in Model 2, an early bar forms but is quickly destroyed. This highlights how gas physics and the presence or absence of star formation and feedback critically affect bar longevity. It also suggests that modifying Model 1 to match the Milky Way’s star formation rate during the bar phase could notably alter the bar’s characteristics.

\subsection{Comparing gas dynamics in the simulations against observations}\label{sec:gas_comparison}

\subsubsection{$(\ell, V_{\rm los})$ diagrams}\label{sec:lvlos}

\begin{figure*}
    \centering
    \includegraphics[width=\textwidth]{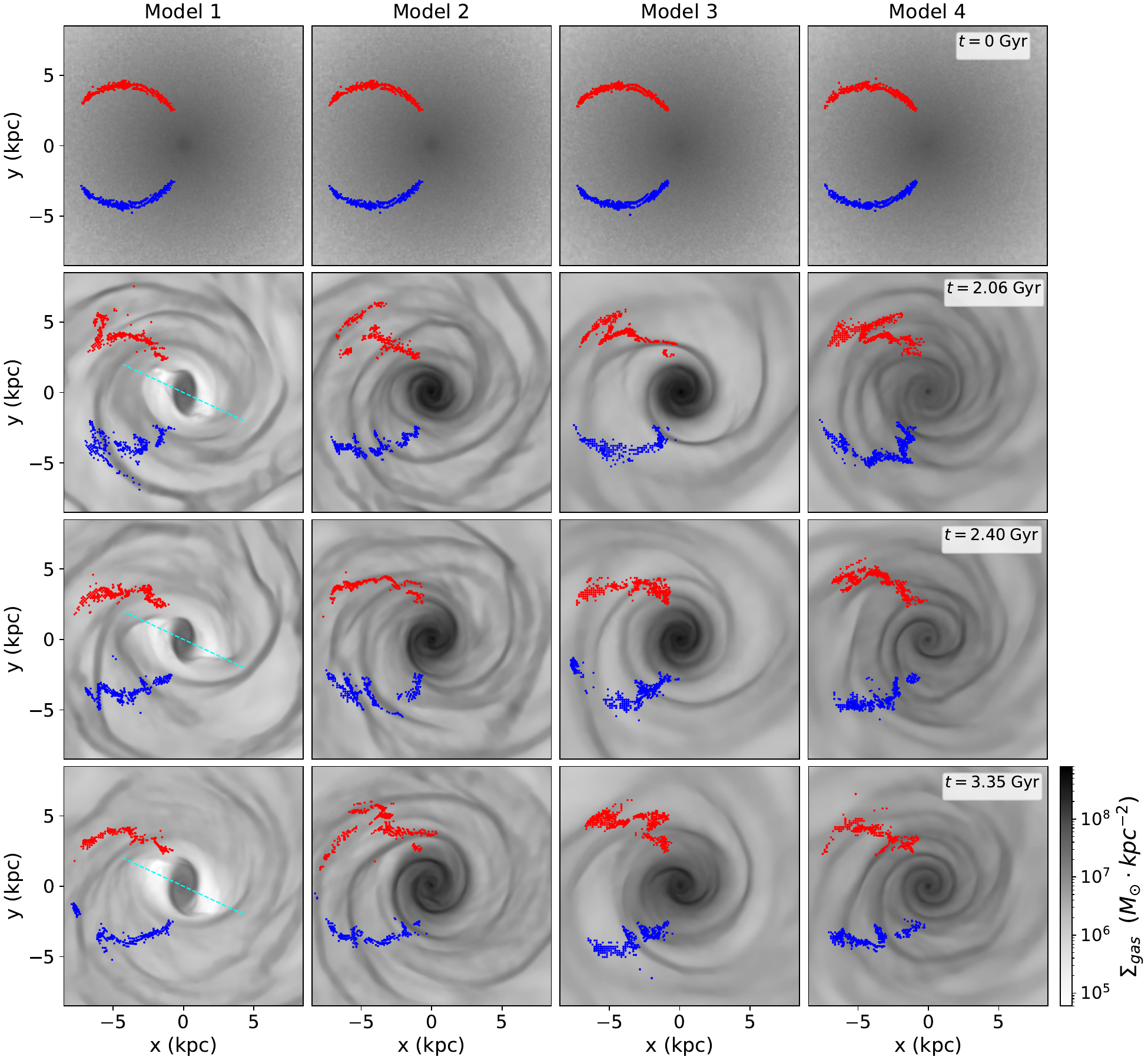}
    \caption{Gas density distributions projected along $z$ for $t = 0, 2.1, 2.4, 3.4$~Gyr for each of our models. As we go from row 1 - 4, we consider $t = 0$~Gyr to $t = 3.4$~Gyr. Similarly, as we go from column 1 - 4, we look at Models 1 - 4. We can see that the presence of a bar for most snapshots in Model 1 leaves a distinct imprint on the gas. No bar is present in any of the selected snapshots for Models 2 - 4. For Models 2 and 3, we see a distinct spiral pattern, which is particularly high in density toward the centre. The spiral pattern that forms in Model 4 is much weaker, with this model forming much less structure overall. The blue and red points mark the location of gas that gives the terminal velocity values in equivalent $(\ell, V_{\rm los})$ diagrams.
    \added[id = anon]{The cyan dashed line is at $25\deg$ to the $x = 0$ axis, which is the angle of the stellar bar.}} 
    \label{fig:xy_allsim_grid}
\end{figure*}

\begin{figure*}
    \centering
    \includegraphics[width=\textwidth]{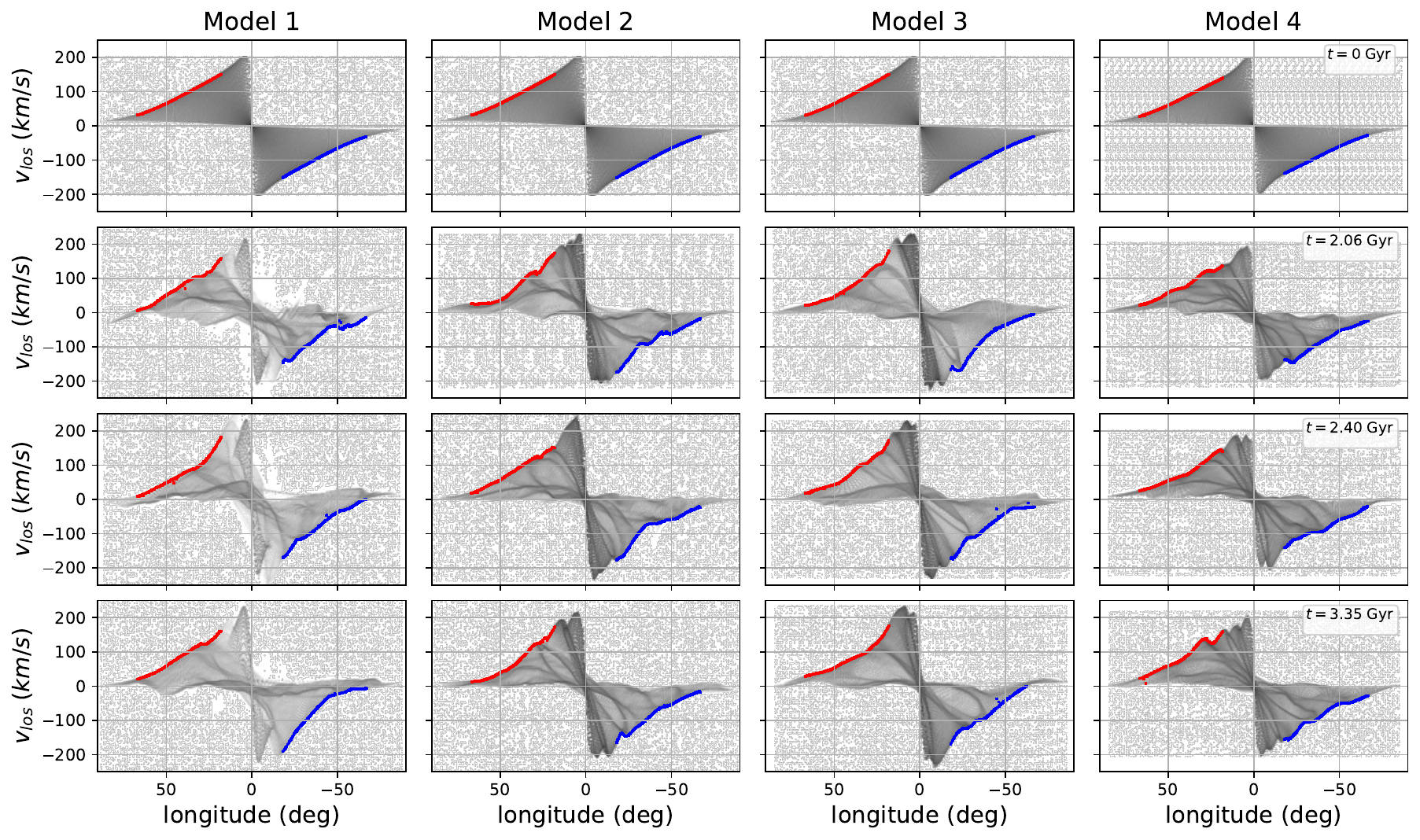}
    \caption{These are the $(\ell, V_{\rm los})$ diagrams that correspond to the same snapshots that were presented in Fig.~\ref{fig:xy_allsim_grid}. The red and blue points mark out the simulated terminal velocity values for these snapshots, which we will later compare against observed data. Red is for Quadrant I data, and blue is for Quadrant IV data. We can see that the terminal velocity values trace along part of the outer envelope of the underlying $(\ell, V_{\rm los})$ distribution.} 
    \label{fig:lvlos_allsim_grid}
\end{figure*}

In Fig.~\ref{fig:xy_allsim_grid}, we present a series of gas surface density maps projected onto $(x,y)$ for Models 1 to 4. Each row corresponds to a different epoch, specifically, from top to bottom: $t =$ 0, 2.06, 2.40, 3.35 Gyr. In Figure \ref{fig:lvlos_allsim_grid}, we present the corresponding density projections onto $(\ell, V_{\rm los})$, following the same arrangement. The selected times have been chosen to facilitate the comparison between the various simulations. Model 1 does have a bar for $t =$ 2.06, 2.40, 3.35 Gyr, and we see in Fig.~\ref{fig:xy_allsim_grid} that this bar leaves a distinct imprint on the gas at these times (an apparent nuclear ring, surrounded by a low gas density region, all where the bar is). \added[id = anon]{For all analysis of barred snapshots in Model 1, the bar is rotated to be at roughly an angle of $25 \deg$ relative to the $x =0$ axis. This means that the semi-major axis of the stellar bar is at roughly a $25 \deg$ angle relative to the Sun-Galactic Centre line. We have marked a dashed cyan line on the barred snapshots of Fig. ~\ref{fig:xy_allsim_grid} to help visualise this. However, the bar angle is much clearer to visualise looking at an $(x,y)$ stellar density projection, rather than a gas density projection. The attachment and detachment of the bar to spiral arms affects the imprint that the stellar bar leaves on the gas, and can appear to make the corresponding gas distribution shift in angle. In reality, while the
\citet{dehnen2023measuring}
code we use to compute bar angle and rotate the bar will not be perfect, we do find that the stellar bar is quite consistently aligned at a $-25 \deg$ across snapshots by visual inspection.} Meanwhile, Models 2 - 4 do not have a bar for these snapshots, and their corresponding $(x,y)$ density projections mostly feature signatures of spiral arms (no bar). To be clear, the bar has already been destroyed by $t = 2.06$~Gyr for Models 2 and 3.

First, we focus on time $t = 0$~Gyr for Model 1 (top left corner of each Fig.~\ref{fig:xy_allsim_grid} and \ref{fig:lvlos_allsim_grid}). The $(x,y)$ distribution of gas is smooth and axisymmetric. The corresponding $(\ell, V_{\rm los})$ plot also has near perfect reflection symmetry. In part, this symmetry in $(\ell, V_{\rm los})$ space is because we have assumed pure circular motion for the sun, with $\vec{V}_{\odot} = (0, 220, 0)$~\kms, when mapping from $(x,y)$ to $(\ell, V_{los})$ ie. we ignore the motion of the Sun relative to the LSR $({v_{\rm \odot, r}, v_{\rm \odot, z} = 0})$. The use of non-zero $v_{\rm \odot, r}$  or $v_{\rm \odot, z}$ would introduce asymmetries in the $(\ell, V_{\rm los})$ diagram, even when the $(x,y)$ density projection is a smooth Gaussian distribution. $\Theta_{\rm 0} = 220$~\kms\ is the value we compute from the rotation curves of the simulations at $R_{\rm 0} = 8.5$~kpc, consistent with \citet{mcclure2016milky}. We note that these choices for $\Theta_{\rm 0}$ and $R_{0}$ do not affect our results in any significant way compared to assuming other sets of reasonable values, e.g. $\Theta_{\rm 0} = 240$~\kms\ at $R_{\rm 0} = 8.2$~kpc.

The initial gas distribution in $(\ell, V_{\rm los})$ space in Model 1 does not feature any significant component in the ``forbidden'' regions  ($V_{\rm los} > 0$ when $\ell < 0$, and $V_{\rm los} < 0$ when $\ell > 0$). Moving to $t = 2.06 - 3.35$~Gyr for Model 1, we can see that a bar is now present. The distribution of gas in $(x,y)$ space in these snapshots can be better understood by considering orbit families of a barred potential \citep{binney1991understanding}. At the centre of a barred potential, gas is believed to follow epicyclic motions about two key underlying stable orbit families: the $x_{1}$ and $x_{2}$ families \citep{contopoulos1977inner}. Between the Corotation Radius (CR) and the Inner Lindblad Resonance (ILR), $x_{1}$ consists of elliptical shaped orbits elongated along the bar such that their major axis is parallel to the bar's major axis. Closer to the ILR, there is an innermost stable $x_{1}$ orbit featuring a `cusp' at both ends \citep{morris1996galactic}. Within this, $x_{1}$ orbits are self-intersecting, with loops at either end. The $x_{2}$ orbit family exists at smaller radii (compared to the $x_{1}$ family), and it is comprised of oval-shaped orbits whose major axis is perpendicular to the major axis of the bar \citep{morris1996galactic}. We can clearly see signatures of these $x_{1}$ and $x_{2}$ orbit families in Model 1 (Fig.~\ref{fig:xy_allsim_grid}, first column), when looking at the $(x,y)$ density distributions of gas for $t =$ 2.06, 2.40, 3.35 Gyr. There is a dense nuclear ring of gas at the centre of the gas distribution, which likely corresponds to gas moving on the $x_{2}$ orbit family. The outline of the low density region, which takes the shape of an ellipse with pointed ends, likely corresponds to the \added[id  = anon]{cusped $x_{1}$ orbit}.

\begin{figure*}
    \centering
    \includegraphics[width=\textwidth]{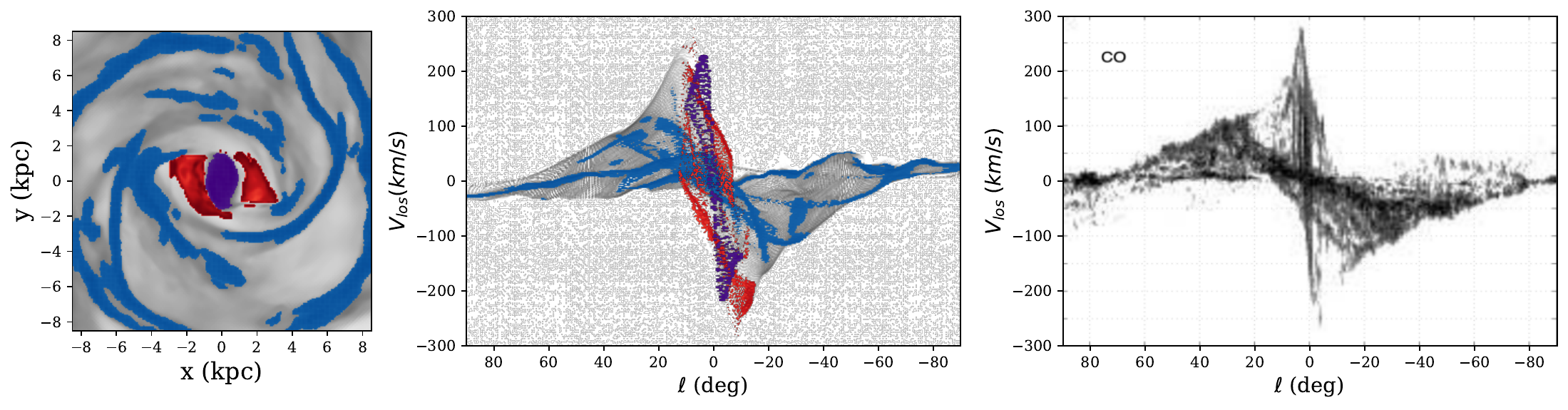}
    \caption{Gas density distribution in $(x,y)$ space (left) and $(\ell, V_{\rm los})$ space (centre) for $t = 2.40$~Gyr of Model 1. We have highlighted different regions of the $(x,y)$ gas density distribution in different colours so that we can see how different features in $(x,y)$ space map across to $(\ell, V_{\rm los})$ diagrams. The nuclear ring region is coloured purple, and maps across to a diagonal shape structure $(\ell, V_{\rm los})$, which is confined to a narrow range of longitudes but extends over a large range of velocities. 
    The low density region in $(x,y)$ space, between the nuclear ring and the cusped $x_{1}$ orbit, is coloured red. We see that this low density gas in the bar region is mostly responsible for the forbidden velocities observed in the $(\ell, V_{\rm los})$ diagram. Finally the spiral arms are blue and often appear as higher density filament-like structures in $(\ell, V_{\rm los})$. The third figure in this panel is the observed CO distribution of gas in the MW. Credit for this third figure, or the observed gas distribution, goes to: \citet{fux19993d}, A\&A, 345, 787, reproduced with permission \copyright  ESO.} 
    \label{fig:isolate_regions}
\end{figure*}

In Fig.~\ref{fig:isolate_regions}, we focus on time $t = 2.40$~Gyr, and isolate different structures in the $(x,y)$ distribution and their representation in $(\ell, V_{\rm los})$ space. The nuclear ring, coloured purple in Fig.~\ref{fig:isolate_regions}, appears as a diagonal column in $(\ell, V_{\rm los})$ space, extending over a large range of $V_{\rm los}$ values, and confined to a short range of longitudes. Gas in the low density region, within the cusped $x_{1}$ orbit, presents as a tilted parallelogram (red) in $(\ell, V_{\rm los})$ space. This component has gas with significant `forbidden' velocities ($V_{\rm los} > 0$ for $\ell < 0$, and $V_{\rm los} < 0$ for $\ell > 0$), which can be explained by gas moving on more \added[id = anon]{elongated} \deleted[id = anon]{elliptical-shaped} orbits in the bar region. \added[id = anon]{In general, higher magnitude forbidden velocities correspond to a stronger bar with more non-circular motions in the corresponding gas distribution.} Furthermore, spiral arms (blue) form dense filaments in $(\ell, V_{\rm los})$ space, which often have loop-like structures. This can best be understood using our simple model, where a spiral can be modelled as a circular structure that does not close on itself, hence producing loops rather than straight lines in $(\ell, V_{\rm los})$ distributions. The gas distributions in Model 2 (Fig.~\ref{fig:lvlos_allsim_grid}), do not feature the signatures of the bar, such as the nuclear ring, cusped $x_{1}$ orbit, or significant forbidden velocities, simply because this simulation does not have a bar for the snapshots included here.

The main analysis in this paper involves comparing terminal velocities of gas in the simulations against observed terminal velocities of gas in the MW. In particular, we use the observed \HI\ terminal velocity data from \citet{mcclure2016milky}, who have measured $V_{\rm t}$ values for longitudes between $|\ell| \approx 18\degree - 67 \degree$. Lines-of-sight at these longitudes look at gas in the region just around the bar. This is shown in Fig.~\ref{fig:quadrants}. As discussed previously, Fig.~\ref{fig:quadrants} presents an $(x,y)$ density projection of gas in a single snapshot of Model 1 for the region within the solar circle. The red and blue dotted lines help us visualise the region of the disc within the  $|\ell| \approx 18\degree - 67 \degree$ longitude range. The red dotted lines correspond to the positive galactic longitudes $\ell = 18 \degree$ and $67 \degree$ in Quadrant I; the blue dotted lines are the equivalent negative galactic longitudes $\ell = -18 \degree$ and $-67 \degree$ in Quadrant IV. As we can see in this diagram, these longitudes do not probe the bar itself, but look at the gas just outside the bar, whose motion is likely to be affected by the bar's properties. On a similar note, in Fig.~\ref{fig:lvlos_allsim_grid}, we have plotted the terminal velocity values on the $(\ell, V_{\rm los})$ distributions, where we can see that the terminal velocity curve traces along the outer envelope of the underlying $(\ell, V_{\rm los})$ distribution. Furthermore, we have coloured the $V_{\rm t}$ curves red and blue, based off whether they are for Quadrants I or IV of the galaxy respectively.

\subsubsection{Gas terminal velocities}\label{sec:terminal_vel}
\begin{figure*}
    \centering
    \includegraphics[width=\textwidth]{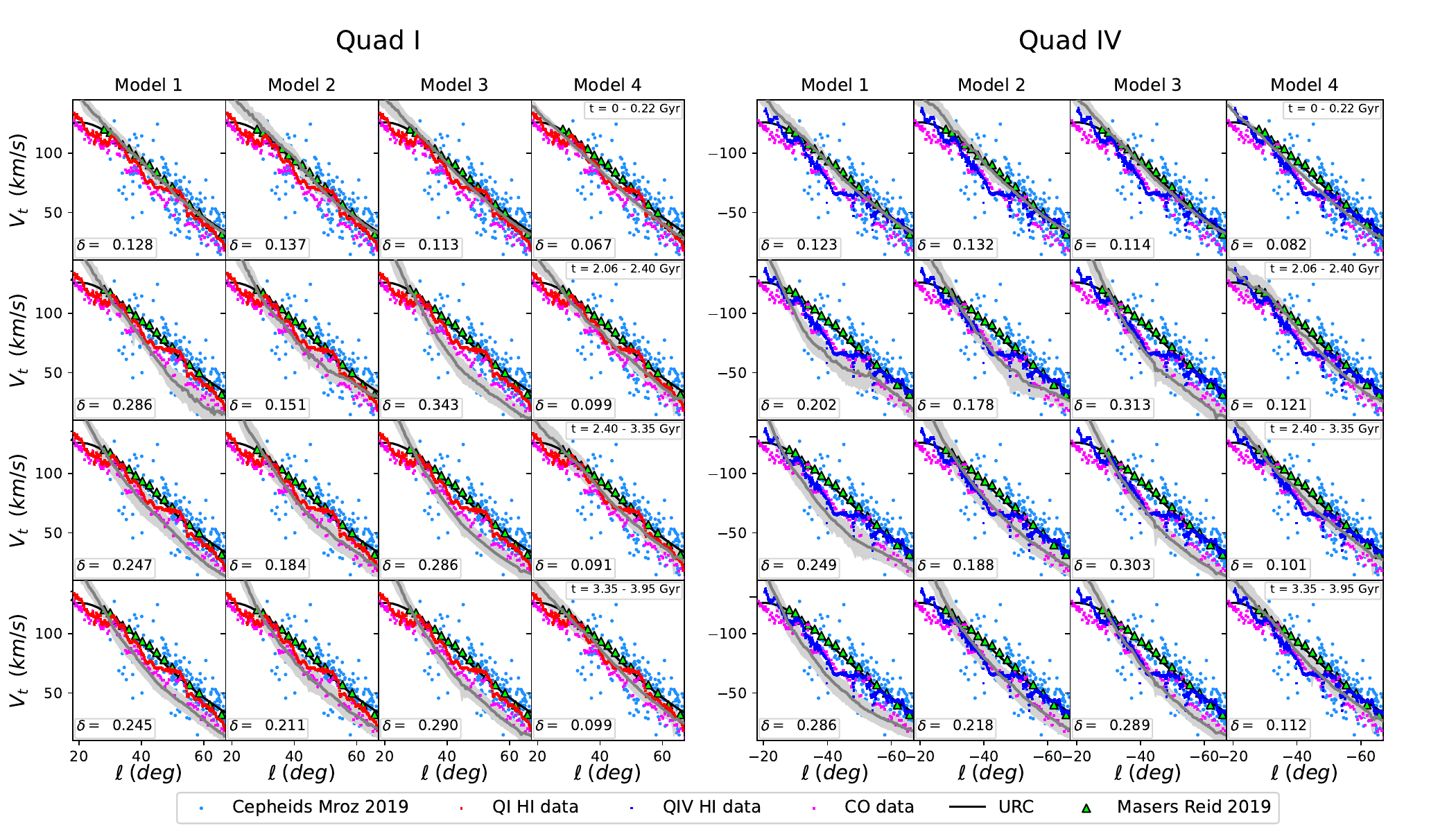}
    \caption{Comparison of simulated versus observed terminal velocity values. The left panel shows results for Quadrant I; right panel for Quadrant IV. Going from the left to right columns, we progress from Models 1 - 4. Going down the rows, we move in time through the simulation, with each row considering a group of snapshots. The first row looks at $t = 0 - 0.24$~Gyr (early on in the models when not much structure has formed). The second to fourth rows consider $t = 2.06 - 3.95$~Gyr 215 - 412, broken into three groups as labeled in the subpanels. The grey curves, with grey shaded region around them, represent the median simulated terminal velocity curves with $\pm 1 \sigma$ uncertainty around them, with the median computed across the respective group of snapshots. The coloured points present different types of observed data, with the various points identified in the legend. The $\delta$ values shown in the bottom left corner of each subpanel quantify the difference between simulated versus observed data.}
    \label{fig:termvel_grid_all}
\end{figure*}

We now compare simulated terminal velocity curves against those estimated from observational data (see Fig.~\ref{fig:termvel_grid_all}). The left panel corresponds to Quadrant I $V_{\rm t}$ values, the right panel to Quadrant IV. For each quadrant, columns 1 to 4 (starting from the left) correspond to Models 1 to 4, respectively, as indicated by the column header.  Simulation time increases row-wise from top to bottom: the first row aggregates the results of individual time steps within $t = 0 - 0.22$~Gyr; the second row, third and fourth row, those in $t = 2.06 - 2.40$~Gyr, $t = 2.40 - 3.35$~Gyr, and $t = 3.35 - 3.95$~Gyr, respectively. We emphasise that these time periods have been selected such that Model 1 always has a bar for $t = 2.06 - 3.95$~Gyr, while Models 2 - 4 do not have a bar at any point in this time period. No model has a bar for the initial time period $t = 0 - 0.22$~Gyr.

In each individual panel, the grey curve represents the median terminal velocity averaged over the corresponding time span (see above); the grey-shaded regions around them indicate the $\pm 1 \sigma$ scatter. The coloured points represent the observed terminal velocity values. These correspond to the data underlying the circular velocity estimates presented in Fig.~\ref{fig:circ_vel_curves} (see figure caption for a list of the respective sources).

To quantify the difference between the (median) simulated terminal velocity curve and the observed terminal velocity curve, we introduce a parameter $\delta$. To calculate it, we divide the longitude range of $|\ell| = 18 \degree - 67 \degree$ into 122 bins, then we compute the median value of all observed data in each bin, and then repeat with the simulated data in each bin. After this, we use the following expression to evaluate $\delta$:
\begin{equation}
    \delta = \sqrt{\frac{1}{N} \sum_{i = 1}^{N} \frac{(\hat{y_{i}} - y_{i})^{2}}{y_{i}^{2}}}
\end{equation}
where $y_{i}$ is the median observed value in each bin, $\hat{y_{i}}$ is the median simulated value in each bin, and N is the total number of observed values used in this computation (equal to the number of bins). 
The lower the value of $\delta$, the better the agreement between simulations and observations.
The value of $\delta$ in each case is shown in the bottom-left corner of the corresponding panel. 

The first key result we find using Figure ~\ref{fig:termvel_grid_all} is that \textit{the simulated terminal curves of Model 1 agree more with observations before a bar forms in the model compared to once a bar is present}. For $t = 0 - 0.22$~Gyr in Model 1, which is very early on before a bar forms, the grey curves from the simulation overlap significantly with the various coloured data points, and the associated $\delta =$ 0.128, 0.123 for Quad I and IV are small. This suggests that the simulated gas dynamics for Model 1 in this time period reproduce the observations well. However, at $t = 2.09 - 4.02$~Gyr in Model 1, at which times a bar is present, the grey curves begin to deviate more significantly from the observed data points; the delta values increase to $\delta =$ 0.286, 0.247, 0.245 in Quad 1 and $\delta =$ 0.202, 0.249, 0.286 for Quad 4. Hence, Model 1 is unable to reproduce the observed terminal velocities well later on in its evolution. At first glance, it seems surprising that Model 1 reproduces observations of the Milky Way gas best at very early times, when the simulation does not morphologically resemble the actual galaxy well at all, but instead the stellar disc is simply a smooth and axisymmetric distribution of stars with no bar and minimal spiral arms. Obviously, Model 1 fits well with observed data at early times because the initial conditions have been selected to achieve this. The state of the model after $4$~Gyr, on the other hand, is difficult to predict since the simulation's evolution is a non-linear process. 

The second key observation we make about Figure \ref{fig:termvel_grid_all} is that \textit{the Model 1 terminal curves at late times ($t = 2.06 - 3.95$~Gyr) deviate more from the observations than the Model 2 terminal curves at late times do}. Comparison between Models 1 versus 2 is valuable because the only difference in the initial conditions between these simulations is that Model 1 has multiphase gas with star formation, while Model 2 features an isothermal gas disc. For $t = 0 - 0.22$~Gyr, the median terminal curve for both simulations are reasonably similar; they overlap with the observed data significantly and the delta values are relatively small. For $t = 2.06 - 3.95$~Gyr, the mean simulated curves for both simulations begin to deviate more from the observations, but this deviation is notably larger for Model 1. The main morphological difference between the simulations in this later time period is that Model 1 does have a bar, while Model 2 does not. Once again, we are observing that the match between the observed versus modelled terminal curve data is at its worst when a bar is present.

\begin{figure*}
    \centering
    \includegraphics[width = \textwidth]{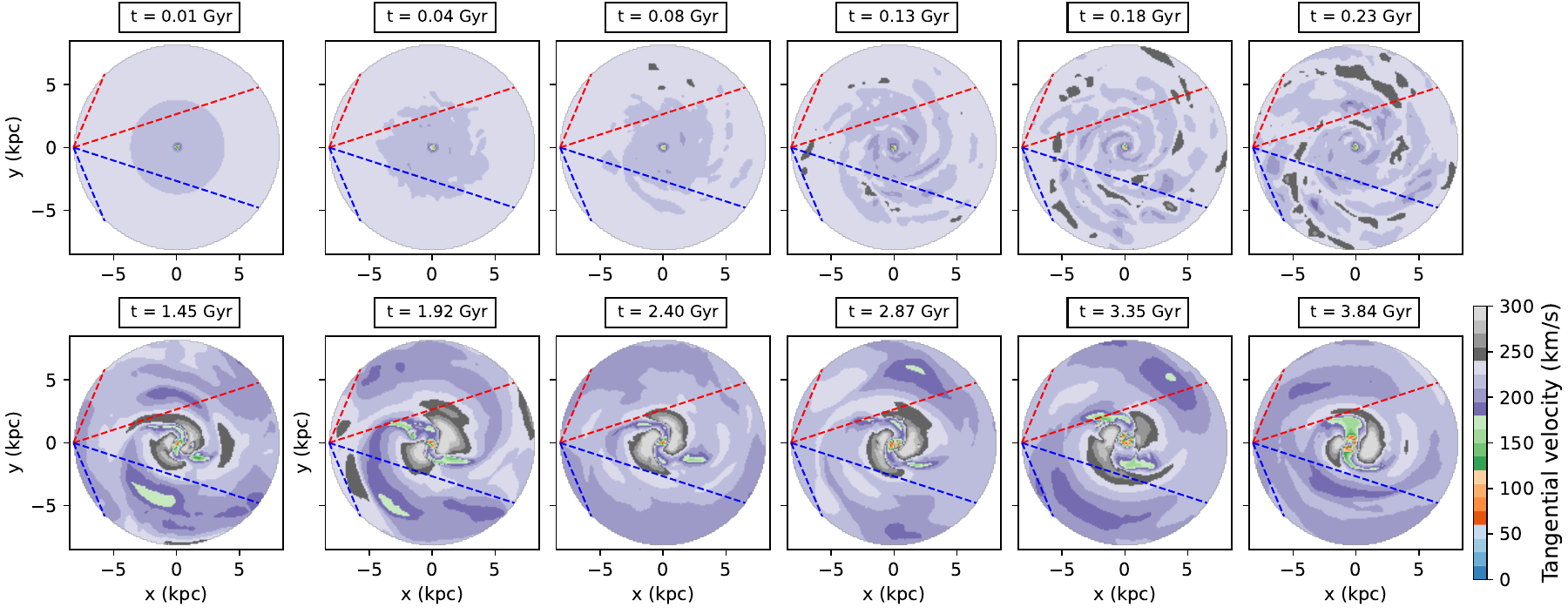}
    \caption{Model 1 tangential velocities of gas in the bar region, and region that we probe for our terminal velocity analysis ($|\ell| = 18\degree - 67 \degree$ and $R < R_{0}$). In the top row, we consider the tangential velocity distribution for various snapshots between $t = 0 - 0.24$~Gyr. This helps us visualise the tangential velocity distribution before the bar or much structure at all has formed in the simulation. In the bottom row, we look at the distribution of tangential velocities for $t = 1.44 - 3.84$~Gyr, at which times we do have a bar. We can see how the presence of a bar significantly changes the tangential velocities of gas in the bar region and inner disc. From there, it is easier to understand how the terminal velocities in the  $|\ell| = 18 \degree - 67 \degree$ longitude range evolve as the bar does.} 
    \label{fig:tang_vel_lrstar}
\end{figure*}

\begin{figure*}
    \centering
    \includegraphics[width = \textwidth]{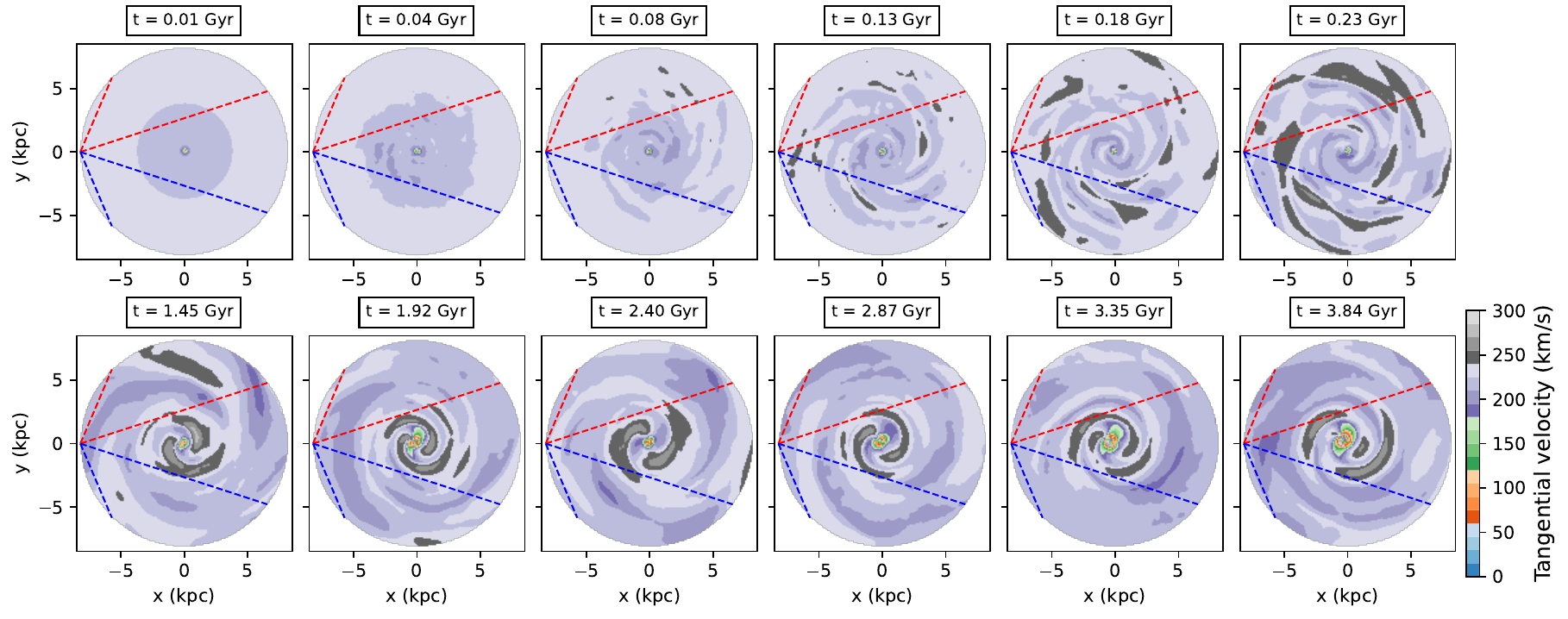}
    \caption{Tangential velocities of gas in Model 2. We consider the same snapshots and regions as was done for Model 1 in Fig.~\ref{fig:tang_vel_lrstar}. This allows us to more directly compare how the tangential velocities in the disc vary across Model 1, where we do consider the tangential velocity distribution for snapshots where a bar forms and evolves, versus Model 2, where we only consider tangential velocity distributions in snapshots without a bar.}
    \label{fig:tang_vel_lr}
\end{figure*}

By inspecting the distribution of in-plane gas velocities in Model 1, we see that \textit{tangential velocities in the $R_{t} \approx 2.6 - 7.8$~kpc radius range decrease as a bar forms}. Our terminal velocity analysis has been done in the $|\ell| = 18\degree - 67 \degree$ longitude range. Using the tangent-point method, this roughly corresponds to the radius range of $R_{t} \approx 2.6 - 7.8$~kpc. As was shown with Fig.~\ref{fig:quadrants}, these lines-of-sight do not look directly at the bar, but rather consider the gas immediately around it. In simple terms, terminal velocities are just in-plane or tangential velocities projected onto relevant lines-of-sight and corrected for the Sun's motion. For this reason, we look more carefully at the tangential velocities of gas in the inner disc $(R < R_{0})$ for each Model 1 and 2 using Figures \ref{fig:tang_vel_lrstar} and \ref{fig:tang_vel_lr}. The shock cavities of the bar region $-$ low gas density regions between the nuclear ring and the cusped $x_{1}$ orbit $-$ are accompanied by very high tangential velocities, while the nuclear ring region features low tangential velocities (mostly greens and oranges in Fig. \ref{fig:tang_vel_lrstar}). Next, we consider the region of gas that is the focus of our terminal velocity analysis: gas immediately around the bar but within the solar circle. As the bar forms, we see the mean tangential velocity in this area decrease: shades of light purple with smaller amounts of grey and medium purple in Figure \ref{fig:tang_vel_lrstar} transition to mostly darker purples with some green also appearing. The decrease of the gas tangential velocities in the $R_{t} \approx 2.6 - 7.8$~kpc radius range is equivalent to the drop in the terminal velocity curves as a bar forms.

Our interpretation of the above results $-$ and a central argument of this paper $-$ is that \textit{the gravitational torques exerted by the bar in Model 1 are too strong compared to those of the Milky Way's bar}. This would explain: (i) why the terminal curves in Model 1 deviate more from the observed data when a bar is present, compared to before a bar forms, and (ii) why, at late times ($t = 2.06 - 3.95$~Gyr), the terminal curves in Model 1 deviate more from observations than those in Model 2 do. Stronger bar torques drive excessive streaming and non-circular gas motions in the region immediately surrounding the bar, which corresponds to the $|\ell| = 18\degree - 67 \degree$ range probed by our terminal velocity analysis. These torques also contribute to the redistribution of tangential velocities within the disc. A weaker bar\added[id = anon]{, with less density contrast between the nuclear ring vs. the shock cavities,} \deleted[id = anon]{would likely reduce the buildup of high tangential velocities in low-density gas zones within the bar,} would likely diminish the decrease in the in-plane gas velocities just outside the bar region. This would, in turn, lessen the increased deviation between simulated versus observed terminal curves that develops in the $|\ell| = 18\degree - 67 \degree$ range after a bar forms in Model 1. Additionally, a weaker bar would funnel less gas toward the centre, resulting in a smaller central mass concentration accumulating in the simulated galaxy, and a more moderate redistribution of mass in the rotation curve.

\subsubsection{Recovering a rotation curve from terminal velocities}\label{sec:rotation_curves}

\begin{figure*}
    \centering
    \includegraphics[width=\textwidth]{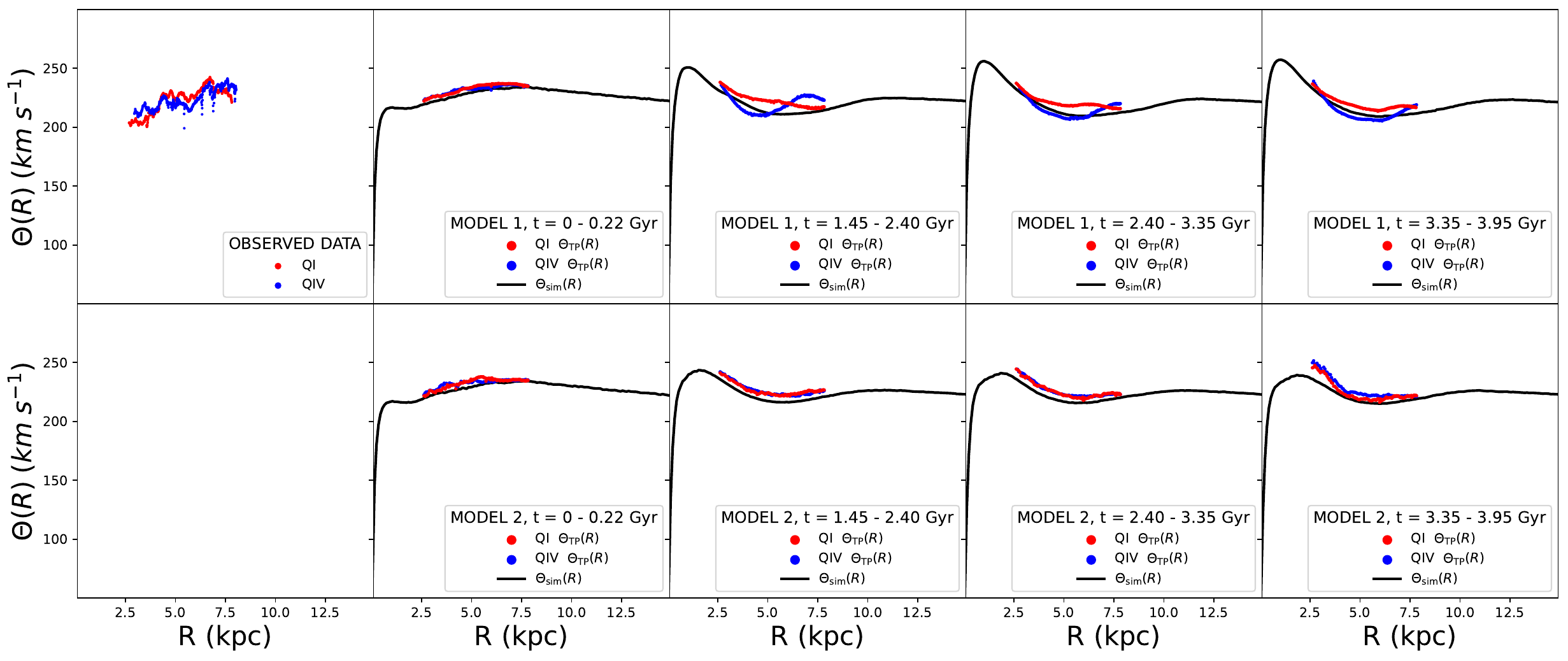}
    \caption{\added[id = anon]{We compare the azimuthally-averaged circular rotation curves of the simulations $(\Theta_{\rm sim}, \text{black curves})$, against the rotation curves we recover using the tangent-point method $(\Theta_{\rm TP}, \text{red and blue curves})$. Specifically, the red points come from applying the tangent-point method to terminal velocity values for Quadrant I of the respective model; blue is the equivalent for Quadrant IV.}}
    \label{fig:rot_curve}
\end{figure*}

Overlaid onto the $(x,y)$ gas distribution displayed in Fig.~\ref{fig:xy_allsim_grid} are the $(x,y)$ coordinates of the gas that contributes most to the terminal velocities (the `envelope' of the $(\ell, V_{\rm los})$ distribution) shown as the red and blue points. For the first snapshot in all the simulations, these $(x,y)$ points lie along a smooth arc. This is consistent with what we saw in Fig.~\ref{fig:terminal_points}, where we considered the case of purely circular motion, and observed that the terminal velocities occurred at tangent points to circular orbits. It is also consistent with the fact that the gas features a near perfect axisymmetric distribution in configuration space at this stage, with no visible structures yet formed.

In contrast, the $(x,y)$ coordinates of the terminal velocity gas points for $t = 2.06 - 3.35$~Gyr no longer follow a perfect arc, and show significant scatter. This is expected because we have non-axisymmetric perturbations, such as a bar and spiral arms, in the simulations. In other words, when the gas does not experience purely circular motion, the terminal velocities will no longer occur at the tangent-point to circular orbits. This implies that the tangent point method for deriving a rotation curve, which is commonly used to estimate $\Theta(R)$ values for the MW, faces many limitations because in reality gas does not experience purely circular motion in the MW.

\added[id = anon]{The limitations in using the tangent-point method to compute circular rotation curves are further illustrated in Fig.~\ref{fig:rot_curve}. The blue and red points are $\Theta_{\text{TP}}(R)$ values, which we recover by applying the tangent-point method to terminal velocities from Quadrant I (red) or IV (blue). On the other hand, the black curves show $\Theta_{\text{sim}}(R)$. This is the azimuthally-averaged circular velocity associated with the $m = 0$ mass distribution, calculated using the \texttt{pynbody} \texttt{v\_circ} profile, and given by: $v_{c}^{2} = R \partial\Phi_{0}/ \partial R$. This quantity is derived by evaluating the radial gravitational acceleration in the $(x, y)$ plane and calculating the corresponding circular velocity from the azimuthally averaged gravitational potential. If the gas in a simulation snapshot experiences purely circular motion, the tangent-point method should successfully recover the actual axisymmetrised circular velocity profile of the snapshot. In other words, for the case of pure circular motion the red and blue points $(\Theta_{\text{TP}}(R))$ should lie perfectly along the black curve $(\Theta_{\text{sim}}(R))$. However, we see in Fig.~\ref{fig:rot_curve} that the tangent-point method does not allow us to consistently recover the circular velocity curve associated with the axisymmetrised potential of simulation snapshots.}

\added[id = anon]{In the top left panel, we present the $\Theta_{\text{TP}}(R)$ values, recovered from applying the tangent-point method to \textit{observed} Quad I and IV \HI\ terminal velocities \citep{mcclure2016milky}. 
Remarkably, the $\Theta_{\text{TP}}(R)$ values recovered from the observed data for Quadrant I (red) versus Quadrant IV (blue) are very similar}, which is by no means obvious or expected in any way, given that the planar gas distribution in the MW is highly axisymmetric.\footnote{Another noteworthy similarity is that between the \HI\ and CO terminal velocity curves; see \citet{mcclure2016milky}, their fig. 2.}

In columns 2-5, we present the results for two of our models. 
\deleted[id = anon]{Here, the red and blue curves are the circular rotation speeds recovered when applying the tangent-point method to terminal velocities from the simulations.} The top row presents results from Model 1; the bottom row are the results from Model 2. Column 2 considers the beginning of each simulation, with $\Theta_{\rm sim}(R)$ values computed as the median across $t = 0 - 0.22$~Gyr. Columns 3 - 5 present $\Theta_{\rm sim}(R)$ results for $t = 2.06 - 3.95$~Gyr, equivalently. For Model 1, we do have a bar for $t = 2.06 - 3.95$~Gyr. Meanwhile, for Model 2, there is no bar for $t = 2.06 - 3.95$~Gyr. 

At the beginning of each model ($t = 0 - 0.22$~Gyr), the $\Theta_{\rm TP}(R)$ values recovered using the tangent-point method (red and blue curves) lie almost perfectly over the actual
\added[id = anon]{azimuthally-averaged} circular rotation curves from the simulation (black curve). This is unsurprising because little structure has formed in the disc at this point so the approximation of gas experiencing circular motion is appropriate. We make two key points about $t = 2.06 - 3.95$~Gyr (columns 2 - 5 of Fig. \ref{fig:rot_curve}): (i) The red and blue curves deviate significantly from each other for Model 1 (top row). In contrast, the red and blue curves agree better with each other for Model 2 (bottom row). \textit{In other words, for $t = 2.06 - 3.95$~Gyr in Fig. \ref{fig:rot_curve}, we recover different $\Theta(R)$ curves with the tangent-point method when using QI versus QIV terminal velocities, and this difference between the QI versus QIV recovered circular speeds is more pronounced for Model 1 compared to Model 2.} (ii) The red and blue curves deviate significantly from the underlying black rotation curve, especially for Model 1. This means that \textit{the $\Theta(R)$ values that we recover using the tangent-point method are not the same as the actual circular rotation curve values for the simulation}. The failure of the tangent-point method in recovering the simulations' actual circular speed curves, combined with the deviation between recovered $\Theta(R)$ values for QI versus QIV (for many snapshots of both models, but for Model 1 especially), suggests that we have significant non-circular and streaming motions in the gas for the $R_{t} \approx 2.6 - 7.8$~kpc radius range in which we do this analysis. These non-circular and streaming effects appear to be stronger in Model 1. We suggest that the bar is responsible for this extra streaming motion that we observe in the recovered rotation curves of Fig. \ref{fig:rot_curve} for Model 1.

Looking at Fig.~\ref{fig:termvel_grid_all}, another key question we ask is: \textit{why do the Model 2 terminal curves for $t = 2.06 - 3.95$~Gyr still deviate significantly from observations, despite the absence of a bar?} This deviation between simulated versus observed terminal curves is not as pronounced for Model 2, compared to Model 1, but the Model 2 $V_{t}$ values are still clearly not consistent with observations. One hypothesis we had to explain this inconsistency between Model 2 $V_{\rm t}$ values and observed data: the Galactic disc of Model 2 may be too reactive. We considered treating the Galactic disc as a two fluid system of stars and gas that gravitationally interact with each other. This is inspired by the work of \citet{jog1984two}, and is explained more thoroughly in Appendix \ref{sec:instabilites}. In this framework, the reactivity of the Galactic disc to perturbations on local scales is governed by the stability of the stars and the gas separately, as well as by the interaction of these two fluids with each other. If we decrease the reactivity of the entire Galactic disc system, the disc should be more resistant to forming clumps on local scales. We also hypothesized that a less reactive disc would exhibit a weaker response \deleted[id = anon]{to substructure formation} \added[id = anon]{to the formation of small structures}, resulting in smaller decreases in terminal velocity values in the $|\ell| = 18 \degree - 67 \degree$ range as the simulation evolved. To test our hypothesis, we first created Model 3. Model 2 has isothermal gas at a temperature of $T = 10^{3} \unit{K}$. In Model 3, this isothermal gas is heated to $T = 10^{4} \unit{K}$. Just as higher temperatures reduce the likelihood of cloud condensation in the atmosphere, we predicted that heating the gas will lower the disc's reactivity and reduce its tendency to form clumps on small scales. Unfortunately, when we look at the terminal velocity curves for Model 3 in Fig.~\ref{fig:termvel_grid_all}, they are not a better match to observations. In fact, Model 3's terminal velocity curves deviate even more significantly from the observations than Model 2's $-$ the reasoning for this is explained in Appendix \ref{sec:instabilites}, but for now we settle with the conclusion that our hypothesis behind Model 3 was not correct. Since the stability of the Galactic disc is governed by both the stars and the gas, this brings us to our motivation for Model 4. For Model 4, we kept the gas at $T = 10^{4} \; \unit{K}$, and we also warmed up the stellar disc, such that the disc has a higher vertical velocity dispersion. In Fig.~\ref{fig:termvel_grid_all}, we see that heating up the stellar disc has significantly reduced the reactivity of the disc to perturbations, and the simulated terminal velocity curves for Model 4 are now highly consistent with observations at all times. Unfortunately, this did not work out quite as we expected. If we look at the circular velocity curves of Fig.~\ref{fig:circ_vel_curves}, the $(x,y)$ gas density plots of Fig.~\ref{fig:xy_allsim_grid}, or even the A$_2$ values computed for the disc using Fourier analysis (Fig.~\ref{fig:A2_fourier}) all for Model 4, we can see that Model 4 hardly forms any structure over its entire duration; we essentially only ever see weak spiral structure develop. It is not surprising that the circular rotation and terminal velocity curves remain similar to their values at $t = 0$~Gyr, and do not deviate significantly from the observed data as time evolves if little structure ever forms in the model's disc. While for Model 4 we have terminal velocity curves that are more consistent with observations, we do not have a Galactic disc that resembles the Milky Way at all (there is no bar, and only weak spiral structure). Overall, we conclude: (i) a successful N-body model of the Milky Way needs to reproduce, after many Gyrs of evolution, both the large-scale morphology (including a bar and spiral arms) of the Galaxy and terminal velocity curves that match observations; and (ii) reducing the reactivity of the Galactic disc is not a sufficient strategy for achieving this. We suggest future steps forward on how to potentially resolve these problems in sections \ref{sec:future}.

Finally, we acknowledge that the terminal velocity curves, to some extent, do reflect the circular rotation curve. Crucially, our circular speed curves deviate from observed data as the simulations evolve. Particularly, in the $R \approx 2.6 - 7.8 \unit{kpc}$ radius range, which corresponds approximately to the $|\ell| = 18 \degree - 67 \degree$ longitude range, the $\Theta(R)$ values drop to values that are too low compared to observed data. We may then expect the $V_{\rm t}$ to also decrease significantly simply because of these changes in the rotation curve. The differences that emerge between our modelled circular rotation curves versus the observed data are definitely a limitation of this work, which is important to tackle for any future N-body MW-like models. Ideally, we want the simulated circular speed curves to match the observed data well, before using terminal velocity curves to constrain bar properties. We believe that our overarching argument, namely that the Model 1 bar is too strong, still holds true though. The separation between the modelled $\Theta(R)$ values recovered using the tangent-point method, for Quadrant I versus IV (as discussed in Section \ref{sec:rotation_curves} and presented in Fig.~\ref{fig:rot_curve}) independently suggest that the simulated bars are too strong. While our comparison of simulated versus observed terminal velocity data is clouded by problems with the rotation curve in Fig.~\ref{fig:termvel_grid_all}, we suggest the effects of the modelled bars being too strong are still present and partially responsible for the deviation between Model 1 versus observed terminal velocity curves.

\subsubsection{Constraining bar properties}\label{sec:constraining_bars}

\begin{figure*}
    \centering
    \includegraphics[height = 20cm]{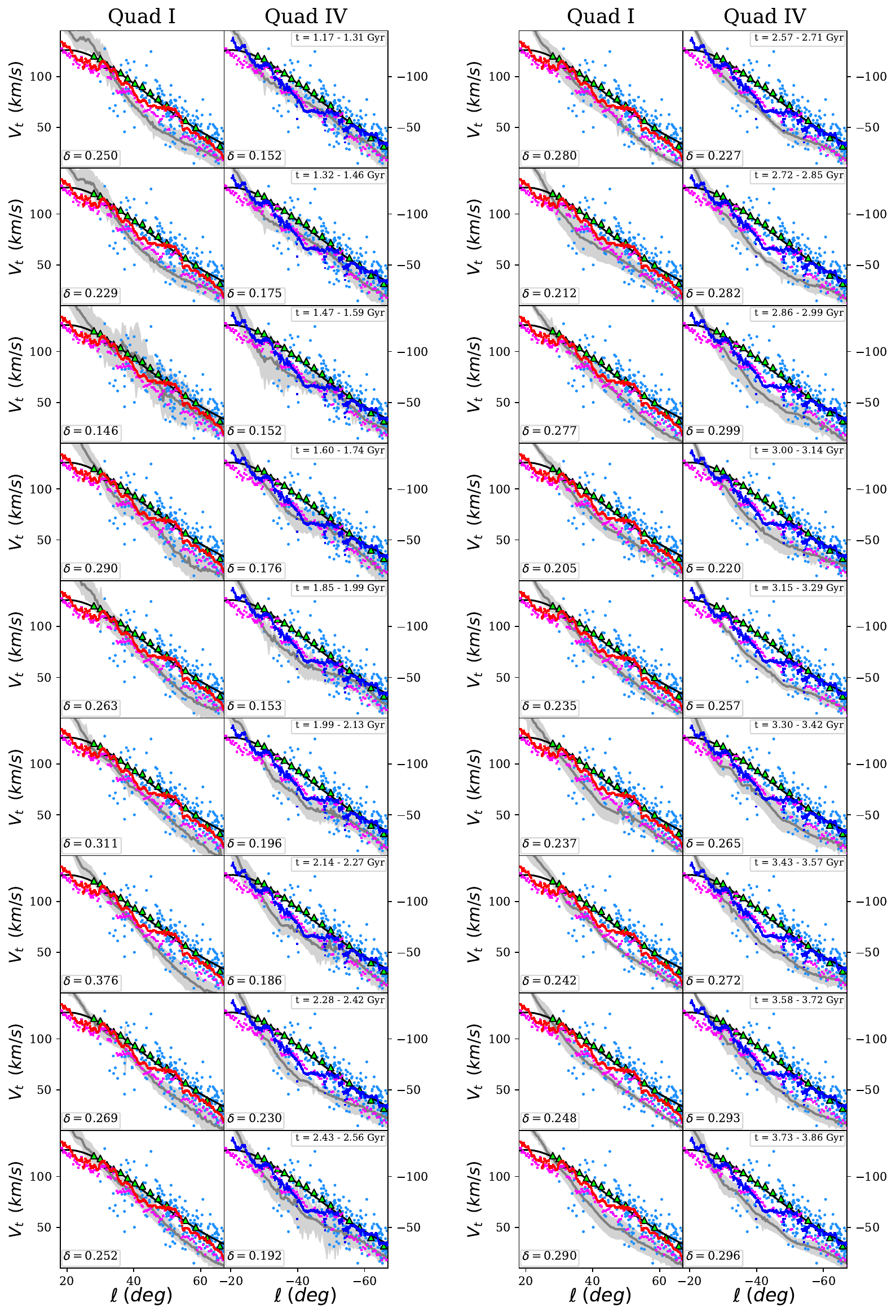}
    \caption{Terminal velocity analysis for $t = 1.17 - 3.86$~Gyr of Model 1. In other words, we compute and display the simulated terminal velocity curves for most snapshots in Model 1 where the bar is fully present. This allows us to more easily see how the simulated terminal velocity curves change as the bar evolves from fast (high $\Omega_{\rm p}$) to slow (low $\Omega_{\rm p}$). In contrast, in Fig.~\ref{fig:termvel_grid_all}, we only considered $t = 2.09 - 4.02$~Gyr in our terminal velocity analysis to make comparison between the various models easier.}
    \label{fig:termvel_allbar}
\end{figure*}

\begin{figure}
    \centering
    \includegraphics[width=0.5\textwidth]{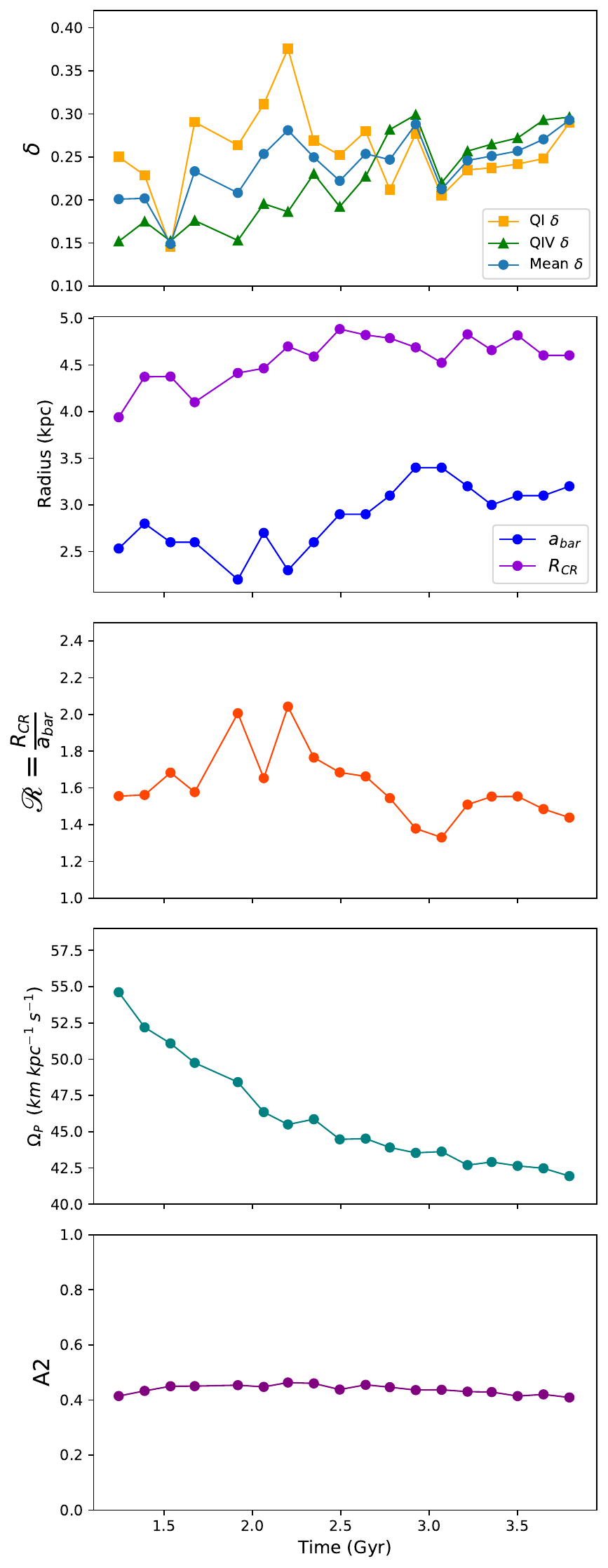}
    \caption{Tracking changes in various bar parameters, alongside the mean $\delta$ values from the terminal velocity curves, as Model 1 evolves. }
    \label{fig:bar_parameters}
\end{figure}

\begin{figure}
    \centering
    \includegraphics[width=0.4\textwidth]
    {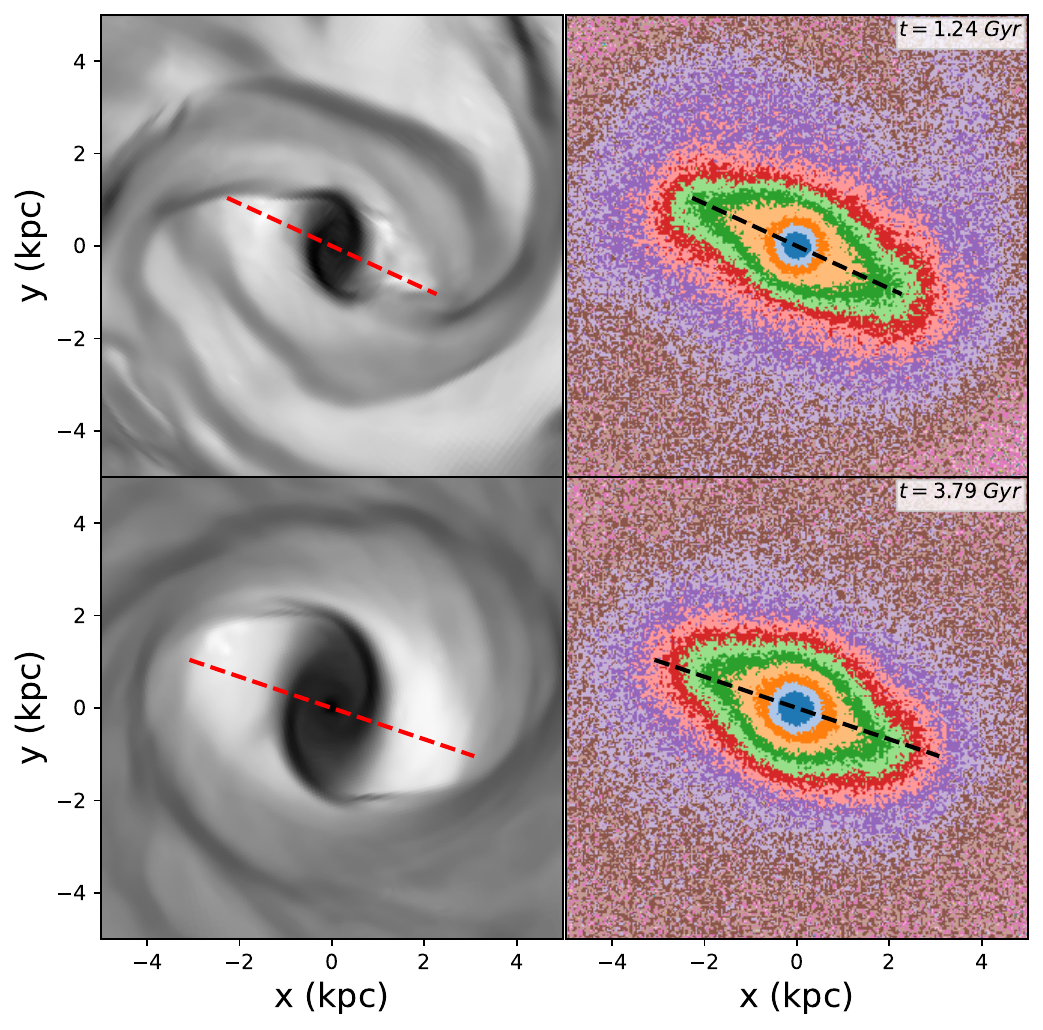}
    \caption{Visualising the method for measuring the bar length. The left panel shows projected $(x,y)$ gas density; the right panel shows projected $(x,y)$ stellar density. We draw a dashed line (red for gas, black for stars) across the semimajor axis of the cusped $x_{1}$ orbit. The location of the cusped $x_{1}$ orbit, one of the key orbits supporting the bar, is easiest to visualise in the $(x,y)$ gas density distribution because it encases a region of very low gas density within the bar region.}
    \label{fig:length_measurement}
\end{figure}

\begin{figure}
    \centering
    \includegraphics[width=0.5\textwidth]{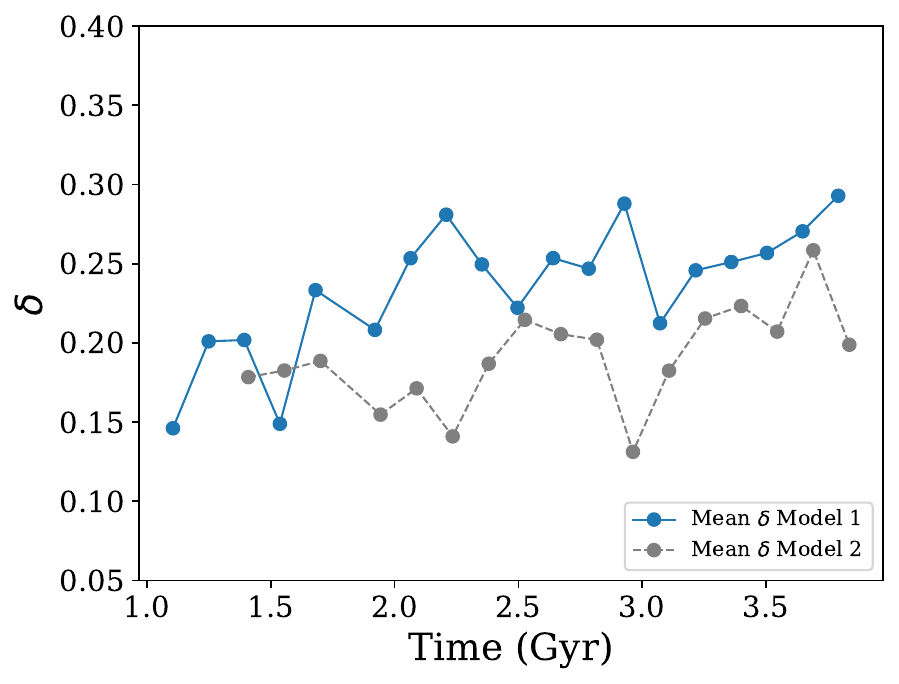}
    \caption{Comparison of mean $\delta$ (defined in Tab. \ref{tab:symbols}) for Models 1 versus 2. We observe that Model 1 mean $\delta$ values tend to be higher than the Model 2 ones for most of the snapshots where Model 1 has a bar (ie. the time range presented here). }
    \label{fig:delta_Model1vs2}
\end{figure}

\begin{figure*}
    \centering
    \includegraphics[width=\textwidth]{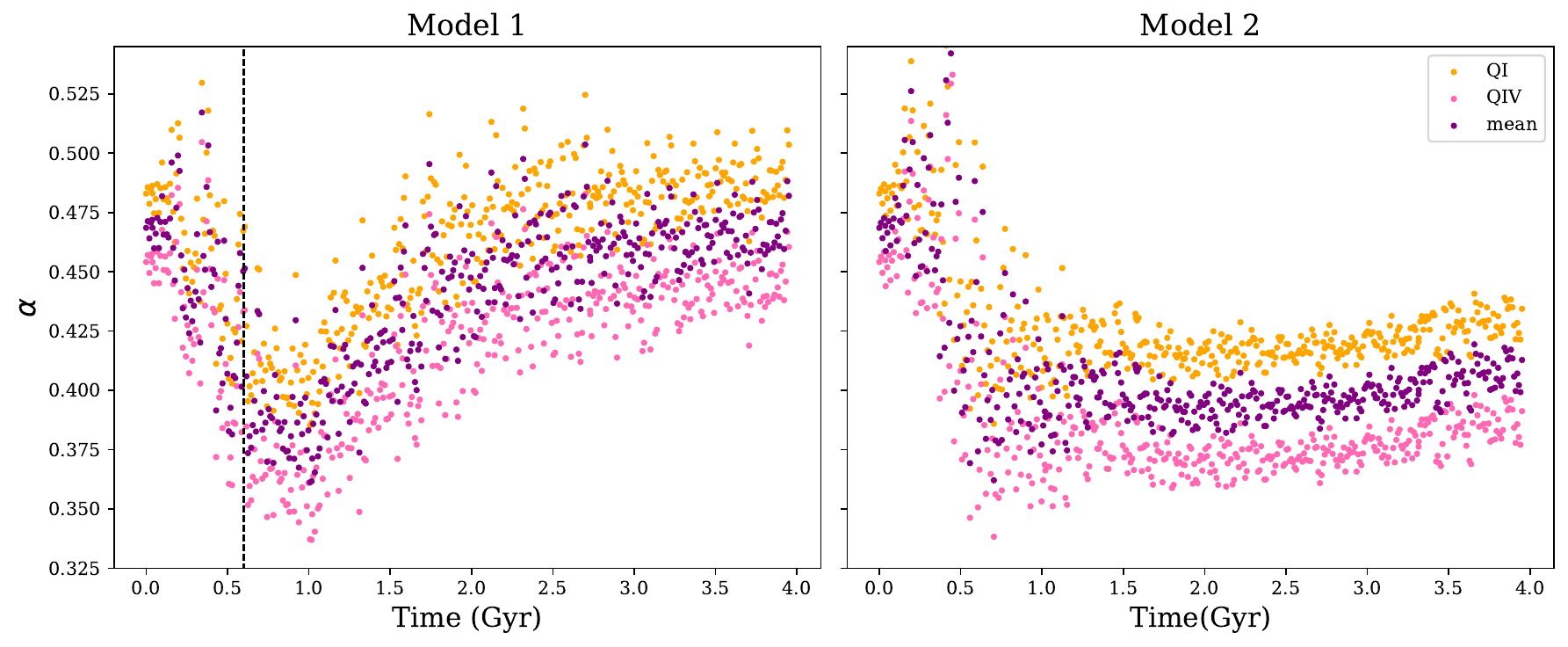}
    \caption{\added[id = anon]{$\alpha$ (defined in Tab. \ref{tab:symbols}) quantifies the difference between simulated azimuthally-averaged circular velocity curves, versus observed values, in the $R_{t} = 2.63 - 7.82$~kpc radius range. Left panel shows results for Model 1; right panel for Model 2. The black dashed line shows the approximate time of bar formation for Model 1.}}
    \label{fig:alpha_bothsims}
\end{figure*}

\begin{figure}
    \centering
    \includegraphics[width=0.4\textwidth]{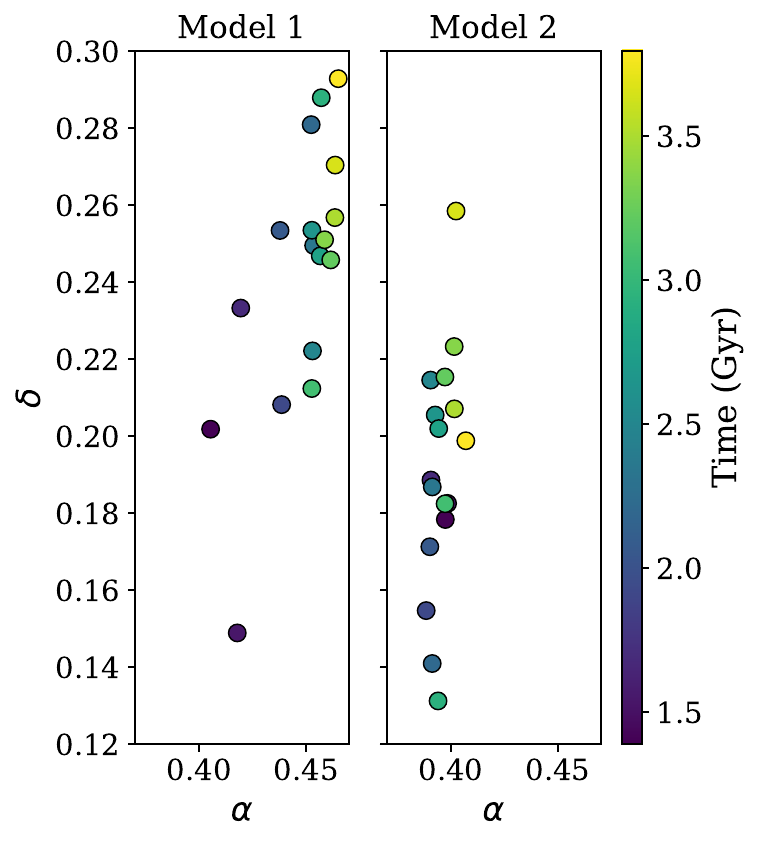}
    \caption{\added[id = anon]{$\alpha$ vs. $\delta$ (both defined in Tab. \ref{tab:symbols}) for the $t = 1.3 - 4.02$~Gyr time period. Each point is the mean value across a group of 15 snapshots. Points are coloured by time in the respective simulation. Left panel is for Model 1; right is for Model 2.}}
    \label{fig:alpha_vs_delta}
\end{figure}

\begin{figure*}
    \centering
    \includegraphics[width=0.8\textwidth]{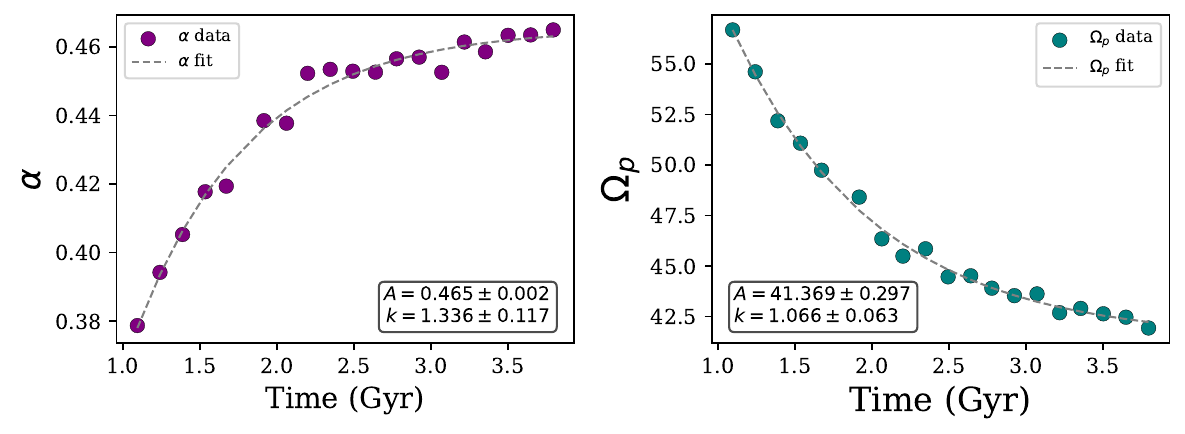}
    \caption{\added[id = anon]{Comparison of $\alpha$ vs. bar pattern speed in the  $t = 1.3 - 4.02$~Gyr, with each point representing a mean value across a group of 15 snapshots. We have performed an exponential fit, $f(t) = A - Be^{-kt}$ to each set of points, where A is the asymptotic value and k is the decay rate constant. }}
    \label{fig:alpha_omega}
\end{figure*}

\begin{figure*}
    \centering
    \includegraphics[width= \textwidth]{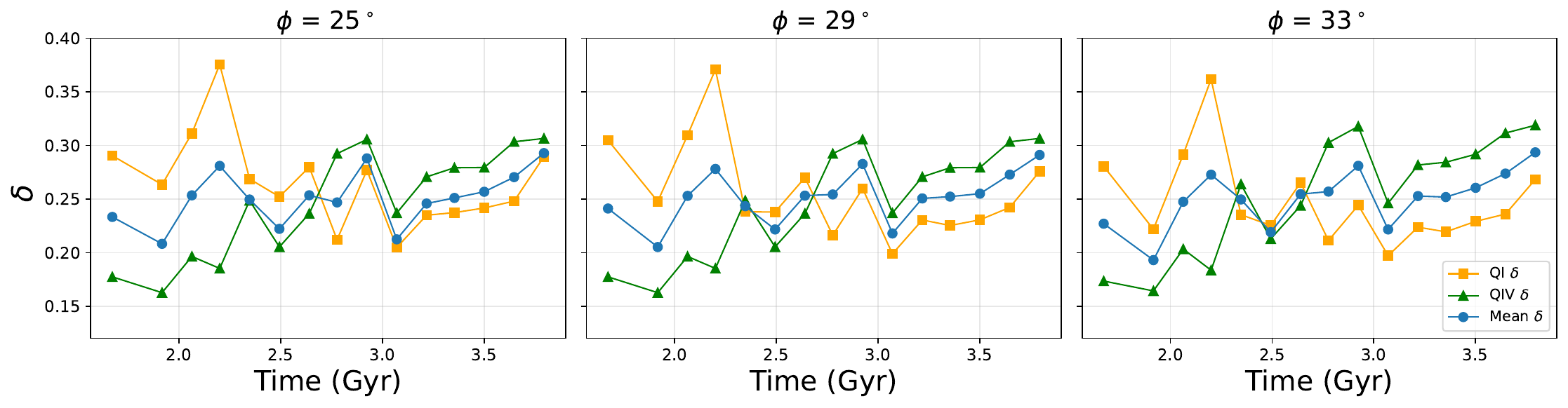}
    \caption{\added[id = anon]{$\delta$ values for different choices of bar angle in Model 1. Left panel is for a bar angle of $\phi = 25 \deg$; middle for $\phi = 29 \deg$; right for $\phi = 33 \deg$. Yellow points are for QI; green for QIV; and blue is an average across both quadrants.}}
    \label{fig:angle_vs_delta}
\end{figure*}

\begin{figure}
    \centering
    \includegraphics[width= 0.45\textwidth]{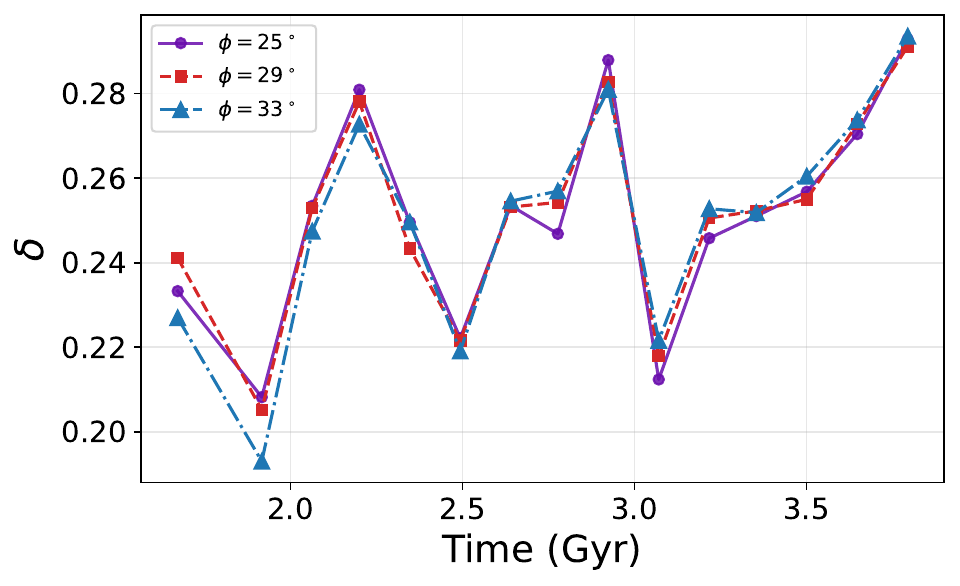}
    \caption{\added[id = anon]{We take the mean $\delta$ curves from Fig. \ref{fig:angle_vs_delta}, and present them together. This allows us to contrast the effect of the three different bar angles $\phi = 25 \deg, 29 \deg, 33 \deg$ on mean $\delta$.}}
    \label{fig:mean_delta_angle}
\end{figure}

In Fig.~\ref{fig:termvel_grid_all}, we only considered times $t= 0 - 0 .22$~Gyr and $t = 2.06 - 3.95$~Gyr for the terminal velocity analysis. This was done because Models 2 and 3 form a bar at some points in the $t = 0.22 - 2.06$~Gyr time period, but this bar does not last long in either simulation. It was easiest to compare the differences in the terminal velocity curves for Models 1 - 4, if we consistently have a bar for all $t = 2.06 - 3.95$~Gyr for Model 1, while Models 1 - 3 do not have a bar at any point in this time period. Nevertheless, Model 1 goes through a significant amount of evolution during $t = 0.22 - 2.06$~Gyr; this is the time period when the bar first forms in Model 1, and it starts off with a very high pattern speed, but then rapidly slows down. We want to consider how the terminal velocity curves change across the entire evolution of the Model 1 bar, starting from the point where the bar first forms and is very fast, and then ending at the time when the bar is significantly slower and longer.  For these reasons, in this section we do the same terminal velocity analysis as before, but we now include earlier times that were previously omitted, even though the Model 1 bar was present in these snapshots. 

Fig.~\ref{fig:termvel_allbar} presents the more in depth terminal velocity analysis for Model 1. The bar in Model 1 first becomes both fully formed and consistently detectable by \citet{dehnen2023measuring}'s code around $t = 1.0$~Gyr. In Fig.~\ref{fig:termvel_allbar}, we take $t = 1.17 - 3.86$~Gyr from Model 1, and divide this set of snapshots into groups of 15. Then we plot the mean simulated terminal velocity curves for these groups of 15 snapshots and step in time through Model 1 as the bar evolves. This way we are more able to directly compare how the terminal velocity curves change as the bar evolves from being very short and fast close to the start of its formation, to slow and longer toward the end of the simulation. In the bottom left corner of each subpanel, we have recorded the $\delta$ value, which quantifies the difference between simulated versus observed terminal velocity data for each group of 15 snapshots. 

Next, in the top panel of Fig.~\ref{fig:bar_parameters}, we plot the evolution of these $\delta$ values from Fig.~\ref{fig:termvel_allbar} across the simulation. This allows us to more clearly observe how $\delta$, or equivalently the difference between simulated versus observed terminal velocity data, changes as Model 1 evolves. Beneath this plot of $\delta$ versus time, we present the evolution of various bar parameters, including: bar semimajor axis length ($a_{\rm bar}$), the corotation radius ($R_{\rm CR}$), the bar rotation rate ($\mathscr{R}$), the pattern speed $\Omega_{\rm p}$, and the bar strength ($A_2$). 

In each case, we have taken the average value of each parameter across the same groups of 15 snapshots as was done for the terminal velocity analysis. We have presented this analysis by averaging values over groups of 15 snapshots, rather than for each individual snapshot, because we have measured the bar semimajor axis $a_{\rm bar}$ length by eye.

\added[id = anon]{The key point we want to make about the bar length values, $a_{\rm bar}$, in Fig. ~\ref{fig:bar_parameters} is that bar length appears to fluctuate with time, and these fluctuations are intrinsic to the bar. They are a result of the bar's actual size oscillating with time, or at least the size of the low density shock cavities in the bar region changing with time.
The fluctuations we observe here are not to be confused with oscillations in A$_2$, or equivalently oscillations in measured bar length that appear when using Fourier analysis to recover $a_{bar}$, since we have used visual inspection to measure bar length in Fig.~\ref{fig:bar_parameters}}.

Originally, we tried measuring the bar major axis length ($a_{\rm bar}$) using the code written by \citet{dehnen2023measuring}, which implements a Fourier analysis method. However, the bar length estimate from this code does not necessarily and consistently provide a satisfactory estimate of the simulated bar's actual length (W. Dehnen, {\em private communication}). This becomes obvious to us when plotting the $a_{\rm bar}$ values from \citet{dehnen2023measuring}'s code over $(x,y)$ stellar and gas density projections of the relevant snapshots. There are some snapshots where we see these estimated bar length values significantly overshooting the bar, such that the bar's estimated length appears to extend into the spiral arm and interarm regions that are clearly well beyond the bar. \added[id = anon]{Since we find Fourier methods to be highly unreliable and often inaccurate in measuring bar length, it is not appropriate to use such methods to constrain the bar length in this study, and hence we choose to use visual inspection instead.} Fig.~\ref{fig:length_measurement} demonstrates how we actually measure the $a_{\rm bar}$ values presented in Fig.~\ref{fig:bar_parameters}. The gas density distributions in Fig.~\ref{fig:length_measurement} have a distinct high density nuclear ring region at the centre of the bar, and there is also a clear low density region between the nuclear ring and the rough outer outline of the bar. The outline of this low density region likely marks the approximate location of the cusped $x_{1}$ orbit. Hence, to estimate the length of the bar major axis, we simply draw a red dashed line across the major axis of the cusped $x_{1}$ orbit. We present examples of this visual inspection method used to measure bar length in Fig.~\ref{fig:length_measurement}. Although a visual inspection method for measuring bar length works well here, we note that such a method is not ideal because it is difficult to apply consistently to all bars that can vary significantly in their properties (eg. cases where the bar is very weak, lopsided, or the low density region encased by the cusped $x_{1}$ orbit is less clear). We note that many of the pre-existing methods for measuring bar length have limitations, and highlight the need for a more formal definition of bar length and a more reliable bar length measurement method \citep{iles2025b}. 

In the $t = 3.79$~Gyr $(x,y)$ gas density distribution of Fig.~\ref{fig:length_measurement}, we also observe that the Model 1 bar has unusually low gas densities towards its ends at this time in the simulation. In fact, most snapshots at later times in Model 1 have low gas density regions near the bar ends. We believe this to be inconsistent with the MW, which most likely has high gas densities towards its ends because this is where the connecting arms, 3-kpc arm and 135-km/s arms are supposed to meet. Some good examples of what we expect these higher gas density bar ends to look like are given by the models of \citet{fux19993d, li2022gas}. The low gas density cavities in some snapshots of our Model 1 bar correspond to zones of outflowing gas. We suggest that the strong torques associated with our Model 1 bar drive very strong shocks, which also result in these shock cavities in the gas being too big.

Returning to Fig.~\ref{fig:bar_parameters}, we observe, on average, that $\delta$ increases as Model 1 evolves. This represents an increasing difference between simulated versus observed terminal velocity data over time in Model 1. As already discussed in Section \ref{sec:terminal_vel}, we argued that this is because of: (i) The Model 1 bar being too strong, which causes excess streaming and non-circular motions of gas in the inner disc, directly around the bar. (ii) Changes in the $(\Theta(R), R)$ circular rotation curve as Model 1 evolves over time (shown in Fig.~\ref{fig:circ_vel_curves}) also contribute to this increasing deviation between simulated versus observed $V_{\rm t}$ data.

In the second panel of Fig.~\ref{fig:bar_parameters}, we observe that, on average, the bar major axis length gradually increases and the corotation radius also gradually moves outward. There is some small fluctuation in the bar length on short timescales such that the bar length does not consistently grow over time, but rather occasionally the bar length will decrease slightly before increasing again. Since the bar length measurements were made by visual inspection, we highlight that these mild oscillations in the bar length as the model evolves are intrinsic to the bar and not a result of the measurement method, since we know that apparent fluctuations in bar length can also be introduced when attempting to measure $a_{\rm bar}$ using Fourier methods. In the third panel of Fig.~\ref{fig:bar_parameters}, we present the bar rotation rate parameter $\mathscr{R} = \frac{R_{CR}}{a_{\rm bar}}$. There in no clear trend in the evolution of $\mathscr{R}$ in Model 1 over time. The median $\mathscr{R}$ value is 1.56. This indicates that the ends of the bar are located a reasonable distance within the corotation radius for most of the bar's lifetime in this model. 

In the fourth subpanel of Fig.~\ref{fig:bar_parameters}, we present the pattern speed, $\Omega_{\rm p}$, of the bar evolving over time. This was computed using the Tremaine-Weinberg method implemented in \citet{dehnen2023measuring}'s code. Many studies identify  $\mathscr{R} < 1.4$ with fast bars, and $\mathscr{R} > 1.4$ with slow bars \citep{athanassoula1992existence, elmegreen1996pattern, debattistA2000constraints}. \added[id = anon]{However, we do not observe any correlation between $\mathscr{R}$ and $\Omega_{\rm p}$, or bar slow down, in Model 1. $\Omega_{p}$ in Model 1 almost continuously decreases, whereas $\mathscr{R}$ fluctuates considerably and shows no clear trend with time.  Additionally, there is no theoretical explanation as to why there should be a correlation between $\mathscr{R}$ and bar slow down. \citet{athanassoula2002formation,athanassoula2013bars} have shown that as the $R_{CR}$ moves outward, $a_{bar}$ can increase in such a way that the ratio $\mathscr{R}$ remains approximately constant. In our case, we suggest that the lack of correlation between $\mathscr{R}$ and $\Omega_{\rm p}$ is partly because of the circular rotation curve changing substantially as the simulation evolves, and this affects the rate at which the $R_{CR}$ moves outward as the bar grows, in comparison to what would happen if the circular rotation curve was flat \citep{debattistA2000constraints}. As has been pointed out by others \citep{font2017kinematic}, and as we observe for Model 1, the rotation rate parameter $\mathscr{R}$ is not always a clear reflection of the underlying bar pattern speed.}

In the final row of Fig.~\ref{fig:bar_parameters}, we plot A$_2$ against time. These A$_2$ values are computed by taking the mean A$_2$ across groups of 15 snapshots, and so are different to the A$_2$ curves displayed in Fig.~\ref{fig:fourier_all}, where the maximum A$_2$ value for each individual snapshot is displayed. For the time period that we have selected here (which is the time period for which the bar is properly formed, has already buckled, and is consistently detectable by \citet{dehnen2023measuring}'s code) we note that $A_{\rm 2}$ remains approximately constant for $\mathrm{\approx 3 \; \unit{Gyr}}$.

In Fig.~\ref{fig:delta_Model1vs2}, we compare the mean $\delta$ values for Model 1 versus Model 2 for the time period where Model 1 has a bar. As just stated, for the points shown in this Figure, Model 1 does have a bar, meanwhile Model 2 does not. We have deliberately omitted the first two data points for the Model 2 $\delta$ values because the Model 2 bar is still dissolving at these points. The idea behind contrasting these two models is to further explore the effect that the bar has on the $\delta$ values, and how the $\delta$ values would differ if a bar was not present (as happens in Model 2). We see that for most times presented, Model 1 has slightly higher $\delta$ values than Model 2. Once again, we argue that this is partly a result of the Model 1 bar being too strong, and partly a reflection of the evolving circular velocity curves.

\subsubsection{Circular Velocity Curve Deviations}

\added[id = anon]{We established that the tangent-point method is particularly useful because it allows us to directly convert between circular velocities $(\Theta(R))$ and terminal velocities $(V_{t})$, for the case of pure circular motion. As Model 1 evolves, we also witnessed its circular velocity profile deviate from observed $\Theta(R)$ values for the actual Milky Way (see Fig. \ref{fig:circ_vel_curves}). Deviations between the simulation's azimuthally-averaged circular velocity curve and the Milky Way's actual $(\Theta(R), R)$ curve will single-handedly create differences between modelled versus observed terminal velocities. Here, we quantify the differences between the simulated versus observed circular velocity profiles in order to examine how much these differences contribute to our calculated $\delta$ values. We define $\alpha$:} 
\added[id = anon]{\begin{equation}
    \alpha = \sqrt{\frac{1}{N} \sum_{i = 1}^{N} \frac{(\hat{y_{i}} - y_{i})^{2}}{y_{i}^{2}}}
\end{equation}}

\added[id = anon]{In this case, $\hat{y_{i}}$ are the $\Theta_{\rm sim}(R)$ values from the simulation, and $y_{i}$ is the mean of all observed circular velocity values shown in Fig. \ref{fig:circ_vel_curves} (coloured points). $\alpha$ is computed within the $R_{t} = 2.63 - 7.82$~kpc radius range, since this is approximately equivalent to the $|\ell| = 18\degree - 67\degree$ longitude range of the terminal velocity analysis. N is the number of bins in radius and is set to 52. In Fig. \ref{fig:alpha_bothsims}, QI $\alpha$ is shown in orange---this is calculated using only the observed circular velocities applicable to QI (ie. QIV $\HI$ terminal velocities from \citet{mcclure2016milky} are omitted). Similarly, QIV $\alpha$ is shown in pink, and uses only observed $\Theta(R)$ values relevant to QIV. The mean $\alpha$ across both quadrants is shown in purple. There is a slight offset between the QI and QIV $\alpha$ values, arising from small differences in the observed circular velocity data used. The vertical scale of Fig.~\ref{fig:alpha_bothsims} makes this offset appear more pronounced.}

\added[id=anon]{The most striking feature of Fig.~\ref{fig:alpha_bothsims} is the difference in $\alpha$ between Models 1 and 2. Early in both models, the simulated circular velocity curves are slightly offset from the observed mean $\Theta(R)$ values in the $R_{t} = 2.63$--$7.82$~kpc range. When the simulations first begin to evolve, the initial $\alpha$ values gradually decrease. In Model 1, the black dashed line marks the approximate time of bar formation. Soon after the bar forms, $\alpha$ reaches a minimum. Then, $\alpha$ starts to increase again as the bar grows fatter and slows down, eventually returning to roughly its initial value. The simulated $\Theta_{\rm sim}(R)$ curve in Model 1 most closely matches the observations shortly after bar formation. Model 2 shows a similar early decline in $\alpha$, reaching its minimum slightly earlier than Model 1. However, the $\alpha$ values for Model 2 never increase back up like they did for Model 1. The Model 2 bar is short-lived, persisting only between $t \approx 0.49$ and $1.20$~Gyr, which approximately corresponds to the period where $\alpha$ is at its lowest. The significant rise in $\alpha$ in Model 1 at later times reflects the impact of its evolving bar, as also seen in the dip of the Model 1 circular velocity curve in Fig.~\ref{fig:circ_vel_curves}. The fact that the circular velocity curves best match the observations shortly after bar formation could be an argument for a younger, fast Galactic bar.}

\added[id=anon]{Figure~\ref{fig:alpha_vs_delta} shows the relationship between $\alpha$ and $\delta$ for the groups of 15 snapshots in the $t = 1.3$--$4.02$~Gyr period considered earlier. Each point represents the mean $\alpha$ and mean $\delta$ for a group of snapshots, with points coloured by time. In Model 2, where no bar is present during this period, $\alpha$ shows only minor fluctuations, as seen in Fig.~\ref{fig:alpha_vs_delta}. However, $\delta$ increases steadily, indicating that changes in $\delta$—or equivalently in the simulated terminal velocity curves—are not simply driven by changes in the $(\Theta_{\rm sim}(R), R)$ curve. By contrast, Model 1 shows significant evolution in both $\alpha$ and $\delta$ during this time. $\alpha$ asymptotes toward $\alpha \approx 0.47$. There is a time period toward the end of the model, where $\alpha$ remains approximately constant, while $\delta$ continues to rise. This suggests that the ``dip'' feature in the Model 1 circular velocity curve eventually stabilises, while the terminal velocity curves continue to evolve beyond this point.}

\added[id=anon]{Figure~\ref{fig:alpha_omega} shows the mean $\alpha$ and $\Omega_{p}$ values for the groups of 15 snapshots in the $t = 1.3$--$4.02$~Gyr period. We fit an exponential function,
$f(t) = A - B e^{-k t}$, to the data, where $A$ is the asymptotic value and $k$ is the decay rate. The fitted values of $A$ and $k$ are shown in Fig.~\ref{fig:alpha_omega}. There is a clear link between $\alpha$ approaching a steady value and the bar pattern speed asymptoting as the bar slows. The circular velocity curve in Model 1 stabilises around the same time that the bar’s slowdown ends. Notably, the fitted decay rate for $\alpha$ ($k \approx 1.35$) is slightly higher than that for $\Omega_{p}$ ($k \approx 1.1$), suggesting that the circular velocity curve stabilises slightly faster than the bar slows.}

\subsubsection{Bar angle and terminal velocities}

\added[id=anon]{Throughout this paper, we have adopted a bar inclination angle of $\phi = 25^\circ$ relative to the $x$-axis, consistent with \citet{fux19993d}, who suggested $\phi = 25^\circ$. However, estimates of the Galactic bar angle vary widely, with the long bar estimated at $\phi = 28^\circ$--$33^\circ$ and the B/P bulge at $\phi = 27^\circ \pm 2^\circ$ \citep{bland2016galaxy}. In Fig.~\ref{fig:angle_vs_delta}, we explore how varying $\phi$ affects the terminal velocity curves through $\delta$. We again use groups of 15 snapshots from $t = 1.3$--$4.02$~Gyr, with each point representing a mean $\delta$ value. We consider $\phi = 25^\circ$, $29^\circ$, and $33^\circ$, as labelled in each subpanel.}

\added[id=anon]{Overall, the $\delta$ trends are similar across the three angles, but there are some important differences. At earlier times, $\delta$ values in Quadrants I and IV differ significantly for all bar angles. However, at later times—especially for $\phi = 25^\circ$—the $\delta$ values for QI and QIV converge. Increasing the bar angle delays this convergence; larger $\phi$ values ($29^\circ$ and $33^\circ$) maintain asymmetries between QI and QIV for longer. In general, we may expect larger bar angles to lead to greater asymmetry between the QI and QIV terminal curves because a more inclined bar will introduce stronger asymmetry about $x = 0$ in $(x,y)$ space.}

\added[id=anon]{Furthermore, we note that the observed $\mathrm{H\textsc{i}}$ terminal velocities from \citet{mcclure2016milky} are highly symmetric between QI and QIV, potentially favouring a smaller bar angle (or a bar that has evolved for a longer time period). Star formation may also contribute to asymmetry. SFR creates a plume of high velocity points that are likely to affect the terminal curves. Early in Model 1, when the SFR is higher, asymmetric features in $(\ell, V_{\rm los})$ space are more pronounced.}

\added[id=anon]{In Fig.~\ref{fig:mean_delta_angle}, we directly compare the mean $\delta$ values for the three bar angles and find them nearly identical. This shows that varying the bar angle has little impact on the overall mean deviation between the terminal curves and observations across both quadrants. The primary effect of changing the bar angle is the variation in asymmetry between Quadrants I and IV.}

\section{Future steps and direction }\label{sec:future}
Building a working model of a complex system lies at the heart of many subfields of physics. There are concerted efforts to understand the Earth's atmosphere, for example, on short and long timescales. The terrestrial ecosystem has immense complexity where increasing amounts of data are becoming available with each passing year, demanding increased sophistication and improvements to our atmospheric models. 

There are strong parallels here with ongoing efforts to understand the Galactic ecosystem. Several groups are attempting to build an N-body `surrogate' model of the Milky Way \citep{tep21v,hortA2024proto}, but all efforts to date fall short. A major pitfall is that, as we have shown, one can construct an equilibrium figure that matches the azimuthally-averaged Milky Way as we know it today. But when the system is allowed to evolve, the baryons can become redistributed through strong, non-axisymmetric dynamical modes leading to spiral-arm and/or bar formation. In fact, a central bar {\it is} observed in the Milky Way but now the azimuthally-averaged properties no longer match those observed. This is precisely the situation that besets our new work.

At the present time, we do not have a theory for how to set up a prescribed bar model relevant to the Milky Way that is an equilibrium figure at the outset. Nonlinear systems typically exhibit sensitive dependence on initial conditions where small changes in the initial conditions can lead to very different outcomes. This is a hallmark of complex systems (e.g. high dimensionality), making it hard to predict which specific conditions will result in a desired outcome. \added[id = anon]{This problem is well recognized in galactic dynamics \citep{binney2011galactic, 2016Aumer, sellwood2009stochasticity, weinberg1998fluctuations}}.

Our goal here has been to produce a `surrogate model' with more realism given the inclusion of a multiphase ISM maintained by star formation and metal production, induced turbulence \citep[cf.][]{renaud2013sub}, and a hot corona with halo accretion and recycling \citep{tep24a}.

This work has identified that either the simulated gas is too responsive compared to observed gas, or the simulated bar generates too much torque compared to the Milky Way's bar, or possibly both are true. Our next steps are (i) to improve the physics of the gas phases by introducing MHD and cosmic-ray heating, (ii) to run the models for the full lifetime of the Milky Way. A longer timespan may weaken the bar and lessen much of the observed gas streaming. The coupling of star formation and supernovae, or halo accretion and recycling, or both, may lead to gas being sufficiently overpressured to be less responsive to dynamical forces. All of these processes, and more, will assist in our understanding of the Galaxy's short-term and long-term evolution.

\section{Conclusions}\label{sec:Conclusions}
One of the main goals of this work is to construct hydrodynamic N-body `surrogate' models of the Milky Way. A first step toward this goal was made by \citet{tep21v}, who presented isolated N-body MW-like simulations of only stars and dark matter (no gas), and assessed the ability of the stellar component in these models to reproduce observations of the actual Galaxy. In this paper, we extend the original work with four new `surrogate' models of the Milky Way. Model 1 is the most sophisticated because it includes multiphase gas, star formation, stellar feedback and chemical enrichment. Models 2 - 4 include an isothermal gas disc.

While \citet{tep21v} focused exclusively on the stellar component, here we compare the gas dynamics of these more advanced N-body models against the actual MW. We extract the simulated cold gas terminal velocities from our models and compare them to the observed Quadrants I and IV defined in Galactic coordinates. We attempt to recover circular rotation speed curves from our terminal velocities, and assess the accuracy of the tangent-point method in allowing us to do this using the simulations. Our focus is on the gas dynamics in the zone directly around the bar since this is the region covered by the observations \citep{mcclure2016milky}.

The key conclusions of this work are summarised below. We cite some relevant results from \citet{tep24a, mcclure2016milky} to help put our work in context. 

\begin{enumerate}

    \item We present four new hydrodynamic N-body ``surrogate'' models of the MW, which have been constructed with the \nexus{} framework \citep{tep24a}. We observe that star formation and accompanying turbulence have a major impact on the bar strength and \added[id = anon]{bar lifetime in simulations \citep[see also][]{bland2024turbulent, weinberg2024impact}}.

    \item The \textit{observed} terminal velocities for QI and QIV \HI, and for CO, are all very similar to each other, despite the large-scale and fine-scale distributions of atomic gas being distinctly different to the dense molecular gas in the Galaxy. When these \textit{observed} terminal velocities are used to recover circular rotation speeds, the circular rotation speeds for QI and QIV \HI, and for CO, are also all very similar to each other \citep{mcclure2016milky}. 

    \item When we take QI versus QIV terminal velocities from snapshots of the \textit{simulations} that contain a bar, and apply the tangent-point method, the recovered circular rotation speeds are distinctly different for QI versus QIV. The recovered rotation speeds are also distinctly different to the actual rotation curves of the relevant snapshots.  

    \item When the simulations form a bar, the associated simulated terminal velocities begin to deviate more from the observed terminal velocities, compared to times before the bar formed. 
    
    \item We argue that the deviation between the actual rotation curve versus recovered rotation (from tangent-point method), and the increased deviation between simulated versus observed terminal velocities after a bar forms, is due to gas streaming in the region of the Galaxy around the bar, which reflects non-circular motions of the gas. 
    
    \item The streaming of the gas in the region around the simulated bar is much stronger than the bar-driven gas streaming in the Galaxy.
    
    \item The simulated gas distribution is almost completely evacuated at the ends of the bar, unlike what is observed in the Galaxy.

    \item In earlier work, many authors have used the amplitude of the $m = 2$ Fourier mode (A$_2$) as a proxy for bar strength. But we recommend the use of the bar's torque as a more physical proxy as this is directly implicated in gas streaming.
    
    \item Given these disparities, we conclude that the torque action of our modelled bars is too strong compared to the Milky Way's bar. This calls for new evolutionary models leading to bars that are either less massive, or have a different structure (e.g. more vertically extended or centrally concentrated), or have a lower pattern speed, or have a faster deceleration, or a combination of these.
\end{enumerate}

\section*{Acknowledgements}

We acknowledge helpful conversations with the Galacticos group at the Sydney Institute for Astronomy and, in particular, we would like to thank Elizabeth Iles. \added[id = anon]{We also thank the anonymous referee for their very useful and constructive feedback.} TTG acknowledges financial support from the Australian Research Council (ARC) through an Australian Laureate Fellowship awarded to JBH. The computations and data storage were enabled by two facilities: (i) the National Computing Infrastructure (NCI) Adapter Scheme, provided by NCI Australia, an NCRIS capability supported by the Australian Government; and (ii) LUNARC, the Centre for Scientific and Technical Computing at Lund University (resource allocations LU 2023/2-39 and LU 2023/12-6).

\section*{Data Availability}

The software data underlying this article will be shared on reasonable request to the corresponding author.


\bibliographystyle{mnras}
\bibliography{mnras_template} 

\begin{thebibliography}{}
\makeatletter
\relax
\def\mn@urlcharsother{\let\do\@makeother \do\$\do\&\do\#\do\^\do\_\do\%\do\~}
\def\mn@doi{\begingroup\mn@urlcharsother \@ifnextchar [ {\mn@doi@} {\mn@doi@[]}}
\def\mn@doi@[#1]#2{\def\@tempa{#1}\ifx\@tempa\@empty \href {http://dx.doi.org/#2} {doi:#2}\else \href {http://dx.doi.org/#2} {#1}\fi \endgroup}
\def\mn@eprint#1#2{\mn@eprint@#1:#2::\@nil}
\def\mn@eprint@arXiv#1{\href {http://arxiv.org/abs/#1} {{\tt arXiv:#1}}}
\def\mn@eprint@dblp#1{\href {http://dblp.uni-trier.de/rec/bibtex/#1.xml} {dblp:#1}}
\def\mn@eprint@#1:#2:#3:#4\@nil{\def\@tempa {#1}\def\@tempb {#2}\def\@tempc {#3}\ifx \@tempc \@empty \let \@tempc \@tempb \let \@tempb \@tempa \fi \ifx \@tempb \@empty \def\@tempb {arXiv}\fi \@ifundefined {mn@eprint@\@tempb}{\@tempb:\@tempc}{\expandafter \expandafter \csname mn@eprint@\@tempb\endcsname \expandafter{\@tempc}}}

\bibitem[\protect\citeauthoryear{{Agertz} et~al.,}{{Agertz} et~al.}{2021}]{age21l}
{Agertz} O.,  et~al., 2021, \mn@doi [\mnras] {10.1093/mnras/stab322}, \href {https://ui.adsabs.harvard.edu/abs/2021MNRAS.503.5826A} {503, 5826}

\bibitem[\protect\citeauthoryear{Aguerri, M{\'e}ndez-Abreu  \& Corsini}{Aguerri et~al.}{2009}]{aguerri2009population}
Aguerri J.,  M{\'e}ndez-Abreu J.,   Corsini E.,  2009, Astronomy \& Astrophysics, 495, 491

\bibitem[\protect\citeauthoryear{Anderson, Debattista, Erwin, Liddicott, Deg  \& Beraldo~e Silva}{Anderson et~al.}{2022}]{anderson2022secular}
Anderson S.~R.,  Debattista V.~P.,  Erwin P.,  Liddicott D.~J.,  Deg N.,   Beraldo~e Silva L.,  2022, Monthly Notices of the Royal Astronomical Society, 513, 1642

\bibitem[\protect\citeauthoryear{Antoja et~al.,}{Antoja et~al.}{2014}]{antoja2014constraints}
Antoja T.,  et~al., 2014, Astronomy \& Astrophysics, 563, A60

\bibitem[\protect\citeauthoryear{Athanassoula}{Athanassoula}{1992}]{athanassoula1992existence}
Athanassoula E.,  1992, Monthly Notices of the Royal Astronomical Society, 259, 345

\bibitem[\protect\citeauthoryear{Athanassoula}{Athanassoula}{2002}]{athanassoula2002formation}
Athanassoula E.,  2002, arXiv preprint astro-ph/0209438

\bibitem[\protect\citeauthoryear{Athanassoula}{Athanassoula}{2013}]{athanassoula2013bars}
Athanassoula E.,  2013, Secular Evolution of Galaxies, 305

\bibitem[\protect\citeauthoryear{Athanassoula, Machado  \& Rodionov}{Athanassoula et~al.}{2013}]{athanassoulA2013bar}
Athanassoula E.,  Machado R.~E.,   Rodionov S.,  2013, Monthly Notices of the Royal Astronomical Society, 429, 1949

\bibitem[\protect\citeauthoryear{{Aumer}, {Binney}  \& {Sch{\"o}nrich}}{{Aumer} et~al.}{2016}]{2016Aumer}
{Aumer} M.,  {Binney} J.,   {Sch{\"o}nrich} R.,  2016, \mn@doi [\mnras] {10.1093/mnras/stw777}, \href {https://ui.adsabs.harvard.edu/abs/2016MNRAS.459.3326A} {459, 3326}

\bibitem[\protect\citeauthoryear{Barbanis \& Woltjer}{Barbanis \& Woltjer}{1967}]{barbanis1967orbits}
Barbanis B.,  Woltjer L.,  1967, Astrophysical Journal, vol. 150, p. 461, 150, 461

\bibitem[\protect\citeauthoryear{{Benjamin} et~al.,}{{Benjamin} et~al.}{2005}]{glimpse}
{Benjamin} R.~A.,  et~al., 2005, \mn@doi [\apjl] {10.1086/491785}, \href {https://ui.adsabs.harvard.edu/abs/2005ApJ...630L.149B} {630, L149}

\bibitem[\protect\citeauthoryear{Binney \& Tremaine}{Binney \& Tremaine}{2011}]{binney2011galactic}
Binney J.,  Tremaine S.,  2011, Galactic Dynamics.
Princeton Series in Astrophysics, Princeton University Press

\bibitem[\protect\citeauthoryear{Binney, Gerhard, Stark, Bally  \& Uchida}{Binney et~al.}{1991}]{binney1991understanding}
Binney J.,  Gerhard O.~E.,  Stark A.~A.,  Bally J.,   Uchida K.~I.,  1991, Monthly Notices of the Royal Astronomical Society, 252, 210

\bibitem[\protect\citeauthoryear{Binney, Gerhard  \& Spergel}{Binney et~al.}{1997}]{binney1997photometric}
Binney J.,  Gerhard O.,   Spergel D.,  1997, Monthly Notices of the Royal Astronomical Society, 288, 365

\bibitem[\protect\citeauthoryear{Bissantz, Englmaier  \& Gerhard}{Bissantz et~al.}{2003}]{bissantz2003gas}
Bissantz N.,  Englmaier P.,   Gerhard O.,  2003, Monthly Notices of the Royal Astronomical Society, 340, 949

\bibitem[\protect\citeauthoryear{Bland-Hawthorn \& Gerhard}{Bland-Hawthorn \& Gerhard}{2016}]{bland2016galaxy}
Bland-Hawthorn J.,  Gerhard O.,  2016, Annual Review of Astronomy and Astrophysics, 54, 529

\bibitem[\protect\citeauthoryear{{Bland-Hawthorn}, {Tepper-Garcia}, {Agertz}  \& {Freeman}}{{Bland-Hawthorn} et~al.}{2023}]{bla23a}
{Bland-Hawthorn} J.,  {Tepper-Garcia} T.,  {Agertz} O.,   {Freeman} K.,  2023, \apj, 947, 80

\bibitem[\protect\citeauthoryear{Bland-Hawthorn, Tepper-Garcia, Agertz  \& Federrath}{Bland-Hawthorn et~al.}{2024}]{bland2024turbulent}
Bland-Hawthorn J.,  Tepper-Garcia T.,  Agertz O.,   Federrath C.,  2024, The Astrophysical Journal, 968, 86

\bibitem[\protect\citeauthoryear{Block, Puerari, Knapen, Elmegreen, Buta, Stedman  \& Elmegreen}{Block et~al.}{2001}]{block2001gravitational}
Block D.,  Puerari I.,  Knapen J.,  Elmegreen B.~G.,  Buta R.,  Stedman S.,   Elmegreen D.~M.,  2001, Astronomy \& Astrophysics, 375, 761

\bibitem[\protect\citeauthoryear{Bovy \& Rix}{Bovy \& Rix}{2013}]{bovy2013direct}
Bovy J.,  Rix H.-W.,  2013, The Astrophysical Journal, 779, 115

\bibitem[\protect\citeauthoryear{Burton \& Gordon}{Burton \& Gordon}{1978}]{burton1978carbon}
Burton W.,  Gordon M.,  1978, Astronomy and Astrophysics, vol. 63, no. 1-2, Feb. 1978, p. 7-27., 63, 7

\bibitem[\protect\citeauthoryear{Buta \& Block}{Buta \& Block}{2001}]{butA2001dust}
Buta R.,  Block D.,  2001, The Astrophysical Journal, 550, 243

\bibitem[\protect\citeauthoryear{Carlberg \& Sellwood}{Carlberg \& Sellwood}{1985}]{carlberg1985dynamical}
Carlberg R.,  Sellwood J.,  1985, Astrophysical Journal, Part 1 (ISSN 0004-637X), vol. 292, May 1, 1985, p. 79-89. Research supported by the Natural Sciences and Engineering Research Council of Canada and Science and Engineering Research Council of England., 292, 79

\bibitem[\protect\citeauthoryear{Carr \& Lacey}{Carr \& Lacey}{1987}]{carr1987dark}
Carr B.,  Lacey C.,  1987, Astrophysical Journal, Part 1 (ISSN 0004-637X), vol. 316, May 1, 1987, p. 23-35. Research supported by the University of Toronto., 316, 23

\bibitem[\protect\citeauthoryear{Chiba \& Sch{\"o}nrich}{Chiba \& Sch{\"o}nrich}{2021}]{chiba2021tree}
Chiba R.,  Sch{\"o}nrich R.,  2021, Monthly Notices of the Royal Astronomical Society, 505, 2412

\bibitem[\protect\citeauthoryear{Chomiuk \& Povich}{Chomiuk \& Povich}{2011}]{chomiuk2011toward}
Chomiuk L.,  Povich M.~S.,  2011, The Astronomical Journal, 142, 197

\bibitem[\protect\citeauthoryear{Clemens}{Clemens}{1985}]{clemens1985massachusetts}
Clemens D.~P.,  1985, Astrophysical Journal, Part 1 (ISSN 0004-637X), vol. 295, Aug. 15, 1985, p. 422-428, 431-436., 295, 422

\bibitem[\protect\citeauthoryear{Combes}{Combes}{2004}]{combes2004role}
Combes F.,  2004, Proceedings of the International Astronomical Union, 2004, 383

\bibitem[\protect\citeauthoryear{Contopoulos \& Mertzanides}{Contopoulos \& Mertzanides}{1977}]{contopoulos1977inner}
Contopoulos G.,  Mertzanides C.,  1977, Astronomy and Astrophysics, vol. 61, no. 4, Nov. 1977, p. 477-485., 61, 477

\bibitem[\protect\citeauthoryear{Dame et~al.,}{Dame et~al.}{1987}]{dame1987composite}
Dame T.,  et~al., 1987, Astrophysical Journal, Part 1 (ISSN 0004-637X), vol. 322, Nov. 15, 1987, p. 706-720. Research supported by the Naturvetenskapliga Forskningsradet., 322, 706

\bibitem[\protect\citeauthoryear{De~Lorenzi, Debattista, Gerhard  \& Sambhus}{De~Lorenzi et~al.}{2007}]{de2007nmagic}
De~Lorenzi F.,  Debattista V.~P.,  Gerhard O.,   Sambhus N.,  2007, Monthly Notices of the Royal Astronomical Society, 376, 71

\bibitem[\protect\citeauthoryear{Debattista \& Sellwood}{Debattista \& Sellwood}{2000}]{debattistA2000constraints}
Debattista V.~P.,  Sellwood J.,  2000, The Astrophysical Journal, 543, 704

\bibitem[\protect\citeauthoryear{Dehnen}{Dehnen}{2000}]{dehnen2000effect}
Dehnen W.,  2000, The Astronomical Journal, 119, 800

\bibitem[\protect\citeauthoryear{Dehnen, Semczuk  \& Sch{\"o}nrich}{Dehnen et~al.}{2023}]{dehnen2023measuring}
Dehnen W.,  Semczuk M.,   Sch{\"o}nrich R.,  2023, Monthly Notices of the Royal Astronomical Society, 518, 2712

\bibitem[\protect\citeauthoryear{{Drimmel} et~al.,}{{Drimmel} et~al.}{2023}]{dri23a}
{Drimmel} R.,  et~al., 2023, \aap, 670, A10

\bibitem[\protect\citeauthoryear{Dwek et~al.,}{Dwek et~al.}{1995}]{dwek1995morphology}
Dwek E.,  et~al., 1995, Astrophysical Journal, Part 1 (ISSN 0004-637X), vol. 445, no. 2, p. 716-730, 445, 716

\bibitem[\protect\citeauthoryear{Elia et~al.,}{Elia et~al.}{2022}]{eliA2022star}
Elia D.,  et~al., 2022, The Astrophysical Journal, 941, 162

\bibitem[\protect\citeauthoryear{Elmegreen}{Elmegreen}{1996}]{elmegreen1996pattern}
Elmegreen B.,  1996, in International Astronomical Union Colloquium. pp 197--206

\bibitem[\protect\citeauthoryear{Englmaier \& Gerhard}{Englmaier \& Gerhard}{1999}]{englmaier1999gas}
Englmaier P.,  Gerhard O.,  1999, Monthly Notices of the Royal Astronomical Society, 304, 512

\bibitem[\protect\citeauthoryear{Ewen \& Purcell}{Ewen \& Purcell}{1951}]{ewen1951observation}
Ewen H.~I.,  Purcell E.~M.,  1951, Nature, 168, 356

\bibitem[\protect\citeauthoryear{Fich, Blitz  \& Stark}{Fich et~al.}{1989}]{fich1989rotation}
Fich M.,  Blitz L.,   Stark A.~A.,  1989, Astrophysical Journal, Part 1 (ISSN 0004-637X), vol. 342, July 1, 1989, p. 272-284., 342, 272

\bibitem[\protect\citeauthoryear{Font et~al.,}{Font et~al.}{2017}]{font2017kinematic}
Font J.,  et~al., 2017, The Astrophysical Journal, 835, 279

\bibitem[\protect\citeauthoryear{Fragkoudi et~al.,}{Fragkoudi et~al.}{2019}]{fragkoudi2019ridges}
Fragkoudi F.,  et~al., 2019, Monthly Notices of the Royal Astronomical Society, 488, 3324

\bibitem[\protect\citeauthoryear{Fuchs}{Fuchs}{2001}]{fuchs2001density}
Fuchs B.,  2001, Astronomy \& Astrophysics, 368, 107

\bibitem[\protect\citeauthoryear{Fujii, B{\'e}dorf, Baba  \& Portegies~Zwart}{Fujii et~al.}{2018}]{fujii2018dynamics}
Fujii M.,  B{\'e}dorf J.,  Baba J.,   Portegies~Zwart S.,  2018, Monthly Notices of the Royal Astronomical Society, 477, 1451

\bibitem[\protect\citeauthoryear{{Fux}}{{Fux}}{1999}]{fux19993d}
{Fux} R.,  1999, \mn@doi [\aap] {10.48550/arXiv.astro-ph/9903154}, \href {https://ui.adsabs.harvard.edu/abs/1999A&A...345..787F} {345, 787}

\bibitem[\protect\citeauthoryear{Grand, Springel, G{\'o}mez, Marinacci, Pakmor, Campbell  \& Jenkins}{Grand et~al.}{2016}]{grand2016vertical}
Grand R.~J.,  Springel V.,  G{\'o}mez F.~A.,  Marinacci F.,  Pakmor R.,  Campbell D.~J.,   Jenkins A.,  2016, Monthly Notices of the Royal Astronomical Society, 459, 199

\bibitem[\protect\citeauthoryear{Gunn, Knapp  \& Tremaine}{Gunn et~al.}{1979}]{gunn1979global}
Gunn J.,  Knapp G.,   Tremaine S.,  1979, Astronomical Journal, vol. 84, Aug. 1979, p. 1181-1188., 84, 1181

\bibitem[\protect\citeauthoryear{H{\"a}nninen \& Flynn}{H{\"a}nninen \& Flynn}{2002}]{hanninen2002simulations}
H{\"a}nninen J.,  Flynn C.,  2002, Monthly Notices of the Royal Astronomical Society, 337, 731

\bibitem[\protect\citeauthoryear{H{\"a}nninen \& Flynn}{H{\"a}nninen \& Flynn}{2004}]{hanninen2004numerical}
H{\"a}nninen J.,  Flynn C.,  2004, Astronomy \& Astrophysics, 421, 1001

\bibitem[\protect\citeauthoryear{{Hernquist}}{{Hernquist}}{1990}]{her90a}
{Hernquist} L.,  1990, \apj, 356, 359

\bibitem[\protect\citeauthoryear{Hilmi et~al.,}{Hilmi et~al.}{2020}]{hilmi2020fluctuations}
Hilmi T.,  et~al., 2020, Monthly Notices of the Royal Astronomical Society, 497, 933

\bibitem[\protect\citeauthoryear{Holmberg \& Flynn}{Holmberg \& Flynn}{2000}]{holmberg2000local}
Holmberg J.,  Flynn C.,  2000, Monthly Notices of the Royal Astronomical Society, 313, 209

\bibitem[\protect\citeauthoryear{Horta et~al.,}{Horta et~al.}{2024}]{hortA2024proto}
Horta D.,  et~al., 2024, Monthly Notices of the Royal Astronomical Society, 527, 9810

\bibitem[\protect\citeauthoryear{Hou \& Han}{Hou \& Han}{2014}]{hou2014observed}
Hou L.,  Han J.,  2014, Astronomy \& Astrophysics, 569, A125

\bibitem[\protect\citeauthoryear{{Iles} et~al.,}{{Iles} et~al.}{2025}]{iles2025b}
{Iles} E.~J.,  et~al., 2025, \pasa, submitted

\bibitem[\protect\citeauthoryear{Jenkins \& Binney}{Jenkins \& Binney}{1990}]{jenkins1990spiral}
Jenkins A.,  Binney J.,  1990, Monthly Notices of the Royal Astronomical Society (ISSN 0035-8711), vol. 245, July 15, 1990, p. 305-317. Research supported by the University of Arizona., 245, 305

\bibitem[\protect\citeauthoryear{Jog \& Solomon}{Jog \& Solomon}{1984}]{jog1984two}
Jog C.~J.,  Solomon P.,  1984, Astrophysical Journal, 276, 114

\bibitem[\protect\citeauthoryear{Kataria \& Shen}{Kataria \& Shen}{2022}]{katariA2022effects}
Kataria S.~K.,  Shen J.,  2022, The Astrophysical Journal, 940, 175

\bibitem[\protect\citeauthoryear{Kataria \& Shen}{Kataria \& Shen}{2024}]{katariA2024importance}
Kataria S.~K.,  Shen J.,  2024, The Astrophysical Journal, 970, 45

\bibitem[\protect\citeauthoryear{Khoperskov, Zasov  \& Tyurina}{Khoperskov et~al.}{2003}]{khoperskov2003minimum}
Khoperskov A.,  Zasov A.,   Tyurina N.,  2003, Astronomy Reports, 47, 357

\bibitem[\protect\citeauthoryear{Kissmann, Kleimann, Fichtner  \& Grauer}{Kissmann et~al.}{2008}]{kissmann2008local}
Kissmann R.,  Kleimann J.,  Fichtner H.,   Grauer R.,  2008, Monthly Notices of the Royal Astronomical Society, 391, 1577

\bibitem[\protect\citeauthoryear{Kormendy \& Kennicutt~Jr}{Kormendy \& Kennicutt~Jr}{2004}]{kormendy2004secular}
Kormendy J.,  Kennicutt~Jr R.~C.,  2004, Annu. Rev. Astron. Astrophys., 42, 603

\bibitem[\protect\citeauthoryear{Kraljic, Bournaud  \& Martig}{Kraljic et~al.}{2012}]{kraljic2012two}
Kraljic K.,  Bournaud F.,   Martig M.,  2012, The Astrophysical Journal, 757, 60

\bibitem[\protect\citeauthoryear{Lacey}{Lacey}{1984}]{lacey1984influence}
Lacey C.~G.,  1984, Monthly Notices of the Royal Astronomical Society, 208, 687

\bibitem[\protect\citeauthoryear{Lacey \& Ostriker}{Lacey \& Ostriker}{1985}]{lacey1985massive}
Lacey C.~G.,  Ostriker J.~P.,  1985, Astrophysical Journal, Part 1 (ISSN 0004-637X), vol. 299, Dec. 15, 1985, p. 633-652., 299, 633

\bibitem[\protect\citeauthoryear{{Lawrence} et~al.,}{{Lawrence} et~al.}{2007}]{2007MNRAS.379.1599L}
{Lawrence} A.,  et~al., 2007, \mn@doi [\mnras] {10.1111/j.1365-2966.2007.12040.x}, \href {https://ui.adsabs.harvard.edu/abs/2007MNRAS.379.1599L} {379, 1599}

\bibitem[\protect\citeauthoryear{Levine, Heiles  \& Blitz}{Levine et~al.}{2008}]{levine2008milky}
Levine E.,  Heiles C.,   Blitz L.,  2008, The Astrophysical Journal, 679, 1288

\bibitem[\protect\citeauthoryear{Li, Shen, Gerhard  \& Clarke}{Li et~al.}{2022}]{li2022gas}
Li Z.,  Shen J.,  Gerhard O.,   Clarke J.~P.,  2022, The Astrophysical Journal, 925, 71

\bibitem[\protect\citeauthoryear{{\L}okas}{{\L}okas}{2021}]{lokas2021lopsided}
{\L}okas E.~L.,  2021, Astronomy \& Astrophysics, 655, A97

\bibitem[\protect\citeauthoryear{Lucey, Pearson, Hunt, Hawkins, Ness, Petersen, Price-Whelan  \& Weinberg}{Lucey et~al.}{2023}]{lucey2023dynamically}
Lucey M.,  Pearson S.,  Hunt J.~A.,  Hawkins K.,  Ness M.,  Petersen M.~S.,  Price-Whelan A.~M.,   Weinberg M.~D.,  2023, Monthly Notices of the Royal Astronomical Society, 520, 4779

\bibitem[\protect\citeauthoryear{McClure-Griffiths \& Dickey}{McClure-Griffiths \& Dickey}{2007}]{mcclure2007milky}
McClure-Griffiths N.,  Dickey J.~M.,  2007, The Astrophysical Journal, 671, 427

\bibitem[\protect\citeauthoryear{McClure-Griffiths \& Dickey}{McClure-Griffiths \& Dickey}{2016}]{mcclure2016milky}
McClure-Griffiths N.,  Dickey J.~M.,  2016, The Astrophysical Journal, 831, 124

\bibitem[\protect\citeauthoryear{McClure-Griffiths, Stanimirovi{\'c}  \& Rybarczyk}{McClure-Griffiths et~al.}{2023}]{mcclure2023atomic}
McClure-Griffiths N.~M.,  Stanimirovi{\'c} S.,   Rybarczyk D.~R.,  2023, Annual Review of Astronomy and Astrophysics, 61, 19

\bibitem[\protect\citeauthoryear{McKee}{McKee}{1990}]{mckee1990three}
McKee C.~F.,  1990, in The Evolution of the Interstellar Medium. pp 3--29

\bibitem[\protect\citeauthoryear{McKee \& Ostriker}{McKee \& Ostriker}{1977}]{mckee1977theory}
McKee C.~F.,  Ostriker J.~P.,  1977, Astrophysical Journal, Part 1, vol. 218, Nov. 15, 1977, p. 148-169., 218, 148

\bibitem[\protect\citeauthoryear{McMillan}{McMillan}{2016}]{mcmillan2016mass}
McMillan P.~J.,  2016, Monthly Notices of the Royal Astronomical Society, p. stw2759

\bibitem[\protect\citeauthoryear{Minchev \& Quillen}{Minchev \& Quillen}{2006}]{minchev2006radial}
Minchev I.,  Quillen A.,  2006, Monthly Notices of the Royal Astronomical Society, 368, 623

\bibitem[\protect\citeauthoryear{{Minniti} et~al.,}{{Minniti} et~al.}{2010}]{VVV_ref}
{Minniti} D.,  et~al., 2010, \mn@doi [\na] {10.1016/j.newast.2009.12.002}, \href {https://ui.adsabs.harvard.edu/abs/2010NewA...15..433M} {15, 433}

\bibitem[\protect\citeauthoryear{{Mitra{\v{s}}inovi{\'c}} \& {Micic}}{{Mitra{\v{s}}inovi{\'c}} \& {Micic}}{2023}]{mit23a}
{Mitra{\v{s}}inovi{\'c}} A.,  {Micic} M.,  2023, \pasa, 40, e024

\bibitem[\protect\citeauthoryear{{Monaghan}}{{Monaghan}}{1992}]{mon92a}
{Monaghan} J.~J.,  1992, \araa, 30, 543

\bibitem[\protect\citeauthoryear{Monari, Famaey, Siebert, Wegg  \& Gerhard}{Monari et~al.}{2019}]{monari2019signatures}
Monari G.,  Famaey B.,  Siebert A.,  Wegg C.,   Gerhard O.,  2019, Astronomy \& Astrophysics, 626, A41

\bibitem[\protect\citeauthoryear{Morris \& Serabyn}{Morris \& Serabyn}{1996}]{morris1996galactic}
Morris M.,  Serabyn E.,  1996, Annual Review of Astronomy and Astrophysics, 34, 645

\bibitem[\protect\citeauthoryear{Mr{\'o}z et~al.,}{Mr{\'o}z et~al.}{2019}]{mroz2019rotation}
Mr{\'o}z P.,  et~al., 2019, The Astrophysical Journal Letters, 870, L10

\bibitem[\protect\citeauthoryear{{Navarro}, {Frenk}  \& {White}}{{Navarro} et~al.}{1997}]{nav97a}
{Navarro} J.~F.,  {Frenk} C.~S.,   {White} S.~D.~M.,  1997, \apj, 490, 493

\bibitem[\protect\citeauthoryear{Ohta, Hamabe  \& Wakamatsu}{Ohta et~al.}{1990}]{ohta1990surface}
Ohta K.,  Hamabe M.,   Wakamatsu K.-I.,  1990, Astrophysical Journal, Part 1 (ISSN 0004-637X), vol. 357, July 1, 1990, p. 71-90., 357, 71

\bibitem[\protect\citeauthoryear{Ou, Eilers, Necib  \& Frebel}{Ou et~al.}{2024}]{ou2024dark}
Ou X.,  Eilers A.-C.,  Necib L.,   Frebel A.,  2024, Monthly Notices of the Royal Astronomical Society, 528, 693

\bibitem[\protect\citeauthoryear{Patsis \& Athanassoula}{Patsis \& Athanassoula}{2000}]{patsis2000sph}
Patsis P.,  Athanassoula E.,  2000, Astronomy and Astrophysics, v. 358, p. 45-56 (2000), 358, 45

\bibitem[\protect\citeauthoryear{P{\'e}rez-Villegas, Portail, Wegg  \& Gerhard}{P{\'e}rez-Villegas et~al.}{2017}]{perez2017revisiting}
P{\'e}rez-Villegas A.,  Portail M.,  Wegg C.,   Gerhard O.,  2017, The Astrophysical Journal Letters, 840, L2

\bibitem[\protect\citeauthoryear{Pfenniger, Saha  \& Wu}{Pfenniger et~al.}{2023}]{pfenniger2023five}
Pfenniger D.,  Saha K.,   Wu Y.-T.,  2023, Astronomy \& Astrophysics, 673, A36

\bibitem[\protect\citeauthoryear{Portail, Wegg, Gerhard  \& Ness}{Portail et~al.}{2017}]{portail2017chemodynamical}
Portail M.,  Wegg C.,  Gerhard O.,   Ness M.,  2017, Monthly Notices of the Royal Astronomical Society, 470, 1233

\bibitem[\protect\citeauthoryear{Quinn, Hernquist  \& Fullagar}{Quinn et~al.}{1993}]{quinn1993heating}
Quinn P.,  Hernquist L.,   Fullagar D.,  1993, Astrophysical Journal, Part 1 (ISSN 0004-637X), vol. 403, no. 1, p. 74-93., 403, 74

\bibitem[\protect\citeauthoryear{Reid et~al.,}{Reid et~al.}{2019}]{reid2019trigonometric}
Reid M.,  et~al., 2019, The Astrophysical Journal, 885, 131

\bibitem[\protect\citeauthoryear{Renaud et~al.,}{Renaud et~al.}{2013}]{renaud2013sub}
Renaud F.,  et~al., 2013, Monthly Notices of the Royal Astronomical Society, 436, 1836

\bibitem[\protect\citeauthoryear{Saha, Tseng  \& Taam}{Saha et~al.}{2010}]{sahA2010effect}
Saha K.,  Tseng Y.-H.,   Taam R.~E.,  2010, The Astrophysical Journal, 721, 1878

\bibitem[\protect\citeauthoryear{Schwarz}{Schwarz}{1981}]{schwarz1981response}
Schwarz M.,  1981, Astrophysical Journal, Part 1, vol. 247, July 1, 1981, p. 77-88., 247, 77

\bibitem[\protect\citeauthoryear{Sellwood \& Debattista}{Sellwood \& Debattista}{2009}]{sellwood2009stochasticity}
Sellwood J.,  Debattista V.~P.,  2009, Monthly Notices of the Royal Astronomical Society, 398, 1279

\bibitem[\protect\citeauthoryear{{Skrutskie} et~al.,}{{Skrutskie} et~al.}{2006}]{2MASS}
{Skrutskie} M.~F.,  et~al., 2006, \mn@doi [\aj] {10.1086/498708}, \href {https://ui.adsabs.harvard.edu/abs/2006AJ....131.1163S} {131, 1163}

\bibitem[\protect\citeauthoryear{Sofue, Honma  \& Omodaka}{Sofue et~al.}{2009}]{sofue2009unified}
Sofue Y.,  Honma M.,   Omodaka T.,  2009, Publications of the Astronomical Society of Japan, 61, 227

\bibitem[\protect\citeauthoryear{Sormani, Binney  \& Magorrian}{Sormani et~al.}{2015a}]{sormani2015gas}
Sormani M.~C.,  Binney J.,   Magorrian J.,  2015a, Monthly Notices of the Royal Astronomical Society, 449, 2421

\bibitem[\protect\citeauthoryear{Sormani, Binney  \& Magorrian}{Sormani et~al.}{2015b}]{sormani2015gas3}
Sormani M.~C.,  Binney J.,   Magorrian J.,  2015b, Monthly Notices of the Royal Astronomical Society, 454, 1818

\bibitem[\protect\citeauthoryear{Spitzer~Jr \& Schwarzschild}{Spitzer~Jr \& Schwarzschild}{1951}]{spitzer1951possible}
Spitzer~Jr L.,  Schwarzschild M.,  1951, Astrophysical Journal, vol. 114, p. 385, 114, 385

\bibitem[\protect\citeauthoryear{Spitzer~Jr \& Schwarzschild}{Spitzer~Jr \& Schwarzschild}{1953}]{spitzer1953possible}
Spitzer~Jr L.,  Schwarzschild M.,  1953, Astrophysical Journal, vol. 118, p. 106, 118, 106

\bibitem[\protect\citeauthoryear{Syer \& Tremaine}{Syer \& Tremaine}{1996}]{syer1996made}
Syer D.,  Tremaine S.,  1996, Monthly Notices of the Royal Astronomical Society, 282, 223

\bibitem[\protect\citeauthoryear{{Tepper-Garcia} et~al.,}{{Tepper-Garcia} et~al.}{2021}]{tep21v}
{Tepper-Garcia} T.,  et~al., 2021, arXiv e-prints, \href {https://ui.adsabs.harvard.edu/abs/2021arXiv211105466T} {p. arXiv:2111.05466}

\bibitem[\protect\citeauthoryear{{Tepper-Garc{\'\i}a}, {Bland-Hawthorn}, {Vasiliev}, {Agertz}, {Teyssier}  \& {Federrath}}{{Tepper-Garc{\'\i}a} et~al.}{2024}]{tep24a}
{Tepper-Garc{\'\i}a} T.,  {Bland-Hawthorn} J.,  {Vasiliev} E.,  {Agertz} O.,  {Teyssier} R.,   {Federrath} C.,  2024, \mnras

\bibitem[\protect\citeauthoryear{Teyssier}{Teyssier}{2002}]{tey02a}
Teyssier R.,  2002, Astronomy \& Astrophysics, 385, 337

\bibitem[\protect\citeauthoryear{Trick}{Trick}{2022}]{trick2022identifying}
Trick W.~H.,  2022, Monthly Notices of the Royal Astronomical Society, 509, 844

\bibitem[\protect\citeauthoryear{Trick, Fragkoudi, Hunt, Mackereth  \& White}{Trick et~al.}{2021}]{trick2021identifying}
Trick W.~H.,  Fragkoudi F.,  Hunt J.~A.,  Mackereth J.~T.,   White S.~D.,  2021, Monthly Notices of the Royal Astronomical Society, 500, 2645

\bibitem[\protect\citeauthoryear{Tuntipong et~al.,}{Tuntipong et~al.}{2024}]{tuntipong2024sami}
Tuntipong S.,  et~al., 2024, Monthly Notices of the Royal Astronomical Society, 533, 4334

\bibitem[\protect\citeauthoryear{Vallenari et~al.,}{Vallenari et~al.}{2023}]{vallenari2023gaia}
Vallenari A.,  et~al., 2023, Astronomy \& Astrophysics, 674, A1

\bibitem[\protect\citeauthoryear{{Vasiliev}}{{Vasiliev}}{2019}]{vas19a}
{Vasiliev} E.,  2019, \mn@doi [\mnras] {10.1093/mnras/sty2672}, \href {http://adsabs.harvard.edu/abs/2019MNRAS.482.1525V} {482, 1525}

\bibitem[\protect\citeauthoryear{V{\'a}zquez-Semadeni}{V{\'a}zquez-Semadeni}{2012}]{vazquez2012there}
V{\'a}zquez-Semadeni E.,  2012, European Astronomical Society Publications Series, 56, 39

\bibitem[\protect\citeauthoryear{Vislosky et~al.,}{Vislosky et~al.}{2024}]{vislosky2024gaia}
Vislosky E.,  et~al., 2024, Monthly Notices of the Royal Astronomical Society, 528, 3576

\bibitem[\protect\citeauthoryear{Wada \& Norman}{Wada \& Norman}{2001}]{wadA2001numerical}
Wada K.,  Norman C.~A.,  2001, The Astrophysical Journal, 547, 172

\bibitem[\protect\citeauthoryear{Wegg \& Gerhard}{Wegg \& Gerhard}{2013}]{wegg2013mapping}
Wegg C.,  Gerhard O.,  2013, Monthly Notices of the Royal Astronomical Society, 435, 1874

\bibitem[\protect\citeauthoryear{Wegg, Gerhard  \& Portail}{Wegg et~al.}{2015}]{wegg2015structure}
Wegg C.,  Gerhard O.,   Portail M.,  2015, Monthly Notices of the Royal Astronomical Society, 450, 4050

\bibitem[\protect\citeauthoryear{Weiland et~al.,}{Weiland et~al.}{1994}]{weiland1994cobe}
Weiland J.,  et~al., 1994, Astrophysical Journal, Part 2-Letters (ISSN 0004-637X), vol. 425, no. 2, p. L81-L84, 425, L81

\bibitem[\protect\citeauthoryear{Weinberg}{Weinberg}{1998}]{weinberg1998fluctuations}
Weinberg M.~D.,  1998, Monthly Notices of the Royal Astronomical Society, 297, 101

\bibitem[\protect\citeauthoryear{Weinberg}{Weinberg}{2024}]{weinberg2024impact}
Weinberg M.~D.,  2024, arXiv preprint arXiv:2403.12138

\bibitem[\protect\citeauthoryear{Weiner \& Sellwood}{Weiner \& Sellwood}{1999}]{weiner1999properties}
Weiner B.~J.,  Sellwood J.,  1999, The Astrophysical Journal, 524, 112

\bibitem[\protect\citeauthoryear{Wilson, Jefferts  \& Penzias}{Wilson et~al.}{1970}]{wilson1970carbon}
Wilson R.,  Jefferts K.,   Penzias A.,  1970, Astrophysical Journal, vol. 161, p. L43, 161, L43

\bibitem[\protect\citeauthoryear{Yu \& Ho}{Yu \& Ho}{2020}]{yu2020statistical}
Yu S.-Y.,  Ho L.~C.,  2020, The Astrophysical Journal, 900, 150

\bibitem[\protect\citeauthoryear{van Donkelaar, Agertz  \& Renaud}{van Donkelaar et~al.}{2022}]{van2022giant}
van Donkelaar F.,  Agertz O.,   Renaud F.,  2022, Monthly Notices of the Royal Astronomical Society, 512, 3806

\makeatother
\end{thebibliography}



\appendix

\section{Gas reactivity and instabilities in two-fluid systems}\label{sec:instabilites}

\begin{figure*}
    \centering
    \includegraphics[width=\textwidth]{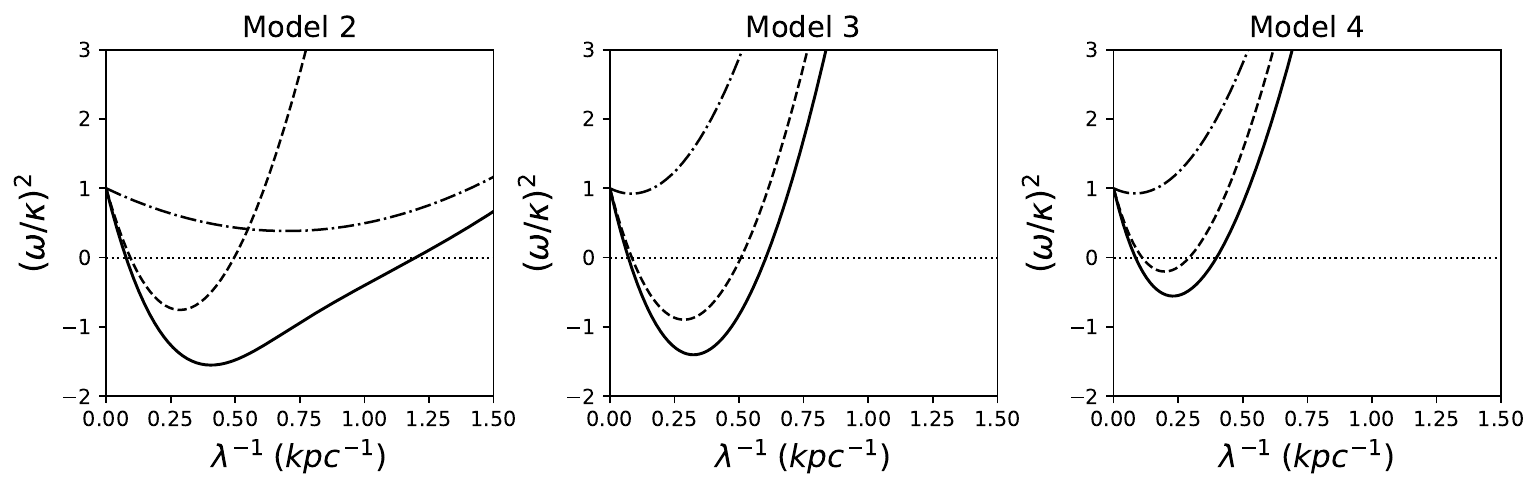}
    \caption{\added[id=anon]{Dispersion relation for Models 2, 3 and 4, in the solar neighbourhood ($R_{0} = 8.2$~kpc) at $t = 0$~Gyr. The dashed black curve is for the stars-alone fluid; dash-dot is for gas alone fluid; solid black curve is for the two-fluid system of stars and gas. The dotted line where $\omega^{2} = 0$ separates regions of stability versus instability. We can see that even when the star and gas fluid systems are each separately stable, the two fluid system of stars plus gas can be unstable.}} 
    \label{fig:dispersion_relation}
\end{figure*}

\begin{table*}
    \centering
    \caption{\added[id=anon]{Parameters measured at $t = 0$~Gyr for Models 2 - 4, within a sphere of radius 0.5~kpc, centered at $R_{0} = 8.2 \unit{kpc}$.}}
    \label{tab:dispersion_params}
    \begin{adjustbox}{max width=\textwidth}
        \begin{tabular}{l
                        S[table-format=1.2]
                        S[table-format=2.1]
                        S[table-format=2.1]
                        S[table-format=2.1]
                        S[table-format=1.3]
                        S[table-format=1.2]
                        S[table-format=1.3]
                        S[table-format=1.3]}
            \toprule
            Model 
            & {Toomre Q} 
            & {$\kappa$ [\si{km.kpc^{-1}.s^{-1}}]} 
            & {$(C_s)_{\text{planar}}$ [\si{km.s^{-1}}]} 
            & {$\Sigma_{s} [\unit{M_{\odot}pc^{-2}}]$}
            & {$h_s$ [\si{kpc}]} 
            & {$C_g$ [\si{km.s^{-1}}]} 
            & {$\Sigma_g [\si{M_{\odot}.kpc^{-2}}]$} 
            & {$h_g$ [\si{kpc}]} \\
            \midrule
            2 & 0.87 & 23.5 & 12.0 & 43.1 & 0.452 & 3.72 & 5.8 & 0.061 \\
            3 & 0.87 & 22.7 & 12.0 & 43.4 & 0.452 & 11.7 & 5.7 & 0.156\\
            4 & 0.88 & 23.3 & 12.5 & 44.5 & 0.912 & 11.7 & 5.8 & 0.191\\
            \bottomrule
        \end{tabular}
    \end{adjustbox}
\end{table*}

\added[id=anon]{
In this section, we approximate a Galactic disc as a two-fluid system of stars and gas, which gravitationally interact with each other. Then, we consider the stability of this system against local axisymmetric perturbations. This analysis is only done for Models 2-4 because we require the gas to be isothermal, and, as we have already stated, Model 1 contains multiphase, star-forming gas.}

\added[id=anon]{\citet{jog1984two} derived a dispersion relation for such a two-fluid Galactic disc system, as follows: i) Writing out the hydrodynamic equations for the two-fluid disc for an axisymmetric case (ii) Linearising these equations when considering the response of the Galactic disc to a weak external perturbation (iii) The trial solution for any perturbed physical quantity of interest (eg. surface densities and velocity dispersions) is written as the magnitude of the respective quantity multiplied by an exponential factor $\exp[i(kr + \omega t)]$ \citep{jog1984two, bland2024turbulent} (iv) 
Finally, by making use of the trial solutions, the dispersion relation for the two fluid system can be derived from the linearised hydrodynamic equations \citep{jog1984two, binney2011galactic}. The final dispersion relation for this two-fluid system, with a correction factor included for scale height is \citep{jog1984two}:}
\added[id=anon]{
\begin{equation}
    \omega^{2}(k) = \frac{1}{2} \left\{ (\alpha_{s}' + \alpha_{g}') 
    - \left[ (\alpha_{s}' + \alpha_{g}')^{2} - 4(\alpha_{s}' \alpha_{g}' - \beta_{s}' \beta_{g}') \right]^{1/2} \right\}
    \label{eq:dispersion_rel}
\end{equation}}

\added[id=anon]{
\begin{equation}
\begin{aligned}
    \alpha_{s}' &= \kappa^{2} + k^{2} C_{s}^{2} - 2 \pi G k \Sigma_{s0} \left\{ \frac{1 - \exp(-k h_{s})}{k h_{s}} \right\}, \\
    \alpha_{g}' &= \kappa^{2} + k^{2} C_{g}^{2} - 2 \pi G k \Sigma_{g0} \left\{ \frac{1 - \exp(-k h_{g})}{k h_{g}} \right\}, \\
    \beta_{s}'  &= 2 \pi G k \Sigma_{s0} \left\{ \frac{1 - \exp(-k h_{s})}{k h_{s}} \right\}, \\
    \beta_{g}'  &= 2 \pi G k \Sigma_{g0} \left\{ \frac{1 - \exp(-k h_{g})}{k h_{g}} \right\}.
\end{aligned}
    \label{equ:dispersion_rel2}
\end{equation}}

\added[id=anon]{Here, $k = 2\pi/\lambda$ is the wavenumber of the perturbation, and $\lambda$ is the wavelength. $\Sigma_{\rm s0}$ and $\Sigma_{\rm g0}$ are the stellar and gas surface densities respectively. $C_{\rm s}$ and $C_{\rm g}$ are sound speeds in the stellar versus gaseous fluids. For stars, $C_{\rm s}$ is set equal to the planar velocity dispersion. $\kappa$ is the epicyclic frequency, $2h_{\rm s}$ and $2h_{\rm g}$ are the scale heights of the star and gas discs.}

\added[id=anon]{Importantly, Equations \ref{eq:dispersion_rel} and \ref{equ:dispersion_rel2} include correction factors for the finite scale heights of the stellar and gas discs. The finite height of the discs reduces the gravitational potential in the $z = 0$ plane. Hence in Equations \ref{eq:dispersion_rel} and \ref{equ:dispersion_rel2}, we multiply the surface densities,  $\Sigma_{\rm s0}$ and $\Sigma_{\rm g0}$, by the respective reduction factors, $\mathcal{R}_\ast = \{[1 - \exp(-k h_{\rm s})]/k h_{\rm s}\}$ and $\mathcal{R}_{\rm gas} = \{[1 - \exp(-k h_{\rm g})]/k h_{\rm g}\}$, in order to take this into account.}

\added[id=anon]{Now, we give a brief interpretation of the dispersion relation given in Equation \ref{eq:dispersion_rel}. $\omega^{2}(k) = \alpha_{\rm s}'$ is the dispersion relation for a stars-alone system. The Toomre Q parameter is derived from this expression, without the reduction factor for scale height taken into account. Similarly, $\omega^{2}(k) = \alpha_{\rm g}'$ is the dispersion relation for a pure gas disc. Finally, there are terms in the overarching two-fluid dispersion relation that result from the gravitational interaction between the two fluids.}

\added[id=anon]{In Fig.~\ref{fig:dispersion_relation}, we plot the normalised angular frequency $(\omega/\kappa)^{2}$ against the inverse wavelength $\lambda^{-1}$ for $t = 0$~Gyr at $R_{0} = 8.2$~kpc, for Models 2 - 4. The parameters used to produce this figure are shown in Table \ref{tab:dispersion_params}. Since the formalism only applies on local scales, we choose to do this analysis for the solar neighborhood. The values in Table \ref{tab:dispersion_params} are computed by taking the mean of the respective parameter in a sphere of radius 0.5~kpc, centered at $R_{0} = 8.2$~kpc, for $t = 0$~Gyr.}

\added[id=anon]{The dispersion relation is key to understanding the stability of our fluid discs against local axisymmetric gravitational perturbations. When $\omega^2 > 0$, perturbations are stable and oscillatory; when $\omega^2 < 0$, they are unstable and grow exponentially. In Fig.~\ref{fig:dispersion_relation}, we mark $(\omega / \kappa)^2 = 0$ with a dotted black line to separate stable and unstable regions. The figure shows the dispersion curves for the stars-alone system (dashed), the gas-alone system (dash-dotted), and the two-fluid system (solid).}

\added[id=anon]{The curves in Fig.~\ref{fig:dispersion_relation} are computed at $t = 0$~Gyr and reflect the initial, smooth, axisymmetric state of the discs—no bar or spiral arms have formed yet. At this stage, the models are not comparable to the Milky Way, and hence we do not expect these dispersion curves to match those proposed for the MW’s solar neighbourhood. Nonetheless, Fig.~\ref{fig:dispersion_relation} helps us understand the motivation behind the design of Models 3 and 4.}

\added[id=anon]{Models 3 and 4 allow us to explore how small changes in local stability affect the long-term evolution of our simulations. The only difference between Models 2 and 3 is that we heat the gas from $T = 10^3 \; \unit{K}$ to $T = 10^4 \; \unit{K}$, increasing the sound speed from $C_{g} = 3.7$ to $11.7$~\si{km.s^{-1}}. Between Models 3 and 4, we heat the stars, increasing $\sigma_z$ from $21$ to $35$~\si{km.s^{-1}} and the stellar scale height at $R_0$ from $0.452$ to $0.912$~kpc. This setup allows us to test the effects of heating either fluid on disc stability and evolution. We anticipate that heating either component increases the system's stability.}

\added[id=anon]{In Fig.~\ref{fig:dispersion_relation}, for $t = 0$ Gyr at $R_{0}$, the stars-alone curves are unstable at some wavelengths in all models. This is consistent with Toomre $Q < 1$, before the correction for scale-height is taken into account. As the discs evolve, increases in stellar planar velocity dispersion and $\kappa$ shift the stars-alone curves into the stable regime, and the models become more Milky Way-like. Comparing models, the stars-alone curve for Model 4 lies higher, with a shorter range of unstable wavelengths than in Models 2 and 3. Similarly, increasing the gas sound speed boosts gas stability. Overall, heating either the gas or the stars also stabilises the two-fluid system against local perturbations.}

\added[id=anon]{How do these changes in gas temperature and stellar vertical velocity dispersion affect the evolution of the models? Naively, an increase in local stability at $t = 0$~ Gyr would likely slow the formation of clumps and condensations, leading one to believe that larger structures, such as bars or spiral arms, would take longer to develop. However, this is not the case. Despite being slightly more stable on local scales, Model 3 forms a stronger, shorter-lived bar compared to Model 2. This outcome is reasonable because the dispersion relation only describes axisymmetric instabilities on \textit{local} scales and does not fully capture global disc dynamics. Our current explanation for the stronger bar in Model 3 is as follows: It has long been known that when we increase the sound speed, the shock loci along the bar become less offset from the bar major axis \citep{englmaier1999gas, patsis2000sph}. As the shock loci shift closer to the bar major axis, the resulting bar becomes thinner and stronger. Thus, increasing the sound speed from $c_{\rm s} = 3.7 \; \unit{km \, s^{-1}}$ in Model 2 to $c_{\rm s} = 11.8 \; \unit{km \, s^{-1}}$ in Model 3 naturally leads to a stronger bar.}

\added[id=anon]{In Fig.~\ref{fig:dispersion_relation}, the two-fluid system of Model 4 appears significantly more stable on local scales than the other models. However, as seen in the $(x,y)$ gas density maps of Fig.~\ref{fig:xy_allsim_grid}, and later in the Fourier A$_2$ amplitudes of Fig.~\ref{fig:A2_fourier}, Model 4 takes a very long time to form any structure. The disc remains so stable that its circular velocity curve hardly evolves over the entire $4.40$~Gyr simulation; no bar forms, and only weak spiral structure ever develops.}


\bsp	
\label{lastpage}
\end{document}